\newcommand{\AP}[3]{Ann.\ Phys.\ {\bf #1},\ #2 (#3)}
\newcommand{\NPA}[3]{Nucl.\ Phys.\ {\bf A#1},\ #2 (#3)}
\newcommand{\NPB}[3]{Nucl.\ Phys.\ {\bf B#1},\ #2 (#3)}

\newcommand{\PLB}[3]{Phys.\ Lett.\ B\ {\bf #1},\ #2 (#3)}
\newcommand{\PR}[3]{Phys.\ Rep.\ {\bf #1},\ #2 (#3)}
\newcommand{\PRL}[3]{Phys.\ Rev.\ Lett.\ {\bf #1},\ #2 (#3)}

\newcommand{\PRC}[3]{Phys.\ Rev.\ C\ {\bf #1},\ #2 (#3)}
\newcommand{\PRD}[3]{Phys.\ Rev.\ D\ {\bf #1},\ #2 (#3)}
\newcommand{\JPG}[3]{J.\ Phys.\ G\ {\bf #1},\ #2 (#3)}

\newcommand{\ZPC}[3]{Z.\ Phys.\ C\ {\bf #1},\ #2 (#3)}

\newcommand{\EPJA}[3]{Eur.\ Phys.\ J.\ A\ {\bf #1},\ #2 (#3)}
\newcommand{\PTP}[3]{Prog.\ Theo.\ Phys.\ {\bf #1},\ #2 (#3)}

\newcommand\q{\theta}

\newcommand{\diracslash}[1]{#1\llap{/\kern2pt}}

\newcommand{\be}{\begin{equation}}
\newcommand{\ee}{\end{equation}}
\newcommand{\bea}{\begin{eqnarray}}
\newcommand{\eea}{\end{eqnarray}}
\newcommand{\ba}[1]{\begin{array}{#1}}
\newcommand{\ea}{\end{array}}

\documentclass[prd,aps,floats,nofootinbib,tightenlines,showpacs]{revtex4-1}
\usepackage{graphicx}
\usepackage{epsfig,graphicx,pstricks}
 \usepackage{wrapfig}
\usepackage{psfrag}
\usepackage{color}
\usepackage{amsmath}
\usepackage{amsfonts}
\usepackage{amssymb}
\usepackage{textcomp}
\usepackage{multirow}

\begin{document}

\title {Chiral symmetry breaking, color superconductivity and equation of state for
magnetized strange quark matter
}
\author{Aman Abhishek}
\email{aman@prl.res.in}
\affiliation{Theory Division, Physical Research Laboratory,
Navrangpura, Ahmedabad 380 009, India}
\author{Hiranmaya Mishra}
\email{hm@prl.res.in}
\affiliation{Theory Division, Physical Research Laboratory,
Navrangpura, Ahmedabad 380 009, India}
%\author{Amruta Mishra}
%\email{amruta@physics.iitd.ac.in}
%\affiliation{Department of Physics, Indian Institute of Technology, New 
%Delhi-110016,India}
%\affiliation{Frankfurt Institute for Advanced Studies,
%Universit\"at Frankfurt, D-60438 Frankfurt, Germany}
%\email{mishra@th.physik.uni-frankfurt.de}

\date{\today} 

\def\be{\begin{equation}}
\def\ee{\end{equation}}
\def\bearr{\begin{eqnarray}}
\def\eearr{\end{eqnarray}}
\def\zbf#1{{\bf {#1}}}
\def\bfm#1{\mbox{\boldmath $#1$}}
\def\hf{\frac{1}{2}}
\def\sl{\hspace{-0.15cm}/}
\def\omit#1{_{\!\rlap{$\scriptscriptstyle \backslash$}
{\scriptscriptstyle #1}}}
\def\vec#1{\mathchoice
        {\mbox{\boldmath $#1$}}
        {\mbox{\boldmath $#1$}}
        {\mbox{\boldmath $\scriptstyle #1$}}
        {\mbox{\boldmath $\scriptscriptstyle #1$}}
}

\begin{abstract}
We investigate the vacuum structure of dense quark matter in strong magnetic fields
at finite temperature and densities in a 3 flavor Nambu Jona Lasinio (NJL)
model including the Kobayashi-Maskawa-t'Hooft (KMT) determinant term using a variational method. The method uses
an explicit structure for the `ground' state in terms of
quark-antiquark condensates as well as diquark condensates.
The mass gap equations  and the superconducting gap equations
are solved self consistently and are used to
compute the thermodynamic potential along with charge neutrality conditions. We also derive the equation of state
for  charge neutral strange quark matter in the presence of strong magnetic fields
which could be relevant for neutron stars.
\end{abstract}


\maketitle

 \section{Introduction}
 The structure of vacuum in quantum chromodynamics (QCD) and its modification under
extreme environment has been a major theoretical and experimental
challenge in current physics \cite{review}. In particular, it is interesting to
study the modification of the structure of ground state at  high 
temperature and/or high baryon densities
as related to the nonperturbative aspects of QCD. This is important not only from a
theoretical point of view, but also for many applications to problems of
quark-gluon plasma (QGP) that could be copiously produced in
relativistic heavy ion collisions as well as for the ultra dense cold
nuclear/quark matter which could be present in the interior of compact stellar objects
like neutron stars.
In addition to hot and dense QCD, the effect of strong magnetic field
on QCD vacuum structure has attracted recent attention. This is motivated
by the possibility of creating ultra strong magnetic fields in non central collisions
at RHIC and LHC. The strengths of the magnetic fields are estimated to be of hadronic
scale \cite{larrywarringa,skokov} of the order of $eB\sim 2 m_\pi^2$ 
($m_\pi^2\simeq 10^{18}$ Gauss) at RHIC, to
about $eB\sim 15 m_\pi^2$ at LHC \cite{skokov}. 
There have been recent calculations both analytic as well as with lattice simulations,
which indicate that QCD phase diagram is affected by 
strong magnetic fields \cite{dima,maglat,fraga}.
% One of the interesting findings
%has been the  chiral magnetic effect.
%Here an electric current of quarks along the magnetic field axis is generated
%if the densities of left and right handed quarks are not equal. At high temperatures
%and in presence of magnetic field such a current can be produced locally.
%The phase structure of dense matter in presence of magnetic field
%along with a non zero chiral density has recently been investigated for two flavor
%PNJL model for high temperatures relevant for RHIC and LHC \cite{fukushimaplb}.
%There have also been many investigations to look into the vacuum structure of QCD
%and it has been recognised that the strong magnetic field acts as a catalyser
%of chiral symmetry breaking \cite{igormag,miranski,klimenko,boomsma}. 

In the context of cold dense matter, compact stars can be strongly
magnetized. Neutron star observations indicate the magnetic field to be of the order
of $10^{12}$-$10^{13}$ Gauss at the surface of ordinary pulsars  \cite{somenath}.
Further, the magnetars which are strongly magnetized
neutron stars, may have  even stronger
magnetic fields of the order of 
$10^{15}-10^{16}$ Gauss \cite{dunc,duncc,dunccc,duncccc,lat,broder,lai}.
Physical upper limit 
on the magnetic field in a gravitationally bound star is $10^{18}$ Gauss
which is obtained by comparing the magnetic and gravitational energies using
virial theorem \cite{somenath}. This limit could be higher for self bound
objects like quark stars \cite{ferrer}. Since the magnetic field strengths
are of the order of QCD scale, this can affect both the thermodynamic as well
as the hydrodynamics of such magnetized matter \cite{armendirk}.
The phase structure of dense matter in presence of magnetic field
along with a non zero chiral density has recently been investigated for two flavor
PNJL model for high temperatures relevant for RHIC and LHC \cite{fukushimaplb}.
There have also been many investigations to look into the vacuum structure of QCD
and it has been recognised that the strong magnetic field acts as a catalyser
of chiral symmetry breaking \cite{igormag,iran,miranski,klimenko,boomsma}.
The effects of magnetic field on the equation of state have been recently studied
in Nambu Jona Lasinio model at zero temperature for three flavors and
the equation of state  has been computed for the cold quark matter \cite{providencia,
amhmbhas} taking into account chiral condensate structure with quark-antiquark pair 
for the ground state.

On the other hand, color superconductivity is now an accepted 
conjectured state of cold and dense quark matter describing Cooper pairing of quarks of different
colors and different flavors \cite{wil,kriscfl}. One can have a rigorous treatment of the phenomenon of such pairing
using asymptotic freedom of QCD
at very high densities. In its simplest form, when masses of the three quarks can be neglected compared to
the chemical potential one can have 
the color flavor locked (CFL) phase\cite{wil,kriscfl}. However, to apply it to neutron star matter, the situation is more complicated
as for the densities expected in the interior of neutron star, the masses of strange quarks cannot be neglected.
Further, many nontrivial complications arise when beta equilibrium and charge neutrality conditions are 
imposed in such systems \cite{andreaskris}.
Since the well known sign problem prevents the first principle lattice simulations at finite chemical
potentials, one has to rely on effective models at this regime of moderate densities. One model that has been
extensively studied in this context has been the Nambu Jona Lasinio (NJL) model with contact interactions \cite{buballarev}. 

Of late, there has been  a lot of attention on the investigation of color superconductivity in presence of magnetic field
\cite{ferrerscmag,igormag,iran,noronah,fukuwarringa}.
Essentially, this is due to its possible application in the astrophysical situations as the densities in compact
star cores are large enough to have possible superconducting phase as well as such compact stars can have 
 strong magnetic field as mentioned above. Let us also mention here that although such systems can be color superconductors,
these phases can be penetrated by a `rotated' long range magnetic field. The corresponding rotated gauge field is a linear
combination of vacuum photon field and the 8-th gluon field\cite{gorbar2000,berges}. These rotated magnetic fields 
are not subjected to
Meissener effect. While the Cooper pair is neutral with respect to the magnetic field, the quark quasi particles
 have well defined charges. Therefore, the pairing phenomenon is affected by the presence of magnetic field. Initially, the effect of magnetic field on superconducting phase has been studied for CFL phase \cite{ferrerscmag} where all the three quarks take part in the pairing
dynamics. However, for realistic densities, such symmetric pairing is disfavored due to large strange quark mass that
leads to large mismatch in the fermi surface. The condition of charge neutrality further complicates the pairing mechanism
leading to gapless modes for homogeneous diquark pairing \cite{igorr,amhm5}. Superconductivity for the two flavor quark matter
in presence of magnetic field has been studied in Ref.s \cite{iran,digal,scoccola} within NJL model. The effect of charge 
neutrality along with the interplay of chiral and superconducting condensates has been analyzed in Ref.s\cite{digal, scoccola} 
in this model. A complete three flavor analysis of magnetized dense quark matter including superconductivity has not been attempted so far. In the present investigation we include the effects of strange quarks that takes part in chiral condensation but not in the
diquark channel in the magnetized quark matter. As we shall see, the strange quarks, similar to vanishing magnetic field case,
play an important role for charge neutral matter and the resulting equation of state. Moreover, with the inclusion of a flavor
mixing interaction term, the strange quark scalar condensate not only affects the light quark condensates but also
the diquark condensates.

We had earlier considered a variational approach to study chiral symmetry breaking
as well as color superconductivity in hot and dense matter with an explicit
structure for the `ground state' \cite{amhm5,hmspmnjl,hmparikh,hmam} with quark-antiquark condensate. The calculations 
were carried out
within NJL model with minimization of free energy density to decide which
 condensate will exist at what density and/or temperature. A nice feature of the 
approach is that the four component quark field operator in the 
chiral symmetry broken phase gets determined from the vacuum structure. 
In the present work, we aim to investigate how the vacuum structure in the 
context of chiral symmetry breaking and color superconductivity gets modified in the presence 
of magnetic field. In the context of chiral symmetry breaking, it was seen that, since the vacuum contains 
quark-antiquark pairs, the Dirac vacuum gets corrections due to the effective magnetic field apart from the
modification of the medium or the fermi sea of quarks. In our analysis we also keep these contributions to the
equation of state. 

 We organize the paper as follows. In  section II, we discuss an ansatz state
with quark-antiquark pairs related to chiral symmetry breaking, diquark and diantiquark pairs for the light
flavors related to color superconductivity in 
in the presence of a magnetic field. 
We then generalize such a state to include the effects of temperature 
and density. In section III, we consider the 3 flavor NJL model along with the so called
the Kobayashi-Maskawa-t'Hooft (KMT) term -- the six fermion determinant interaction term which breaks U(1) 
axial symmetry as in QCD. We use this Hamiltonian and calculate its 
expectation value with respect to the ansatz state
to compute the energy density as well the thermodynamic potential 
for this system. We minimize the thermodynamic potential to determine
the the ansatz functions and the resulting mass gap equations.
These coupled mass and superconducting gap equations are solved and we discuss the results in section IV. We discuss here the results with and without constraints of charge neutrality. 
Finally we summarize and conclude in section V. In the appendix we give some details of the derivation of the
evaluation of expectation values of the order parameters.

\section{ The ansatz for the ground state}

Let us first consider the ground state structure relevant for chiral symmetry breaking
in presence of strong magnetic field \cite{amhmbhas}. We shall then modify the same relevant for
color superconductivity.
To make the notations clear, we first write down the field operator
expansion for quarks with a current quark mass $m$ and charge $q$ in the momentum space in the 
presence of a constant magnetic field $\zbf B$.
We take the field direction to be along the z-axis. 
 We choose the gauge such that the electromagnetic vector potential
is given as $A_\mu(\vec x)=(0,0,Bx,0)$. The quark field operator expansion in presence of
constant magnetic field is given as
is given as \cite{amhmbhas,kausik}

\be
\psi(\zbf x) = \sum_n\sum_r\frac{1}{2\pi}\int{d\zbf p_{\omit x}\left[q_r^0(n,\zbf p_{\omit x})
U_r^0(x,\zbf p_{\omit x},n) + \tilde q_r^0(n,-\zbf p_{\omit x})V_r^0(x,-\zbf p_{\omit x},n)\right]
e^{i\zbf p_{\omit x}\cdot\zbf x_{\omit x}}}.
\label{psiexp}
\ee
The sum over $n$ in the above expansion runs from 0 to infinity.
In the above, 
$\zbf p_{\omit x}\equiv (p_y,p_z)$, and, $r=\pm 1$ denotes
the up and down spins.
We have suppressed 
the color and flavor indices of the quark field operators. 
The quark annihilation and antiquark creation operators,
$q_r^0$ and $\tilde q_r^0$,
 respectively, satisfy the quantum algebra
\begin{equation}
 \lbrace q_r^0(n,\zbf p_{\omit x}),q_{r^\prime}^{0\dag}(n^\prime,\zbf p_{\omit x}^\prime)
\rbrace =
\lbrace \tilde q_r^0(n,\zbf p_{\omit x}),\tilde q_{r^\prime}^{0\dag}(n^\prime,\zbf p_{\omit x}^\prime)\rbrace =
\delta_{rr^\prime}\delta_{nn^\prime}\delta(\zbf p_{\omit x}-\zbf p_{\omit x}^\prime).
\label{acom}
\end{equation}

In the above, $U_r$ and $V_r$ are the four component spinors
for the quarks and antiquarks respectively. The explicit forms of the spinors
for the fermions with mass $m$ and electric charge $q$ are given by
\begin{subequations}
\begin{eqnarray}
U_{\uparrow }^0(x,\vec p_{\omit x},n)&=&
\left(\begin{array}{c}
\cos \frac{\phi_0}{2}\left(\Theta(q)I_n + \Theta(-q)I_{n-1}\right)\\
0\\
\hat p_z\sin \frac{\phi_0}{2}\left(\Theta(q)I_n+\Theta(-q)I_{n-1}\right)\\
-i\hat p_\perp\sin \frac{\phi_0}{2}\left(\Theta(q)I_{n-1}+\Theta(-q)I_{n}\right)\\
\end{array}\right) \\
 U_{\downarrow}^0(x,\vec p_{\omit x},n)& =& 
\left(\begin{array}{c}
0 \\ 
\cos\frac{\phi_0}{2}\left(\Theta(q)I_{n-1}+\Theta(-q)I_n\right) \\
i\hat p_\perp\sin \frac{\phi_0}{2}\left(\Theta(q)I_n-\Theta(-q)I_{n-1}\right)
\\
-\hat p_z\sin\frac{\phi_0}{2}\left(\Theta(q)I_n-\Theta(-q)I_{n-1}\right)\\
\end{array}\right) \\
V_{\uparrow}^0(x,-\zbf p_{\omit x},n) &=& 
\left(\begin{array}{c}
\hat p_\perp\sin\frac{\phi_0}{2}
\left(\Theta(q)I_n-\Theta(-q)I_{n-1}\right)
\\ 
i\hat p_z\sin\frac{\phi_0}{2}
\left(\Theta(q)I_{n-1}+\Theta(-q)I_{n}\right)
\\
0\\
i\cos\frac{\phi_0}{2}
\left(\Theta(q)I_{n-1}+\Theta(-q)I_{n}\right)
\\
\end{array}\right)\\
V_{\downarrow}^0(x,-\vec p_{\omit x},n) &=&
\left(\begin{array}{c}
i\hat p_z\sin\frac{\phi_0}{2}
\left(\Theta(q)I_{n}+\Theta(-q)I_{n-1}\right)
\\ 
\hat p_\perp\sin\frac{\phi_0}{2}
\left(\Theta(q)I_{n-1}-\Theta(-q)I_{n-1}\right)
 \\
-i\cos\frac{\phi_0}{2}
\left(\Theta(q)I_{n}+\Theta(-q)I_{n-1}\right)
\\
0\\
\end{array}\right).
\end{eqnarray}
\label{UVs}
\end{subequations}
In the above, the energy of the n-th Landau level is given as
 $\epsilon_n=\sqrt{m^2+p_z^2+2n|q|B}
\equiv\sqrt{m^2+|\zbf p^2|}$ with $\zbf p^2=p_z^2+\zbf p_\perp^2$ so that 
$p_\perp^2=2n|q|B$, $\hat p_z=p_z/|\zbf p|$, $\hat p_\perp=2n|q|B/|\zbf p|$. In Eq.s (\ref{UVs}), $\cot\phi_0=m/|\zbf p|$.
Clearly, for vanishing masses $\phi_0=\pi/2$.
The functions
$I_n'$s (with $n\ge 0$) are functions of $\xi=|qB|(x-p_y/|qB|)$ and are given as
\be
I_n(\xi)=c_n\exp\left(-\frac{\xi^2}{2}\right)H_n(\xi)
\label{inxi}
\ee
where, $H_n(\xi)$ is the Hermite polynomial of
 the nth order and $I_{-1}=0$.  The normalization
constant $c_n$ is given by
\begin{equation*}
 c_n = \sqrt{\frac{\sqrt{|q|B}}{n!2^n\sqrt{\pi}}}
\end{equation*}
The functions $I_n(\xi)$ satisfy the orthonormality condition
\begin{equation}
 \int{d\xi I_n(\xi)I_m(\xi)} = \sqrt{|q|B}\delta_{n,m}
\label{orthoI}
\end{equation}
so that the spinors are properly normalized.
The detailed derivation of these spinors and some of their
properties are presented in the appendix of Ref.\cite{amhmbhas}.

With  the field operators now defined in terms of the annihilation and the 
creation operators
in presence of a constant magnetic field,
one can write down an ansatz for the ground state as in Ref.\cite{amhmbhas}.
 The ground state taken as a squeezed
coherent state involving quark and antiquarks pairs. Explicitly,
\cite{amhm5,hmam,hmparikh,amhmbhas} 
\begin{equation} 
|\Omega\rangle= {\cal U}_Q|0\rangle.
\label{u0}
\end{equation} 
Here, ${\cal U}_Q$ is an  unitary operator  which creates
quark--antiquark  pairs from the vacuum $|0\rangle$ which in annihilated by the
quark/antiquark annihilation operators given in Eq.(\ref{psiexp}). Explicitly,
the operator, ${\cal U}_Q$ is given as \cite{amhmbhas}
\begin{equation}
{\cal U}_Q=\exp\left(
\sum_{n=0}^{\infty} \int{d\vec p_{\omit x} 
{q_r^{0i}}^\dag(n,\vec p_{\omit x})a_{r,s}^i(n,p_z)h^i
(n,\vec p_z})
\tilde q_s^{0i}(n,-\vec p_{\omit x}) -h.c.\right)
\label{ansatz}
\end{equation}
In the above ansatz for the ground state, the function $h^i(n,p_z)$ is a real function 
describing the quark-antiquark
condensates related to the vacuum realignment for chiral symmetry breaking to be obtained from a minimization
of the thermodynamic potential. 
In the above equation, the spin dependent structure  $a_{r,s}^i$ is given by 
\begin{equation}
 a_{r,s}^i=\frac{1}{|\zbf p_i|}\left[-\sqrt{2n|q_i|B}\delta_{r,s}-ip_z\delta_{r,-s}\right]
\end{equation}
with $|\zbf p_i| = \sqrt{p_z^2+2n|q_i|B} $ denoting the magnitude 
of the three momentum of the quark/antiquark of $i$-th flavor 
(with electric charge $q_i$) in presence of a magnetic field.
 Summation over three colors is understood in the
exponent of ${\cal U}_Q$ in Eq. (\ref{ansatz}). 
Clearly, a nontrivial $h_i(n,p_z)$ breaks the chiral
symmetry.

It is easy to show that the transformation of the ground state as in Eq.(\ref{u0})
is a Bogoliubov transformation. With the ground state transforming as Eq.(\ref{u0}),
any  operator $O^0$ in the $|0\rangle$ basis  transforms as 
\be
O={\cal U}_Q O^0{\cal U}_Q ^\dag
\ee
and, in particular, one can transform the creation and annihilation operators of Eq.(\ref{psiexp})
to define the transformed  operators as above satisfying the same anticommuation relations as in Eq.(\ref{acom}).

\be
 \psi(\zbf x) = \sum_n\sum_r\frac{1}{2\pi}\int{d\zbf p_{\omit x}
\left[q_r(n,\zbf p_{\omit x})
U_r(x,n,\zbf p_{\omit x}) + \tilde q_r(n,-\zbf p_{\omit x})V_r(x,n,-\zbf p_{\omit x})\right]
e^{i\zbf p_{\omit x}\cdot\zbf x_{\omit x}}},
\label{psip}
\ee
with $q_r|\Omega\rangle=0={\tilde q_r}^{\dag}|\Omega\rangle$. In the above, we 
have suppressed the flavor and color indices.
 It is easy to see that the $U,V$ spinors are given by exactly similar 
to spinors $U_0,V_0$ in Eq.(\ref{UVs}) but with the shift of the function  $\phi_0\rightarrow \phi=\phi_0-2 h$
with the function $h(\zbf k)$ to be determined by a minimization of free energy. As we shall see later,
it is more convenient to vary $\phi(\zbf k)$ rather than $h(\zbf k)$. 
Let us note that with Eq.(\ref{psip}),
the four component quark field operator gets defined in terms of the
vacuum structure for chiral symmetry breaking given 
through Eq.(\ref{u0}) and Eq.(\ref{ansatz}) \cite{amspm,spmindianj} in presence of the magnetic field.

The chiral order parameter in the condensate vacuum $|\Omega\rangle$ can be evaluated explicitly
using the field operator expansion given in Eq.(\ref{psip}) and is given by\cite{amhmbhas}
(for $i$-th flavor)
\be
I_s^i=\langle\Omega|\bar \psi^i\psi^i|\Omega\rangle=-\frac{N_c}{(2\pi)^2}\sum_n\alpha_n|q_iB|\int dp_z \cos\phi^i
\label{isi}
\ee
This expression for the quark-antiquark condensate is exactly the same form
as derived earlier in the absence of the magnetic field \cite{hmspmnjl,hmam}
once one realizes that in presence of quantizing magnetic field with discrete
Landau levels, one has \cite{digal}
\begin{equation*}
\int \frac{d\zbf p}{(2\pi) ^3}\rightarrow 
\frac{|qB|}{(2\pi)^2}\sum_{n=0}^\infty\alpha_n\int dp_z.
\end{equation*}

Next, we would like to generalize the ansatz of Eq.(\ref{u0}) with quark-antiquark pairs in presence
of magnetic field, to include quark-quark pairs for the description of the ground state as
 relevant for color superconductivity.
However, few comments in this context are in order. It is known that in presence of
color superconductivity, the diquark is electro-magnetically charged and  the usual magnetic field will have a Meissener effect.
However, a linear combination of the photon field and the gluon field given by $\tilde A_\mu=\cos \alpha A_\mu-\sin\alpha G_\mu^8$,
still remains massless and  is unscreened. For two flavor color superconductivity, $\cos\alpha=g/\sqrt{g^2+e^2/3}
\sim 1/20$ \cite{gorbar2000}.
The electron couples to this rotated gauge field by the coupling $\tilde e=e \cos(\alpha)$.The quark field couples to the rotated gauge field through its rotated charge $\tilde Q$. In units of $\tilde e$, the rotated charge matrix in the flavor- color space is given by 
\be
\tilde Q=Q_f\otimes\bfm 1_c-\bfm 1_f\otimes \frac{T_c^8}{2\sqrt{3}}
\ee.
Thus, the $\tilde e$ charges of  red and green u quarks is $1/2$; red and green down and strange quarks is $-1/2$.
The blue u-quark has $\tilde Q$ charge as +1, while the blue d and s quarks are $\tilde Q $ chargeless.
We shall take the rotated U(1) magnetic field along the $z-$axis and spatially constant as before without
the absence of superconductivity. The ansatz for the ground state with quark-antiquark condensate is now 
taken as, with $i$ being the flavor index,
\be
|\Omega\rangle_\chi=\exp\sum_{flav}(B_i^\dagger-B_i)|0\rangle.
\label{vacchi}
\ee
The flavor dependent quark-antiquark pair creation operator for u-quark ($i=1$) is given as, with $a=1,2,3$
being the color indices for red,blue and green respectively
\be
B_u^\dagger= 
\sum_{a=1}^3\sum_{n=0}^{\infty} \int d\vec p_{\omit x} 
{q_r^{1a}(n,\vec p_{\omit x})^\dag a_{r,s}^1(n,p_z)f^{1a}
(n,\vec p_{\omit x})
\tilde q_s^{1a}(n,-\vec p_{\omit x})}
\label{upap}
\ee
while, for the down and strange quarks ($i$=2,3) the same is given as
\bearr
B_i^\dagger &=&
\sum_{a=1}^2\sum_{n=0}^{\infty} \int d\vec p_{\omit x} 
{q_r^{ia}(n,\vec p_{\omit x})^\dag a_{r,s}^i(n,p_z)h^{ia}
(n,\vec p_{\omit x})
\tilde q_s^{1a}(n,-\vec p_{\omit x})} \nonumber\\
&+&\int d\zbf p q_r^{i3}(\zbf p)^\dag\bfm(\sigma \cdot \zbf {\hat p})_{rs} h^i(\zbf p)\tilde q_s^{i3}(-\zbf p).
\label{dspap}
\eearr

The difference between the pair creation operator in Eqs.(\ref {upap}) and (\ref{dspap}) lies on the
contribution of the blue color. While the up blue quark has $\tilde Q$ charge, the blue quarks of down and strange 
 quark are $\tilde Q$
neutral.

Next, we write down the ansatz state for having quark-quark condensates which is given by
\be
|\Omega\rangle=U_d|\Omega\rangle_\chi\equiv\exp(B_d^\dag-B_d)|\Omega\rangle_\chi.
\label{ansatzdq}
\ee
In the above, $B_d^\dag$ is the diquark (and di-antiquark) creation operator given as
\be
B_d^\dag=
\sum_n\int dp_{\omit x}\left[q_r^{ia}(n,p_{\omit x})^\dag r f(n,p_z)q_{-r}^{jb}(n,-p_{\omit x},p_z)+
i\tilde q_r^{ia}(n,p_{\omit x})^\dag r f_1(n,p_z)\tilde q_{-r}^{jb}(n,p_{\omit x})^\dag \right]\epsilon^{ij3}\epsilon^{3ab}.
\label{bdiquark}
\ee
In the above, $i,j$ are the flavor indices , $a,b$ are the color indices and $r=\pm {1/2}$ are the spin indices.
The levi civita tensor ensures that the operator is antisymmetric in color and flavor space along with the fact that
only $u,d$ quarks with red and green colors take part in diquark condensation. The blue u,d quarks as well as the strange quarks
(all the three colors) do not take part in the diquark condensation. The functions $f(n,p_z)$ and $f_1(n,p_z)$
are condensate functions associated with quark-quark and antiquark-antiquark condensates respectively. These
functions are assumed to be independent of color and flavor indices. We shall give a post facto justification
for this that these function depend upon the average energy and average chemical potentials
of the quarks that condense.

To include the effects of temperature and density we next write
 down the state at finite temperature and density 
$|\Omega(\beta,\mu)\rangle$  through
a thermal Bogoliubov transformation over the state $|\Omega\rangle$ 
using the thermofield dynamics (TFD) method as described in Ref.s \cite{tfd,amph4,amhmbhas}.
This is particularly useful while dealing with operators and expectation values.
We write the thermal state as
\begin{equation} 
|\Omega(\beta,\mu)\rangle={\cal U}_{\beta,\mu}|\Omega\rangle={\cal U}_{\beta,\mu}
{\cal U}_Q |0\rangle,
\label{ubt}
\end{equation} 

where ${\cal U}_{\beta,\mu}$ is given as
\begin{equation*}
{\cal U}_{\beta,\mu}=e^{{\cal B}^{\dagger}(\beta,\mu)-{\cal B}(\beta,\mu)},
\label{ubm}
\end{equation*}
with 

\be
{\cal B}^\dagger(\beta,\mu) = 
\sum_{n=0}^{\infty} \int \Big [ d\vec k_{\omit x} 
q_r^{ia}(n,k_{\omit x})^\dagger \theta_-^{ia}(k_z,n, \beta,\mu)
\underline q_r^{ia} (n,k_{\omit x})^\dagger +
\tilde q_r^{ia} (n,k_{\omit x}) \theta_+^{ia}(k_z,n,\beta,\mu)
\underline { \tilde q}_r^{ia}(n,k_{\omit x})\Big ].
\label{bth}
\ee

In Eq.(\ref{bth}),
the underlined operators are the operators in the extended Hilbert space
 associated with thermal doubling in TFD method, and, 
 the color flavor dependent ansatz functions $\theta_{\pm}^{ia}(n,k_z,\beta,\mu)$
are related to quark and antiquark distributions as can be seen through
the minimization of the thermodynamic potential.

 All the functions in the ansatz in Eq.(\ref{ubt})
are  to be obtained by minimizing the
thermodynamic potential.
We shall carry out this minimization
in the next section. However, before carrying out the minimization
procedure, let us  
focus our attention to the expectation values of some known operators 
to show that with the above variational ansatz for the `ground state' given in
Eq.(\ref{ubt}) these reduce to the already known expressions in the appropriate
limits.
 
 Let us first consider the expectation value of the chiral order
parameter. The 
expectation value for chiral order parameter for the $i$-th flavor is given as
\be
I_s^i= \langle\Omega(\beta,\mu)|\bar\psi_i\psi_i|\Omega(\beta,\mu)\rangle=\sum_{a=1}^3 I_s^{ia}
\label{isi}
\ee
These expectation values can be evaluated easily once we realize that the state $|\Omega(\beta,\mu)\rangle$ as in Eq.(\ref{ubt})
is obtained through successive Bogoliubov transformations on the state $|0\rangle$ as in Eq.(\ref{vacchi}), Eq.(\ref{ansatzdq}).
The details of evaluation for the different order parameters  is relegated to the appendix.
Explicitly, for the quarks that take part in superconductivity 
\be
I_s^{ia}=-\sum_n\alpha_n \frac{|q^{ia}B|}{(2\pi)^2}\int d p_z\cos\phi^{ia}\left(1-F^{ia}-F_1^{ia}\right)\quad\quad(i,a=1,2)
\label{isudsc}
\ee
where, $\alpha_n=(2-\delta_{n,0})$ is the degeneracy factor of the
 $n$-th Landau level (all levels are doubly degenerate except the lowest Landau level). Further,
\be
F^{ia}=\sin^2\theta_-^{ia}+\sin^2 f \left(1-\sin^2\theta_-^{ia}-|\epsilon^{ij}|\epsilon^{ab}|\sin^2\theta_-^{jb}\right)
\label{Fia}
\ee
arising from the quarks which condense and
\be
F_1^{ia}=\sin^2\theta_+^{ia}+\sin^2 f_1 \left(1-\sin^2\theta_+^{ia}-|\epsilon^{ij}|\epsilon^{ab}|\sin^2\theta_+^{jb}\right)
\label{F1ia}
\ee
arising from antiquarks which condense. Thus, the scalar condensates arising from quarks that take part in superconductivity
depend both on the condensate functions in quark-antiquark channel ($\phi^{i}$) as well as 
in quark-quark channel ($f,f_1$). Further, the thermal functions $\sin^2\theta^{ia}_\pm$, as we shall see later, will be related to the number density distribution functions.

Next, for the non-superconducting blue up  quarks, the contribution to the scalar condensate is given by
\be
I_s^{13}=-\sum_n\alpha_n\frac{|q^{1,3}|B}{(2\pi)^2}\int dp_z\cos\phi^{13}\left(1-\sin^2\theta_-^{13}-\sin^2\theta_+^{13}\right).
\label{isu3}
\ee
Let us note that in the limit of vanishing of the color superconducting condensate functions ($f,f_1\rightarrow 0$), 
the contributions given in Eq.(\ref{isudsc}) reduce to Eq.(\ref{isu3}) as they should \cite{amhmbhas}.

Similarly, scalar condensate contribution from the charged strange quarks (red, green) is given by
\be
I_s^{3a}=-\sum_n\alpha_n\frac{|q^{3a}|B}{(2\pi)^2}\int dp_z\cos\phi^{3a}\left(1-\sin^2\theta_-^{3a}-\sin^2\theta_+^{3a}\right).
\quad\quad(a=1,2)
\label{is3a}
\ee
 Finally, for the uncharged quarks i.e. blue down and blue strange quarks, the contributions to the scalar condensates
are given by, for flavor $i$  ($i$=2,3)
\be
I_s^{i3}=-\frac{2}{(2\pi)^3}\int d\zbf k \cos\phi^i \left (1-\sin^2\theta_-^{i3}-\sin^2\theta_+^{i3}\right)
\label{isi3}
\ee

Next, we write down the condensate in the superconducting channel which is given as
\bearr
I_D &=& \langle\bar\psi_{c}^{ia} \gamma^5 \psi^{jb} \rangle \epsilon^{ij} \epsilon^{3ab}\nonumber \\ 
    &=& \frac{2}{(2\pi)^2}\sum_n \alpha_n |q_i B| \int dp_z \cos \left(\frac{\phi_1 - \phi_2}{2}\right)\bigg[ \sin 2f\left(1-\sin^2 \theta_{-}^{1}-\sin^2 \theta_{-}^{2}\right)\nonumber \\ &+& \sin2f_1 \left(1-\sin^2 \theta_{+}^{1}-\sin^2 \theta_{+}^{2}\right)\bigg]
\label{id}
\eearr
Let us note that the superconducting condensate also depends upon the  chiral condensate functions $\phi(p_z)$ through the
function $\cos\left(\frac{\phi_1-\phi_2}{2}\right)$ apart from the thermal distribution functions $\sin^2\theta^{ia}_\pm$.
 Further, this dependence vanishes when the u and d quark 
scalar condensates or equivalently the corresponding masses of the quarks are equal.

The other quantity that we wish to investigate is the axial fermion current 
density that is induced at finite chemical potential including the effect of 
temperature. The expectation value of the 
axial current density is given by
\begin{equation*}
\langle j_5^3\rangle \equiv \langle\bar{\psi_i^a}\gamma^3\gamma^5\psi_j^a\rangle.
\end{equation*}
Using the field operator expansion Eq.(\ref{psip}) and Eq.(\ref{UVs})
for the explicit forms for the spinors, we have for the $i$-th flavor
\be
\langle j_5^{i3}\rangle=
 \sum_n\frac{N_c}{(2\pi)^2} \int{dp_{\omit x}\left(I_n^2-I_{n-1}^2\right)\left(\sin^2
\theta_-^i-\sin^2\theta_+^i\right)}.
\label{j35}
\ee

Integrating over $dp_y$ using the orthonormal condition of Eq.(\ref{orthoI}), 
all the terms in the above sum for the Landau levels cancel out except for the
 zeroth Landau level so that,
\be
\langle j_5^{i3}\rangle=
\frac{N_c|q_i|B}{(2\pi)^2}\int{dp_z\left[\sin^2\theta_-^{i0}-\sin^2\theta_+^{i0}\right]}.
\label{j35p}
\end{equation}
which is identical to that in Ref.\cite{metlitsky} once we identify the functions
$\sin^2\theta_\mp^{i0}$ as the particle and the 
 antiparticle distribution functions for the zero modes (see e.g.
Eq.(\ref{them}) in the next section).

\section{Evaluation of thermodynamic potential and gap equations }
\label{evaluation}
\begin{center}
\begin{table}[h!]
\begin{tabular}{|c| c| c|}
 \hline
Quark & e-charge & $\tilde{e}$-charge \\
\hline
u-red & $\frac{2}{3}$ & $\frac{1}{2}$ \\
u-green & $\frac{2}{3}$ & $\frac{1}{2}$ \\
u-blue & $\frac{2}{3}$ & 1 \\
d-red & -$\frac{1}{3}$ & $\frac{1}{2}$ \\
d-green & -$\frac{1}{3}$ & $\frac{1}{2}$ \\
d-blue & -$\frac{1}{3}$ & 0 \\
s-red & -$\frac{1}{3}$ & $\frac{1}{2}$ \\
s-green & -$\frac{1}{3}$ & $\frac{1}{2}$ \\
s-blue & -$\frac{1}{3}$ & 0 \\
\hline
\end{tabular}
\caption{Table: List of quarks and their electromagnetic and rotated charges}
\label{table1}
\end{table}
\end{center}

As has already been mentioned, we shall consider in the present investigation,
the 3-flavor 
Nambu Jona Lasinio model including the  Kobayashi-Maskawa-t-Hooft (KMT)
determinant interaction. The corresponding
Hamiltonian density is given as \cite{buballarev,amhm5,amhmbhas,amhmdet}
\bearr
{\cal H} & = &\psi^ \dagger(-i\bfm \alpha \cdot \bfm \Pi)
+\gamma^0 \hat m )\psi\nonumber\\
&-&G_s\sum_{A=0}^8\left[(\bar\psi\lambda^A\psi)^2-
(\bar\psi\gamma^5\lambda^A\psi)^2\right]\nonumber\\
&+&K\left[{ det_f[\bar\psi(1+\gamma_5)\psi]
+det_f[\bar\psi(1-\gamma_5)\psi]}\right]\nonumber\\
&-& G_D\left[(\bar\psi\gamma^5\epsilon\epsilon_c\psi^C)(\bar\psi^C\gamma^5\epsilon\epsilon_c\psi)\right]
\label{ham}
\eearr
where $\psi ^{i,a}$ denotes a quark field with color `$a$' 
$(a=r,g,b)$, and flavor `$i$'
 $(i=u,d,s)$, indices. $\bfm \Pi=-i(\zbf\nabla-i\tilde e \bfm{\tilde A}\tilde Q)$ is the canonical
momentum in presence of the  rotated U(1) gauge field $\tilde A_\mu$. When there is no 
superconductivity $A_\mu=\tilde A_\mu$ which is the usual massless photon field with the 
coupling to the quark field being given the electromagnetic charge
$e Q_f$ where, $Q_f$ is diagonal matrix $(2/3,-1/3,-1/3)$. As mentioned in the previous section,
when  superconducting gap is  non vanishing, the massless gauge field is given by 
$ \tilde A_\mu=\cos\alpha A_\mu-\sin\alpha G_\mu^8$, 
where, $\cos\alpha=g/\sqrt{g^2+e^2/3}$. We have taken here the standard convention of $SU(3)_c$ generators in 
the adjoint representation \cite{gorbar2000}.
%E.V. Gorbar, \PRD{62,014007,2000}
The $\tilde Q$ charges of the quarks are given in Table-I.
It may also be relevant here to mention that, while we are taking into account combination of the
photon and gluon field which is massless, the other orthogonal massive component, is either Meissener screened
or nucleated into vortices \cite{sedrakianalford}.
%{sedrakianalford} M. Alford and A. Sedrakian, J Phys. G 37, 075202, 2010

 The matrix of current quark masses is given by
$\hat m$=diag$_f(m_u,m_d,m_s)$ in the flavor space.
We  shall assume  in the present investigation, isospin
symmetry with $m_u$=$m_d$.  
In Eq. (\ref{ham}), $\lambda^A$, $A=1,\cdots 8$ denote the Gellmann matrices
acting in the flavor space and
$\lambda^0 = \sqrt{\frac{2}{3}}\,1\hspace{-1.5mm}1_f$,
$1\hspace{-1.5mm}1_f$ as the unit matrix in the flavor space.
The four point interaction term $\sim G_s$ is symmetric in $SU(3)_V\times
SU(3)_A\times U(1)_V\times U(1)_A$. In contrast, the determinant term
$\sim K$ which for the case of three flavors generates a six point
interaction which breaks $U(1)_A$ symmetry. If the mass term is
neglected, the overall symmetry is $SU(3)_V\times SU(3)_A \times U(1)_V$. This
spontaneously breaks to $SU(3)_V \times U(1)_V$ implying
the conservation of the baryon number and the flavor number. The current
quark mass term introduces additional explicit breaking of chiral symmetry
leading to partial conservation of  the axial  current.
The last term in Eq.(\ref{ham}) describe a scalar diquark interaction in the color
antitriplet and flavor antitriplet channel. Such a form of four point interaction can arise e.g.
by Fierz transformation of a four point vector current-current interaction having quantum numbers of a single gluon exchange.
In that case the diquark coupling $G_D$ is related to the scalar coupling as $G_D=0.75 G_s$.

Next we evaluate the expectation value of the kinetic term in Eq.(\ref{ham}) which is
given as

\be
T=\langle\Omega(\beta,\mu)|
\psi^{ia \dag}(-i\vec\alpha\cdot\zbf\nabla-  \tilde{q}^{ia} Bx\alpha_2)\psi^{ia}|\Omega(\beta,\mu)
\rangle. \equiv\sum_{ia}T^{ia}
\label{T}
\ee
In the above the sum over the colors a and flavors $i$  is understood. The color flavor dependent
charges $\tilde q^{ia}$ for the quasi particles  is given in Table I.
To evaluate this, for non vanishing $\tilde q$ charges, we use Eq. (\ref{psip}) and the results of spatial
derivatives on the functions $I_n(\xi)$ ($\xi=\sqrt{|q_i|B}(x-{p_y}/(|q_i|B)))$. 
\begin{equation*}
\frac{\partial I_n}{\partial x} = \sqrt{|q^{ia}|B}\left[-\xi I_n 
+ \sqrt{2n}I_{n-1}\right], 
\end{equation*}
\be
\frac{\partial I_{n-1}}{\partial x} = 
\sqrt{|\tilde q^{ia}|B}\left[-\xi I_{n-1} + \sqrt{2(n-1)}I_{n-2}\right].
\ee
Using above, straightforward but somewhat tedious manipulations leads to the
contribution arising from the quarks that take part in superconductivity,
i.e. for color, flavor indices $i,a=1,2$,
\be
T^{ia}=-\sum_{n=0}^\infty \alpha_n\frac{|\tilde e B|}{2 (2\pi)^2}\int dp_z (m_i\cos\phi_i+|p_i|\sin\phi_i)(1-F^{ia}-F_1^{ia})
\quad\quad(i,a=1,2).
\label{tsc}
\ee
where, we have defined $|p_i|^2= p_z^2+2n|\tilde q B|$, ($\tilde q=\tilde e/2$). Here,
the quark-antiquark condensate effects are encoded in the function $\phi_i$ while diquark and di-antiquark condensate
effects are encoded in the functions $F^{ia}$ and $F_1^{ia}$ respectively as given in Eq.(\ref{Fia} and Eq.(\ref{F1ia}).

 For the blue u-quark, which is charged but does not take part in diquark condensation the corresponding contribution
to the kinetic term is given by
\be
T^{13}=-\sum_{n=0}^\infty\alpha_n\frac{|\tilde eB|}{(2\pi)^2}\int dp_z (m_1\cos\phi_1+|p_1|\sin\phi_1)
(1-\sin^2\theta_-^{13}-\sin^2\theta_+^{13})
\ee
 The contribution of the charged strange quarks (with charges $\tilde e/2$) to the kinetic energy  is
given by, with $a=1,2$,
\be
T^{3a}=-\sum_{n=0}^\infty\alpha_n\frac{|\tilde eB|}{2(2\pi)^2}\int dp_z (m_3\cos\phi_3+|p_3|\sin\phi_3)
(1-\sin^2\theta_-^{3a}-\sin^2\theta_+^{3a}).
\ee

Finally, the contribution from the $\tilde e $ -charge neutral quarks (blue d and blue s) is given as
\be
T^{i3}=-\int \frac{d\zbf p}{(2\pi)^3}\left(m_i\cos\phi_i+p\sin\phi_i\right)\left(
1-\sin^2\theta_-^{i3}-\sin^2\theta_+^{i3}\right)\quad \quad (i=2,3).
\ee

The contribution to the energy density from the 
the quartic interaction  term in Eq. (\ref{ham}), 
using Eq. (\ref{isi}) turns out to be, 
\begin{equation}
{V_S}\equiv -G_s \langle \Omega(\beta,\mu)|
\sum_{A=0}^8\left[(\bar\psi\lambda^A\psi)^2-
(\bar\psi\gamma^5\lambda^A\psi)^2\right]
| \Omega(\beta,\mu)\rangle
=-2 G_S\sum_{i=1,3}{I_s^i}^2,
\label{vs}
\end{equation}
where, $I_s^i=\langle\bar\psi_i\psi_i\rangle$ is the scalar quark-antiquark condensate given in Eq.(\ref{isi}).
Further, in the above, we have used the properties of the Gellman matrices $
\sum_{A=0}^8\lambda_{ij}^A\lambda_{kl}^A = 2\delta_{il}\delta_{jk}$.

Next, let us discuss the 
contribution from the six quark determinant interaction term to the
energy expectation value. There will be six terms in the expansion of the determinant, each involving three pairs of quark operators of different flavors. These are to be `contracted' in all possible manner while taking the expectation value.  
This means in the present context of having quark-antiquark and diquark condensates, 
one can contract a $\psi$ with a $\bar\psi$ or $\psi$ with a $\psi$ . The former leads to 
condensates having quark-antiquark condensates $I_s^{(i)}$ while the latter leading to diquark condensates $I_D$. 
Further, for the case of quark-antiquark condensate contributions, the contracting $\psi$ and $\bar\psi$
having the same color will lead to the dominant contribution while contracting similar operators with different colors
 will lead to a $N_c$ suppressed 
contribution. Next coming to contributions arising from the diquarks,
terms which are proportional to strange quark-antiquark condensate $\langle\bar ss\rangle$ will be dominant.
These will have the contractions of strange quark-antiquarks having the same color. The rest four terms will be
suppressed atleast by a factor $N_c$. Explicitly these two terms are given by
$\sim\sum_h\bar sO^hs\left[\bar u\hat O^hu\times(\bar d\hat O^hd)-\bar u\hat O^hd\times(\bar d\hat O^hu)\right]$,
where $h=\pm$ and $\hat O^\pm=(1\pm\gamma_5)$. When contracted diquark wise, both the terms give identical contributions,
except that the contribution of the second term will be of opposite sign as compared to the first term. This is a
consequence of flavor antisymmetric nature of the diquark condensates. This leads to
\be
V_{det} =
+K\langle{ det_f[\bar\psi(1+\gamma_5)\psi]
+det_f[\bar\psi(1-\gamma_5)\psi]}\rangle\nonumber\\
=\frac{1}{3}|\epsilon_{ijk}|I_s^{(i)}I_s^{(j)}I_s^{(k)}+\frac{K}{4}I_s^{(3)}I_D^2
\label{vdet}
\ee

Next, the contribution from the diquark interaction is given by
\be
V_D= - \langle G_D\left[(\bar\psi\gamma^5\epsilon\epsilon_c\psi^C)(\bar\psi^C\gamma^5\epsilon\epsilon_c\psi)\right]\rangle
= -G_DI_D^2
\label{vd}
\ee
where, the diquark condensate $I_D$ is already defined in Eq.(\ref{id}).

To calculate the thermodynamic potential (negative of the pressure), we also have to specify the chemical potentials
relevant for the system. Here, we shall be interested in the form of quark matter that might be present in compact stars that are
older than few minutes so that chemical equilibration for weak interaction is satisfied. The relevant chemical potentials 
in such case are the baryon chemical potential $\mu_B=3\mu_q$, the chemical potential $\mu_E$ associated 
with the electromagnetic charge, and, the color potentials $\mu_3$ and $\mu_8$. The chemical potential is a matrix
that is diagonal in color and flavor space and is given by
\be
\mu_{ij,ab}=(\mu\delta_{ij}+Q_{ij}\mu_E)\delta_{ab}+(T_{ab}^3\mu_3+T_{ab}^8\mu_8)\delta_{ij}
\label{muia}
\ee

Since, red and green color of a given flavor of quark is degenerate and the diquark is in blue direction in the
color space , we can assume $\mu_3=0$. As mentioned earlier the flavor space charge $Q\equiv diag(2/3,-1/3,-1/3)$ which couples to the electromagnetic field $A_\mu$.

The thermodynamic potential is then given by using Eq.s(\ref{T}),(\ref{vs}),(\ref{vdet}),(\ref{vd}) and
 with $s$ being the entropy density,
\be
\Omega=T+V_S+V_{det}+V_D-\langle\mu N\rangle -\frac{1}{\beta}s,
\label{Omega}
\ee

where we have introduced
\be
\langle \mu N\rangle=\langle {\psi^{ia}}^\dagger\mu_{ij,ab}\psi^{jb}\rangle
=\sum_{i,a}\mu^{ia}\rho^{ia}
\ee
where, $\rho^{ia}$ is the vector density $\rho^{ia}=\langle{\psi^{ia}}^\dagger\psi^{ia}\rangle$. For the superconducting
quarks this is given by
\be
\rho^{ia}=\sum_n\frac{\alpha_n\tilde eB}{2(2\pi)^2}\int d p_z \left(F^{ia}-F_1^{ia}\right)\quad\quad (i,a=1,2)
\ee
while, for the blue u quark, the same is given by
\be
\rho^{13}=\sum_n\frac{\alpha_n\tilde eB}{(2\pi)^2}\int d p_z \left(\sin^2\theta_-^{13}-\sin^2\theta_+^{13}\right).
\ee
For the charged strange quarks, this density is given by
\be
\rho^{3a}=\sum_n\frac{\alpha_n\tilde eB}{2 (2\pi)^2}\int d p_z \left(\sin^2\theta_-^{3a}-\sin^2\theta_+^{3a}\right) \quad
\quad (a=1,2)
\ee

For the $\tilde e $-uncharged quarks (blue down and blue strange) , the vector density is given by

\be
I_v^{i3}=\frac{2}{(2\pi)^3}\int d\zbf p\left(\sin^2\theta_-^{i3} - \sin^2\theta_+^{i3}\right).\quad\quad (i=2,3)
\ee

Finally, for the entropy density $s=\sum_{i,a}s^{ia}$ where, $s^{ia}$ is the entropy density for
quarks of flavor $i$ and color $a$. For the $\tilde e$-quarks, with charge $\tilde q^{ia}$, the phase space is
Landau quantized and we have the entropy density given as \cite{tfd}
\begin{equation}
s^{ia} = -\sum_n\frac{\alpha_n|q^{ia}|B}{(2\pi)^2}\int{dp_z
\lbrace
(\sin^2\theta_-^{ia}\ln{\sin^2\theta_-^{ia}} +
\cos^2\theta_-^{ia}\ln{\cos^2\theta_-^{ia}}) + (-\rightarrow +)\rbrace}.
\label{entropyia}
\end{equation}

On the other hand, for the uncharged (blue down and blue strange) quarks, the entropy density is given by
\be
s^{i3}=-\frac{2}{(2\pi)^3}\int d\zbf p
\lbrace
(\sin^2\theta_-^{i3}\ln{\sin^2\theta_-^{i3}} +
\cos^2\theta_-^{i3}\ln{\cos^2\theta_-^{i3}}) + (-\rightarrow +)\rbrace \quad\quad(i=2,3).
\label{entropyia1}
\ee

Thus, the thermodynamic potential is now completely defined in terms of the condensate functions $\phi^i$, $f(k)$ and the
thermal distribution functions $\theta_{\mp}^{ia}$ which will be determined through a functional extremisation
of the thermodynamic potential. Minimizing the thermodynamic potential with respect to the quark-antiquark condensate
function $\phi_i(p)$ i.e. $\delta \Omega/\delta\phi_i=0$ leads to,
\be
\cot\phi^{ia}=\frac{(m_i-4G_sI_s^i+K\epsilon^{ijk}I_s^jI_s^k+K/4 I_D^2\delta_{i3})}{|p_{ia}|}\equiv\frac{M_i}{|p_{ia}|}
\label{cotphi}
\ee
where,  as earlier, we have defined $|p_{ia}|=\sqrt{p_z^2+2n|q_{ia}| B}$ and we have defined the constituent 
quark mass $M_i=m_i-4 G_s I_s^{(i)}+K|\epsilon_{ijk}|I_s^{(i)}I_s^{(j)}I_s^{(k)}
+K/4 I_D^2\delta^{i3}$. These expressions are actually self consistent  equations for the constituent quark masses as 
scalar condensate $I_s^{(i)}$ as given in Eq.(\ref{isi}) involve $M_i$ through their dependence on $\phi_i$. Explicitly,
these mass gap equations are given as
\be
M^u=m^u-4G_sI_s^{(u)}+2KI_s^{(d)}I_s^{(s)},
\label{massu}
\ee
\be
M^d=m^d-4G_sI_s^{(d)}+2KI_s^{(u)}I_s^{(s)},
\label{massd}
\ee
\be
M^s=m^s-4G_sI_s^{(s)}+2KI_s^{(d)}I_s^{(u)}+\frac{K}{4}I_D^2,
\label{masss}
\ee

Let us note that while the color and flavor dependence on the quark-antiquark condensate functions $\phi^{ia}$ arises {\em only}
from the momentum $|p_{ia}|=\sqrt{p_z^2+2n|\tilde q_{ia}|B}$ through the 
color flavor dependent $\tilde q$ charges, the constituent quark masses are color singlets and 
are given by the solutions of the self consistent equations Eq.(\ref{massu})-Eq.(\ref{masss}). 
Further, the flavor mixing determinant interaction makes the masses of quark of a given flavor dependent
upon the condensates of the other flavor quarks. This
apart, the strange quark mass explicitly depends upon the diquark condensates through this determinant interaction.
 Note that for the two flavor
superconductivity as considered here, the strange quark mass is 
affected explicitly by the superconducting gap given by
the last term on the right hand side Eq.(\ref{masss}). Of course, there is
implicit dependence on the superconducting gap in the second term through the functions
 $F$ and $F_1$ (given in Eq.s (\ref{Fia}) and (\ref{F1ia})).
Further, when chiral symmetry is restored for the light
quarks i.e., when the scalar condensates for the non strange quarks vanish, 
still, the determinant term gives rise to a  density dependent
dynamical strange quark mass \cite{amhmdet}. Such a mass generation
is very different from the typical mechanism of quark mass generation
through quark--antiquark condensates \cite{steiner}.

In a similar manner,  minimizing the thermodynamic potential with respect to the diquark function $f(k)$ 
and di-antiquark function $f_1(k)$ i.e. $\frac{\delta \Omega}{\delta f(k)}=0$ and $\frac{\delta \Omega}{\delta f_1(k)}=0$
leads to
\be
\tan2 f(k)=\frac{2(G_D-\frac{K}{4} I_s^{(3)})I_D}{\bar\epsilon_n-\bar\mu}\cos(\frac{\phi_1-\phi_2}{2})
\equiv\frac{\Delta}{\bar\epsilon_n-\bar\mu}\cos(\frac{\phi_1-\phi_2}{2});\quad\quad\tan2 f_1(k)=
\frac{\Delta}{\bar\epsilon_n+\bar\mu}\cos(\frac{\phi_1-\phi_2}{2})
\label{2scgap}
\ee
where, we have defined the superconducting gap  $\Delta $ as
\be
\Delta=2\left(G_D-\frac{K}{4}I_s^{(3)}\right)I_D
\label{scgap}
\ee
and, $\bar\epsilon=(\epsilon_n^u+\epsilon_n^d)/2$ , 
$\bar\mu=(\mu^{ur}+\mu^{dg})/2=\mu+1/6\mu_E+1/\sqrt 3\mu_8$, where, we have used Eq.(\ref{muia}) for the chemical potentials. 
Further, $\epsilon_n^i $ is the n$^{th}$ Landau level energy for the i$^{th}$ flavor with constituent quark mass $M_i$
 given as $\epsilon_n^i=\sqrt{p_z^2+2n|q_i|B+M_i^2}$.
%Thus the superconducting gap depends upon the masses of the quarks that condense
%and is independent of the masses when the masses of the condensing quarks are the same.
It is thus seen that the diquark condensate functions depend upon
the {\em average} energy and the {\em average} chemical potential
of the quarks that condense. We also note here that the 
diquark condensate functions depends upon the masses of the two quarks which
condense through the function $\cos \big ((\phi_1-\phi_2)/2\big )$.
The function $\cos\phi_i=M_i/\epsilon_n^i$, 
can be different for u,d quarks,
when the charge neutrality condition is imposed. 
Such a normalization factor is always there when
the condensing fermions have different masses as has been noted in Ref.
\cite{aichlin} in the context of CFL phase. 

Finally, the minimization of the thermodynamic potential with respect to the
thermal functions $\theta_{\pm}^{ia}(\zbf k)$ gives
\be
\sin^2\theta_\pm^{ia}=\frac{1}{\exp(\beta(\omega_{i,a}\pm\mu_{ia}))+1},
\label{them}
\ee
 \noindent Various $\omega^{ia}$'s $(i,a\equiv {\rm {flavor,color}})$
are explicitly  given as
\begin{mathletters}
\be
\omega_{n \pm}^{11} =
\omega_{n \pm}^{12}=\bar\omega_{n \pm} +\delta\epsilon_n\pm \delta_\mu\equiv \omega_{n \pm}^u
\label{omgpmu}
\ee
\be
\omega_{n \pm}^{21} = \omega_{n \pm}^{22} =
\bar\omega_{n \pm}-\delta\epsilon_n\mp\delta_\mu\equiv\omega_{n \pm}^d
\label{omgpmd}
\ee
for the quarks participating in condensation. Here,
$\bar{\omega_{n \pm}}=\sqrt{(\bar \epsilon_n\pm \bar\mu)^2+\Delta^2 \cos^2(\phi_1-\phi_2)/2}$.
Further, $\delta\epsilon_n= (\epsilon_n^u-\epsilon_n^d)/2$ is half the energy difference between the quarks which condense
in a given Landau level 
and $\delta \mu=(\mu_{ur}-\mu_{dg})/2= \mu_E/2$ is half the difference between the chemical potentials
of the two condensing quarks.
 For the charged quarks which do not participate in the superconductivity,
\be
\omega_{n \pm}^{ia}=\epsilon^i_n{\pm}\mu^{ia}.
\label{disps}
\ee

\end{mathletters}
 In the above, the upper sign corresponds to
antiparticle excitation energies while the lower sign corresponds to the particle excitation energies.

Let us note that when the charge neutrality conditions are not imposed, the masses of u and d quarks will be almost the
 same but for the effect of the (rotated) magnetic field as the magnitude of the charges
for red and green quarks are the same and that of the blue color is different. Since the chemical potentials of all the quarks are the same when charge neutrality is not imposed, 
all the four quasi particles
taking part in diquark condensation  will have (almost) the
same energy $\bar\omega_{n-}$. On the other hand, when charge neutrality condition is imposed, it is clear from the
dispersion relations given in Eq.(\ref{omgpmu}), (\ref{omgpmd}) that  
it is possible to have zero modes, i.e., $\omega^{ia}=0$
depending upon the values of $\delta\epsilon_n$
and $\delta \mu$. So, although we shall have nonzero order
parameter $\Delta$, there will be fermionic zero modes or the 
gapless superconducting phase \cite {abrikosov, krischprl}. 

Substituting the solutions for the quark-antiquark condensate function $\phi^i$ of Eq.(\ref{cotphi}), we have the
solutions for the different quark-antiquark condensates i.e. $I_s^{ia}$ given by, using
equations  Eq.(\ref{isudsc}), Eq.(\ref{isu3}) and Eq.(\ref{is3a}),
\be
I_s^{ia}=-\sum_n\frac{\alpha_n}{(2\pi)^2}(\tilde eB/2)\int dp_z\frac{M_i}{\sqrt{p_z^2+2n(\tilde e B/2)+M_i^2}}
\left(1-F^{ia}-F_1^{ia}\right)
\quad\quad(i,a=1,2)
\label{isia1}
\ee
\be
I_s^{13}=-\sum_n\frac{\alpha_n}{(2\pi)^2}(\tilde eB)\int dp_z\frac{M_1}{\sqrt{p_z^2+2n(\tilde e B)+M_1^2}}
\left(1-\sin^2\theta_-^{13}-\sin^2\theta_+^{13}\right)
\label{is131}
\ee
\be
I_s^{3a}=-\sum_n\frac{\alpha_n}{(2\pi)^2}(\tilde eB/2)\int dp_z\frac{M_3}{\sqrt{p_z^2+2n(\tilde e B/2)+M_3^2}}
\left(1--\sin^2\theta_-^{3a}-\sin^2\theta_+^{3a}\right)
\quad\quad(a=1,2)
\label{is3a1}
\ee
for the $\tilde e $ charged quarks while for the uncharged quarks (blue d and blue strange quarks),
\be
I_s^{i3}=-\frac{2}{(2\pi)^3}\int d\zbf p\frac{M_i}{i\sqrt{\zbf p^2+M^2_i}}\left(1-\sin^2\theta_-^{i3}-\sin^2\theta_+^{i3}\right)
\quad\quad (i=2,3)
\label{isi31}
\ee
Similarly, substituting the solutions for the diquark /di-antiquark condensate functions from Eq.(\ref{2scgap})
in Eq. (\ref{id}), we have, with the usual notations,$\bar\xi_{n\pm}=\bar\epsilon_n\pm\bar\mu$ and
$\bar\omega_{n\pm}=\sqrt{\xi_{n\pm}^2+\Delta^2\cos^2(\phi_1-\phi_2)/2}$,
\be
I_D =  
    \frac{2}{(2\pi)^2}\sum_n \alpha_n |\tilde eB/2| \int dp_z \Delta \cos^2 \left(\frac{\phi_1 - \phi_2}{2}\right)
\bigg[ \frac{1}{\bar\omega_{n-}}\left(1-\sin^2 \theta_{-}^{1}-\sin^2 \theta_{-}^{2}\right)+ \frac{1}{\bar\omega_{n+}} 
\left(1-\sin^2 \theta_{+}^{1}-\sin^2 \theta_{+}^{2}\right)\bigg]
\label{id1}
\ee
Thus Eq.s(\ref{massu})- (\ref{masss}) for the mass gaps, Eq.(\ref{scgap}) for the superconducting gap
and Eq.s (\ref{isia1})-(\ref{id1}) define the self consistent mass 
gap equation for the $i$-th quark flavor and the superconducting gap .

Next we discuss the thermodynamic potential.
We substitute the solutions for the condensate functions Eq.(\ref{cotphi}), Eq.(\ref{2scgap})
in the expression for the thermodynamic potential Eq.(\ref{Omega}) and use the gap equations Eq.s(\ref{massu})-(\ref{masss})
and Eq.(\ref{scgap}). The thermodynamic potential is then given by
\be
\Omega_q=\Omega_{1/2}^{sc}+\Omega_{1/2}^s+\Omega_0+\Omega_1+4 G_s\sum_i{I_s^i}^2-4KI_s^uI_s^dI_s^s+\frac{\Delta^2}{4G_D'}
-\frac{K}{4}{I_s^s}I_D^2
\label{thpot}
\ee
where, we have defined, an effective diquark coupling $G_D'=G_D-\frac{K}{4I_s^s}$ in presence of
the determinant term which mixes the flavors. Let us now discuss each of the terms in Eq.(\ref{thpot}).The first term
is the contribution from the quarks that take part in superconductivity i.e. the red and blue, u,d quarks. This
 contribution is given by
\bearr
\Omega_{1/2}^{sc}&=&-2\sum_n\frac{\alpha_n(\tilde eB/2)}{(2\pi)^2}
\int(\epsilon_n^u+\epsilon_n^d) dp_z\nonumber\\
 &+& 2\sum_n\frac{\alpha_n(\tilde eB/2)}{(2\pi)^2}\int \left((\bar\xi_{n-}+\bar\xi_{n+})
-(\bar\omega_{n-}+\bar\omega_{n+})
\right)\nonumber\\
&-2& \sum_n\sum_{i=u,d}\frac{ 2\alpha_n(\tilde eB)/2}{(2\pi)^2\beta}\int dp_z\left[\log(1+\exp(-\beta(\omega_{n-}^i-\mu_{ir})))
+\log(1+\exp(-\beta(\omega_{n+}^i+\mu_{ir})))\right]
\nonumber\\
&\equiv&\Omega_{1/2,0}^{sc}(T=0,\mu=0)+\Omega_{1/2,med}^{sc}(T,\mu)
\label{Omgsc}
\eearr
where, we have separated the contribution of the medium $\Omega_{1/2,med}^{sc}$ from $T=0,\mu=0$ contribution.
Similarly, the ($\tilde e$) charged
strange quark contribution
to the thermodynamic potential is given by
\bearr
\Omega_{1/2}^{s}&=&-2\sum_n\frac{\alpha_n(\tilde eB)/2}{(2\pi)^2}\int (\epsilon_n^s)\nonumber\\
&-& \sum_n\sum_{a=1,2}\sum_{s=\pm 1}
\frac {\alpha_n(\tilde eB)/2}{(2\pi)^2\beta}\int dp_z\left[\log(1+\exp(-\beta(\omega_{3a}+s\mu_{ia})\right]\nonumber\\
&\equiv&\Omega_{1/2,0}^{s}+\Omega_{1/2,med}^{s}
\label{Omegas}
\eearr
 The term $\Omega_1$ in Eq.(\ref{thpot}) arises from the blue colored u- quark with charge $\tilde e$ and is given as
\be
\Omega_1=-\sum_n\frac{\alpha_n(\tilde eB)}{(2\pi)^2}\int (\epsilon_n^u)\nonumber\\
- \sum_n\sum_{s=\pm 1}
\frac {\alpha_n(\tilde eB)}{(2\pi)^2\beta}\int dp_z\left[\log(1+\exp(-\beta(\omega_{33}+s\mu_{33})\right]
\equiv\Omega_{1,0}^{u}+\Omega_{1,med}^{u}
\label{Omegau3}
\ee
 Finally, the $\tilde e$ uncharged quarks' contributions to the thermodynamic potential $\Omega_0$ is given by
\be
\Omega_0=-2\sum_{i=2,3}\int \frac{d\zbf p}{(2\pi^3)}\epsilon^i(\zbf p) -\frac{2}{(2\pi)^3\beta}\int d\zbf p
\sum_{s=\mp 1}\left[\log(1+\exp(-\beta(\omega_{23}+s\mu_{33})\right]
\label{thpotch0}
\ee

Now, all the zero temperature and zero chemical potential contributions of the thermodynamic potential in Eq.s(\ref{Omgsc})-
(\ref{thpotch0}) are ultraviolet divergent. This divergence also gets transmitted 
to the gap equations  through the quark-antiquark as well as diquark condensates in
equations Eq.(\ref{isia1}), Eq.(\ref{is131}),Eq.(\ref{is3a1})and Eq.(\ref{id1}).
For the chargeless case, these can be rendered finite through a regularization with a sharp cut off in the magnitude of
three momentum as is usually done in the NJL models. However, it is also seen that a sharp cutoff in the presence of magnetic field for charged particles suffers from cut-off artifacts since the continuous momentum dependence in two  spatial dimensions are 
replaced by sum over discrete Landau levels. To avoid this, some calculations use a smooth
parametrisation for the cutoff as e.g. in Ref.\cite{fukushimaplb}
. In the present work  however we follow the elegant
procedure that was followed in Ref. \cite{providencia} by adding and subtracting
a vacuum (zero field) contribution to the  thermodynamic potential which is also
divergent. This manipulation makes e.g. the Dirac vacuum contribution in presence of magnetic field
to a physically more appealing form by separating the same to a zero field vacuum contribution
and a finite field contribution written in terms of Riemann-Hurwitz $\zeta$ function. The vacuum  contribution
to the energy density 
arising from  a charged quark can be written as \cite{amhmbhas,providencia},
\bearr
 &&-\sum_{n=0}^{\infty}
\frac{\alpha_n|q_iB|}{(2\pi)^2}\int{dp_z}\sqrt{p_z^2+2n|q_i|B+M_i^2}\nonumber\\
&=& -\frac{2}{(2\pi)^3}\int{d\zbf p\sqrt{\zbf p^2 + M_i^2}}\nonumber\\
 &-& \frac{1}{2\pi^2}|q_iB|^2\left[\zeta^\prime(-1,x_i) 
- \frac{1}{2}(x_i^2-x_i)\ln{x_i} + \frac{x_i^2}{4}\right],
\label{t1}
\eearr
where, we have defined the dimensionless quantity,
$x_i = \frac{M_i^2}{2|q_iB|}$, i.e. the mass parameter in units of the magnetic field.
Further, $\zeta^\prime(-1,x)=d\zeta(z,x)/dz|_{z=1} $ is the derivative of the 
Riemann-Hurwitz zeta function \cite{hurwitz}.

Using Eq.(\ref{t1}), the quark-antiquark condensate of  ($\tilde q$) charged quarks can be written as
\bearr
\langle \bar\psi^{ia}\psi^{ia}\rangle &=& -\frac{2}{(2\pi)^3}\int d\zbf p \frac{M_i}{\sqrt{\zbf p^2+M_i^2}}\nonumber\\
&-& \frac{M_i|q_iB|}{2\pi^2}\left[x_i(1-\ln x_i)+\ln\Gamma(x_i)+\frac{1}{2}\ln(\frac{x_i}{2\pi})\right]+{I_s^{ia}}_{med}\nonumber\\
&\equiv& {I_s^{ia}}_{vac}+{I_s^{ia}}_{field}+{I_s^{ia}}_{med}
\label{is}
\eearr
The first term, ${I_s^{ia}}_{vac}$ can be explicitly evaluated with a cutoff $\Lambda$ as

\be
{I_s^{ia}}_{vac}=\frac{M_i}{2\pi^2}\left[\Lambda \sqrt{\Lambda^2+M_i^2}-M_i^2\log\left(\frac{\Lambda+\sqrt{\Lambda^2+M_i^2}}{M_i}
\right)\right].
\label{isivac}
\ee

The medium contribution to the scalar condensate from the  superconducting part is 
\be
{I_s^{ia}}_{med}=\sum_n\frac{\alpha_n(\tilde e B/2)}{(2\pi)^2}\int dp_z\frac{M_i}{\epsilon_n^i}\left (F^{ia}-F_1^{ia}\right),
\ee
while, for the non superconducting  blue  u-quarks,
\be
{I_s^{13}}_{med}=\sum_n\frac{\alpha_n(\tilde e B)}{(2\pi)^2}\int dp_z\frac{M_1}{\epsilon_n^1}
\left(\sin^2\theta_-^{13}-\sin^2\theta_+^{13}\right).
\ee
Similarly, the contribution of the medium to the ($\tilde q$) charged strange quark-antiquark condensate is
\be
{I_s^{3a}}_{med}=\sum_n\frac{\alpha_n(\tilde e B/2)}{(2\pi)^2}\int dp_z\frac{M_3}{\epsilon_n^3}
\left(\sin^2\theta_-^{3a}-\sin^2\theta_+^{3a}\right)\quad (a=1,2)
\ee

In what follows, we shall focus our attention to zero temperature calculations.
Using the relation
 $\lim_{\beta \rightarrow\infty}\frac{1}{\beta}
\ln(1+\exp(-\beta\omega))=-\omega\theta(-\omega)$ and using Eq.s(\ref{Omgsc}), Eq.(\ref{t1}),
we have the zero temperature thermodynamic potential for the color superconducting quarks given as

\be
\Omega_{1/2}^{sc}(T=0,\mu,B)= \Omega_{1/2,0}^{sc}(T=0,\mu=0)+\Omega_{1/2,med}^{sc}(T=0,\mu)
\ee
with,

\be
\Omega_{1/2,0}^{sc}(T=0,\mu=0)=-2 \times 2 \sum_{i=u,d} G(\Lambda,M_i)-2\sum_{i=u,d}F(x_i,B)
\label{omgsc1}
\ee
where we have defined the function $G(\Lambda,M)$ as 
\bearr
G(\Lambda,M)&=&\frac{1}{(2\pi)^3}\int \sqrt{\zbf p^2+M^2} d\zbf p\nonumber\\
&=&\frac{1}{16\pi^2}\left[\Lambda\sqrt{\Lambda^2+M^2}(2\Lambda^2+M^2)-M^4\log\left(\frac{\Lambda+\sqrt{\Lambda^2+M^2}}{M}
\right)\right].
\eearr
The prefactors in the first term correspond to color and spin degeneracy factors while the same in the second term
correspond to the color degeneracy factor.
The magnetic field dependent function, $F(x_i,B)$ with $x_i=M_i^2/|q_iB|$,
\be
 F(x_i)=\frac{1}{2\pi^2}|q_iB|^2\left[\zeta^\prime(-1,x_i) 
- \frac{1}{2}(x_i^2-x_i)\ln{x_i} + \frac{x_i^2}{4}\right],
\label{fieldvac}
\ee

The medium contribution from the superconducting quarks is given as 
\bearr
\Omega_{1/2,med}^{sc}(T=0,\mu) &=& 2  \sum_{n=0}^{n_{max}} \frac{\alpha_n (\frac{\tilde{e}B}{2})}{(2\pi^2)} 
\int_0^{p_{zn}^{max}} dp_z \left[\bar{\xi}_{n-}+\bar{\xi}_{n+}-(\bar{\omega}_{n-}+\bar{\omega}_{n+})\right]\nonumber\\
&+&
2  \sum_{n=0}^{n_{max}} \sum_{i=u,d} \frac{\alpha_n(\frac{\tilde{e}B}{2})}{2\pi^2} \int_0^{p_{zn}^{max}} 
dp_z i\left[\omega_{n-}^i\theta (-\omega_{n-}^i)+\omega_{n+}^i\theta (-\omega_{n+}^i)\right]  .
\label{omgsc2}
\eearr
The three momentum cutoff $\Lambda$ for the magnitude of momentum in the absence of magnetic field leads to the
sum over the Landau level upto $n_{max}=\frac{\Lambda^2}{\tilde{e}B}$. Futher, the positivity of the magnitude of $p_z$,
restricts the cutoff in $|p_z|$ as 
$p_{z,max}^n=\sqrt{\Lambda^2-n\tilde{e}B}$  for a given value of $n$ of the Landau level.

The contribution of the blue up quark to the thermodynamic potential $\Omega_1=\Omega_{1,0}+\Omega_{1,med}$ with 
\be
\Omega_{1,0}(T=0,\mu=0)=-2 G(\Lambda,M_u)- F(x_u,B)
\label{omgub1}
\ee
and 
\be
\Omega_{1,med}(T=0,\mu)=\sum_{n=0}^{n_{max}^u} \frac{\alpha (\tilde{e}B)}{(2\pi^2)}  
\left[\mu_{ub}\sqrt{\mu_{ub}^2-M_{nu}^2}+M_{nu}^2 \log\left(\frac{\mu_{ub}+\sqrt{\mu_{ub}^2-M_{nu}^2}}{M_{nu}}\right)\right]
\label{omgub2}
\ee
where $M_{nu}=\sqrt{M_u^2+2 n \tilde{e}B}$ is the $n^{th}$ Landau level mass for up quark and 
$n_{max}^u=Int\left[\frac{\mu_{ub}^2-M_u^2}{2\tilde eB}\right] $
is the maximum number of Landau level consistent with the zero temperature distribution function.

The $\tilde{e}$ charged strange quark contribution to the thermodynamic potential 
$\Omega_{1/2}^s=\Omega_{1/2,0}^s+\Omega_{1/2,med}^s$, with
\be
\Omega_{1/2,0}^s(T=0,\mu=0)=-2\times 2 G(\Lambda,M_s)- 2 F(x_s,B)
\label{omgsrg1}
\ee
and 
\be
\Omega_{1/2,med}(T=0,\mu)= 2  \sum_{n=0}^{n_{max}^s} \frac{\alpha (\frac{\tilde{e}B}{2})}{(2\pi^2)} 
 \left[\mu_{sr}\sqrt{\mu_{sr}^2-M_{ns}^2}+M_{ns}^2 \log\left(\frac{\mu_{sr}+\sqrt{\mu_{sr}^2-M_{ns}^2}}{M_{ns}}\right)\right]
\label{omgsrg2}
\ee
where, $M_{ns}=\sqrt{M_s^2+2 n \tilde{e}B}$ is the $n^{th}$ Landau level mass for the s-quarks.
Further, the sum over the Landau levels is restricted to $n_{max}^s=Int\left[\frac{\mu_{sr}^2-M_s^2}{\tilde eB}\right]$
 arising from the distribution function at zero temperature $\theta(\mu-\epsilon_n)$ .

For the uncharged quarks, i.e. blue down and strange quarks we have, $\Omega_0=\Omega_{0,0}+\Omega_{0,med}$ with
\be
\Omega_{0,0}(T=0,\mu=0)=-2  \sum_{i=d,s} G(\Lambda,M_i)
\label{omgds01}
\ee
and  for the medium part, with $p_{fi}=\sqrt{\mu_i^2-M_i^2}$,
\be
\Omega_{0,med}(T=0,\mu)=2  \sum_{i=d,s} H_i(\mu_{i3},p_{fi}).
\label{omgds02}
\ee

In the above $H_i$ is the medium contribution from a single charge less flavor given as 
\be
H_i(\mu,p_f)= \frac{1}{16\pi^2}\left[p_{fi}\mu_{i}(p_{fi}^2+\mu_i^2)-M_i^4 \log \left(\frac{\mu^i+p_{fi}}{M^i}\right)
\right]
\label{H1}
\ee

Next, we write down the expressions for the condensates at zero temperature, that are needed to compute the 
thermodynamic potential in Eq.(\ref{thpot}). This is already given by Eq.(\ref{is}). Here, we write down
explicitly the zero temperature limit for the same.
 The scalar condensate for,say, u-quarks is given as
\be
I_s^u={I_s^u}_{vac}+{I_s^{ur}}_{med}+{I_s^{ug}}_{med}+{I_s^{uba}}_{med}+\sum_{a=1}^3 I_s^{field-u}(x_{ua})
\ee
The vacuum contribution ${I_s^u}_{vac}$ is already given in Eq.(\ref{isivac}).

The scalar condensate medium contribution from the superconducting red up and green up quarks are given as
\be
{I_s^{ur}}_{med}={I_s^{ug}}_{med}=-\sum_{n=0}^{n_{max}}\frac{\alpha_n(\tilde e B/2)}{(2\pi)^2}\int dp_z \frac{M_u}{\epsilon_n^u}\left(F^{ur}-F_1^{ur}\right)
\ee
The expressions for the distribution functions $F^{ia}$ and $F_1^{ia}$ is already given in Eq.s (\ref{Fia})-(\ref{F1ia})
in terms of the diquark condensate functions and the thermal distribution functions.In the zero temperature limit,
the distribution functions for e.g. u- quarks become
\be
F^{ur}= \frac{1}{2}\left(1-\frac{\bar\xi_{n-}}{\bar\omega_{n-}}\right)\left(1-\theta(-\omega^d)\right)
\label{fur}
\ee
and
\be
F_1^{ur}=\frac{1}{2}\left(1-\frac{\bar\xi_{n+}}{\bar\omega_{n+}}\right).
\label{f1ur}
\ee

The blue up quark contribution to the scalar condensate is given by
\be
{I_s^{ub}}_{med}=
- \sum_{n=0}^{n_{max}^u}\frac{2M\alpha_n\tilde e B}{(2\pi)^2}\log\left(\frac{p_z^{max}+\sqrt{{p_z^{max}}^2+M_{nu}^2}}{M_{nu}}\right)
\ee
As in Eq.(\ref{omgub2}) here we have defined  the n-th Landau level mass for the blue up quark as
$M_{nu}^2=M_u^2+2n|\tilde e B|$. The magnetic field contribution to the scalar condensate for the up quarks of a
given color `$a$' is given by
\be
I_s^{field-u}(x_{ua})=
- \frac{M_u|q_aB|}{2\pi^2}\left[x_a(1-\ln x_a)+\ln\Gamma(x_a)+\frac{1}{2}\frac{x_a}{2\pi}\right]
\ee
where, $x_a=M_u^2/2|q_aB|$ and $q_a=1/2\tilde e $ for red and green colors and $\q_a=\tilde e$ for blue color up quarks.

In an identical manner, the scalar condensates for the down and strange quarks $I_s^d, I_s^s$ can be written down with appropriate
changes for the  charges and the masses. The diquark condensate $4I_D$ is given in Eq.(\ref{id1}) where the zero temperature
limit can be taken by replacing the distribution functions $\sin^2\theta^i=\theta (-\omega^i)$, $(i=u,d)$. 
Thus  the thermodynamic potential gets completely defined for the quark matter in presence of magnetic field.

In the context of neutron star matter,
the quark phase that could be present in the interior,
consists of the u,d,s quarks as well as electrons, 
in weak equilibrium 
\begin{subequations}
\be
d\rightarrow u+e^-+\bar\nu_{e^-},
\end{equation}
\be
s\rightarrow u+e^-+\bar\nu_{e^-},
\ee
and,
\begin{equation}
s+u \rightarrow d+u,
\end{equation}
\end{subequations}
leading to the relations between the chemical potentials 
$\mu_u$, $\mu_d$, $\mu_s$, $\mu_E$
as
\be
\mu_s=\mu_d=\mu_u+\mu_E.
\ee
 The neutrino chemical potentials are taken to be zero as they can diffuse out
of the star. So there are {\em two} independent chemical potentials needed
to describe the matter in the neutron star interior which we take to be the
quark chemical potential $\mu_q$
and the electric charge chemical
potential, $\mu_e$ in terms of which the chemical potentials are given by
$\mu_s=\mu_q-\frac{1}{3}\mu_e =\mu_d$, $\mu_u=\mu_q+\frac{2}{3}\mu_e$ and
$\mu_E=-\mu_e$. In addition, for description of the charge neutral matter, there is a
further constraint for the chemical potentials through the following relation for the
particle densities given by
\be
Q_E=\frac{2}{3}\rho_u-\frac{1}{3}\rho_d-\frac{1}{3}\rho_s-\rho_E=0.
\label{neutrality}
\ee
The color neutrality condition corresponds to
\be
Q_8=\frac{1}{\sqrt{3}}\sum_{i=u,d,s}\left(\rho^{i1}+\rho^{i2}-2\rho^{i3}\right)=0
\label{colneutrality}
\ee
In the above, $\rho^{ia}$ is the number density for quarks of flavor $i$ and color $a$.
In particular, the number densities of the condensing quarks are given as
\be
\rho^{ia}=\frac{1}{(2\pi)^2}\sum_n\frac{\tilde eB}{2}\int dp_z (F^{ia}-F_1^{ia}),(i,a=1,2)
\ee
where, $F^{ia}, F_1^{ia}$ are defined in Eq.s (\ref{Fia}), and Eq.(\ref{F1ia}) respectively in terms of
the condensate functions and e.g. for zero temperature is given explicitly in Eq. (\ref{fur}) for up red quarks.
For the blue colored quarks, the same for the up blue quarks is given by
\be
\rho^{ub}=\frac{1}{2\pi^2}\sum_{n=0}^{n_{max}^u}\alpha_n\tilde eB\sqrt{\mu_{ub}^2-M_u^2-2n\tilde eB}
\ee
while, for the $\tilde e$ uncharged d quarks,
\be
\rho^{db}=\frac{(\mu_{db}^2-M_d^2)^{3/2}}{3\pi^2}
\ee

For the charged strange quarks the number densities are given by
\be
\rho^{sr}=\rho^{sg}=\frac{1}{(2\pi)^2}\sum_{n=0}^{n_{max}^s}\alpha_n\tilde eB\sqrt{\mu_{sr}^2-M_s^2-n\tilde eB}
\ee
while, for the $\tilde e$ uncharged blue strange quarks,
\be
\rho^{sb}=\frac{(\mu_{sb}^2-M_s^2)^{3/2}}{3\pi^2}
\ee
The electron number density is given by
\be
\rho_E=2 \frac{1}{2\pi^2}\sum_n^{n_{maxe}}\alpha_n(\tilde eB\sqrt{\mu_E^2-2n\tilde eB})
\ee

 To discuss the pressure in the context of matter in the core of the neutron star, one also have to add the
contribution of the electrons to the thermodynamic potential. Since we shall describe the system as a function of $\tilde e B$,
we shall take the approximations $\tilde e\sim e$, $A_\mu\sim \tilde A_\mu$ to a good approximation as the mixing angle is small.
 The corresponding thermodynamic potential for the electrons is given by
\be
\Omega_e=\frac{\tilde e}{4\pi^2}\sum_{n=0}^{n_{max}^e}\alpha_n\left[\mu_E^2-2n\tilde e B\log 
\left(\frac{\mu_E+\sqrt{(\mu_e^2-2n\tilde e B)}}{\sqrt{2n \tilde e B}}\right)\right].
\label{thpote}
\ee
where, $n_{max}^e=\frac{\mu_E^2}{2|\tilde e B|}$.
Thus the total thermodynamic potential or the negative of the pressure is given as
\be
\Omega=\Omega_q+\Omega_e
\label{totomg}
\ee
The thermodynamic potential (Eq. (\ref{totomg})), the mass and superconducting gap equations Eq.(\ref{massu}),Eq.(\ref{massd})
,Eq.(\ref{masss}) and Eq.(\ref{scgap}),
along with  the charge neutrality conditions, Eq.(\ref{neutrality}), Eq.(\ref{colneutrality}) 
are the
basis for our numerical calculations for various physical situations that we shall
 discuss in in detail in the following section.

\section{results and discussions}
%-------------------------------------------------------------------------------------------------------
We begin the discussions with the parameters of the NJL model. The model parameters are the three current masses 
of quarks, namely m$_u$,m$_d$ and m$_s$ and the couplings G$_s$, G$_d$ and the determinant coupling K. 
This apart, one additional parameter, the momentum cut off $\Lambda$, is also required to regularize 
the divergent integrals which are characteristic of the four point interaction of NJL models. 
Except for the diquark coupling G$_d$, there are several parameter 
%[ P. Rehberg, S.P. Klevansky, and J. Hu ̈fner, Phys. Rev. C 53 (1996) 410., T. Hatsuda and T. Kunihiro, Phys. Rep. 247 (1994) 221., M. %Lutz, S. Klimt, and W. Weise, Nucl. Phys. A 542 (1992) 521.] 
sets for the couplings derived from fitting of the meson spectrum and chiral condensate
\cite{hatkun,lkw,rehberg}. The diquark coupling is not known from fitting since one does not have a 
diquark spectrum to fit with. Fierz transforming quark-antiquark term gives the relation G$_d$=0.75 G$_s$.
 Although not precise, many other references use this value. 
The parameters used in our calculations
are m$_u$=5.5 MeV, m$_d$=5.5 MeV, m$_s$=140.7 MeV for the current quark masses, the momentum
cutoff $\Lambda=602.5 MeV$ and the couplings G$_s$ $\Lambda^2$ =1.835 and K$\Lambda^5$=12.36 as have been
chosen in Ref.\cite{rehberg}.
After choosing the light current quark mass m$_u$=m$_d$=5.5 MeV, the remaining four parameters are
 chosen to fit vacuum values of pion decay constant f$_\pi$, masses of pion, kaon $\eta^\prime$. With this 
set of parameters the $\eta$ meson mass is underestimated by about 6 percent and leads to 
 u and d constituent mass in vacuum to be about 368 MeV. The strange mass is about 549 MeV at zero 
temperature and density. The determinant interaction is responsible for U(1)$_A$ anomaly and 
getting the correct eta mass. Further, this interaction also mixes the various gap equations and affects 
the superconducting gap significantly as we shall see. However, we must point out that there is a 
large discrepancy in the determination of this six fermion interaction coupling K. E.g. in Ref.\cite{hatkun}
the parameter K$\Lambda^5$ differs by as large as 30 percent as compared to the value chosen here. 
This discrepancy is due to the difference in the treatment of $\eta$' 
mesons with a high mass\cite{buballarev}. Infact, this leads to an unphysical imaginary part for 
the corresponding polarization diagram in the $\eta$' meson channel. This is unavoidable because NJL is not confining 
and is unrealistic in this context. Within the above mentioned limitations of the model and 
the uncertainty in the value of the determinant coupling, we proceed with the present parameter 
set which has already been used for phase diagram of dense matter in the Refs.\cite{ruester,buballarev} and
 for neutron star matter in Ref.\cite{leupold}.

              We begin our discussion for the simpler case where the charge neutrality conditions are not 
imposed. In this case, the electrical and color charge chemical potential are set to zero so that all 
the quarks have same potential $\mu_q$. In this case we have to solve four gap equations, 
three for the constituent masses Eq.s(\ref{massu},\ref{massd},\ref{masss}) and the fourth for the superconducting gap 
Eq.s(\ref{scgap},\ref{id1}). For given values of quark chemical potential and magnetic field we solve 
the gap equations self consistently. Few comments regarding solving these gap equations may be in order.
 We solve the gap equations at T=0. For non-vanishing magnetic fields, all the landau levels for 
the medium part up to a cutoff, n$_{max}$=$\frac{\sqrt{\mu^2-M_i^2}}{2\tilde{e}B}$ for each flavor i, 
are taken into account. Near the $\mu_c$, the critical chemical potential, there can be multiple solutions 
for the gap equations. We have chosen the solutions which have the lowest thermodynamic potential. 

\begin{figure}
\vspace{-0.4cm}
\begin{center}
\begin{tabular}{c c }
\includegraphics[width=8cm,height=8cm]{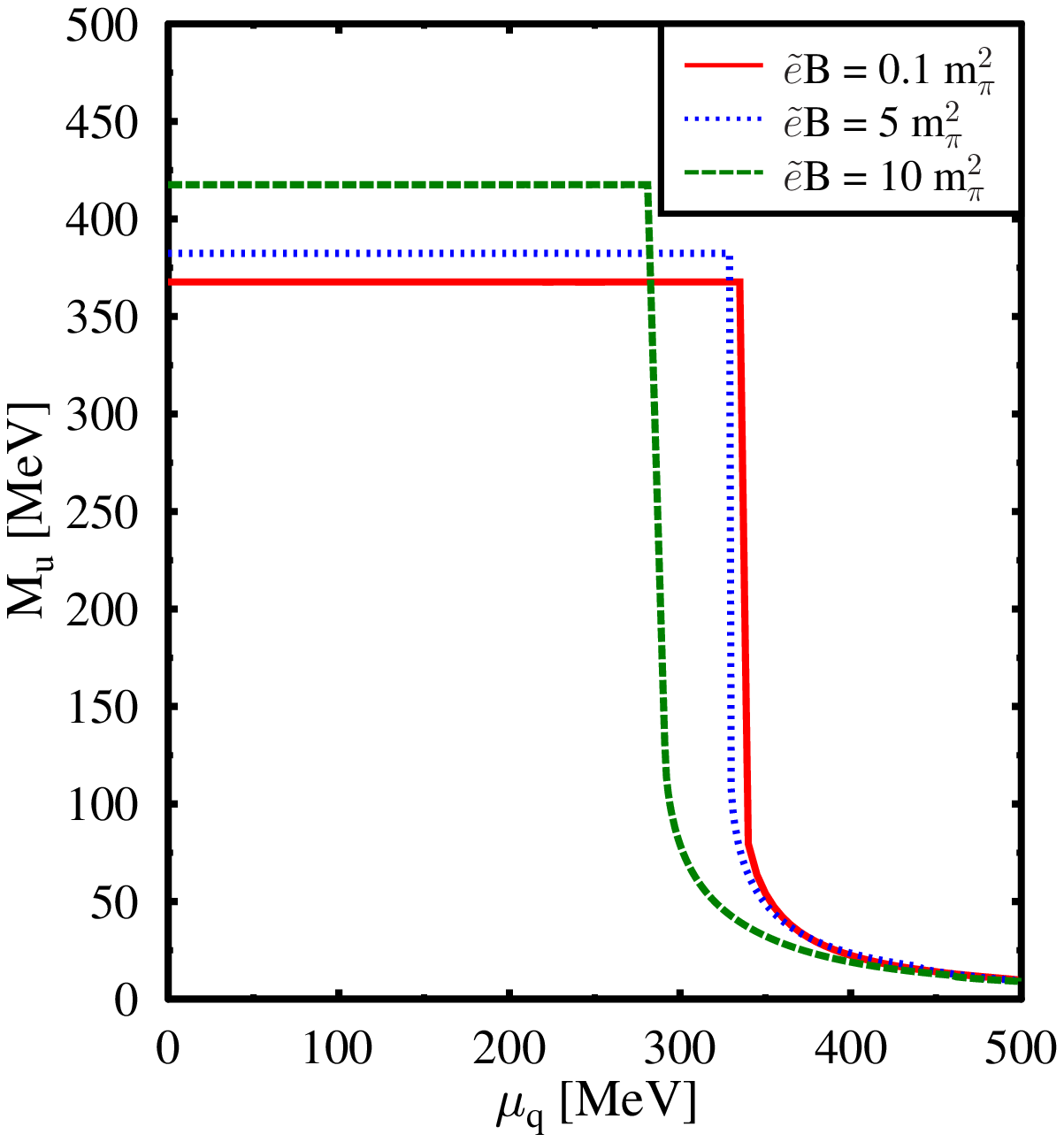}&
\includegraphics[width=8cm,height=8cm]{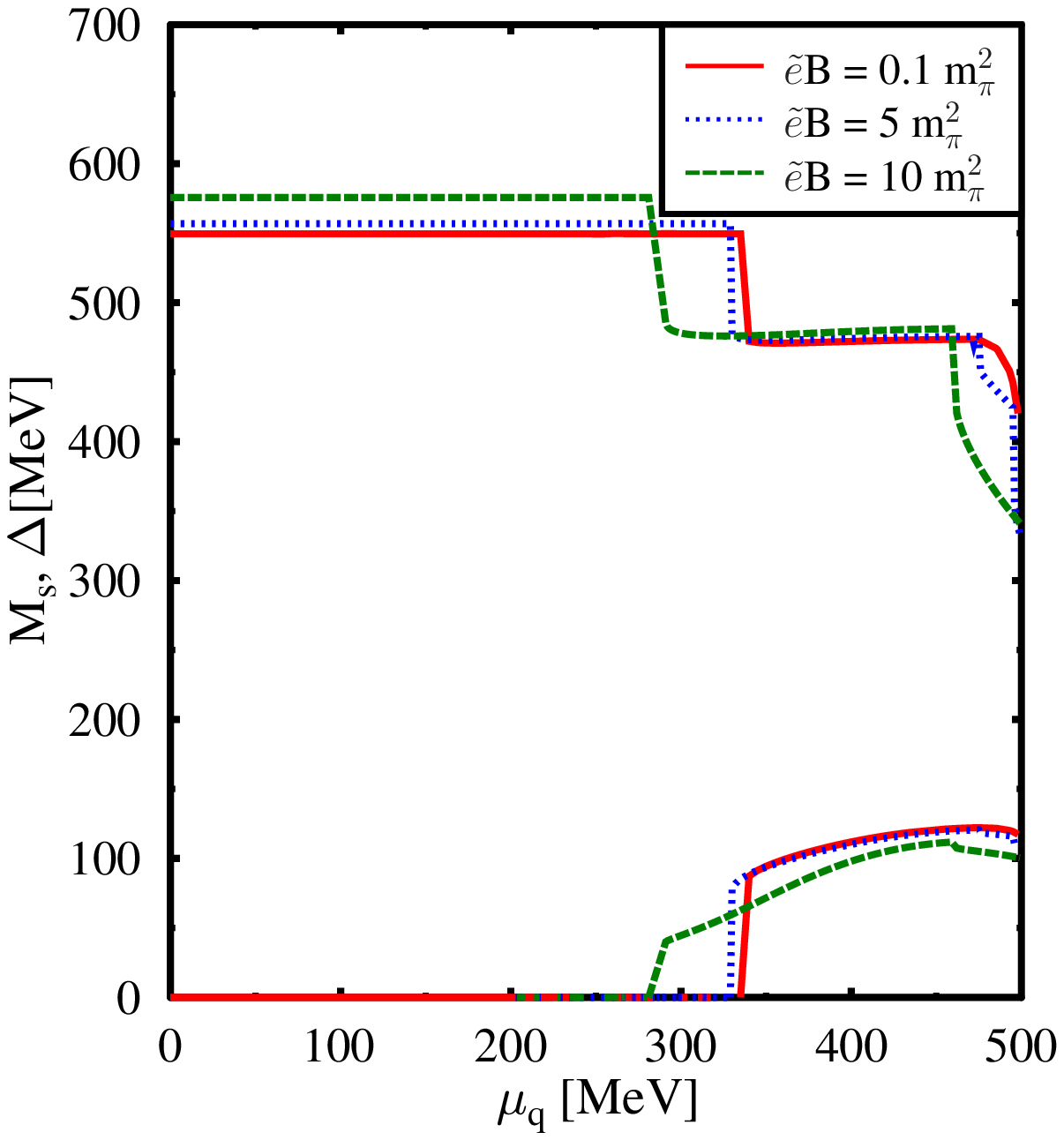}\\
Fig. 1-a & Fig.1-b
\end{tabular}
\end{center}
\caption{Constituent quark masses and superconducting gap when charge neutrality 
conditions are not imposed.
Fig.1-a shows the M$_u$  at zero temperature 
 as a function of quark chemical potential for different values of the magnetic field.
Fig. 1-b  shows the same for the strange quark mass M$_s$ and the superconducting gap.
}
\label{fig1}
\end{figure}

      In Fig.\ref{fig1} ,  we have shown the variation of the masses as a function of quark chemical potential $\mu_q$ 
for three different values of magnetic fields, $\tilde{e}$B=0.1$ m_\pi^2$,5 m$_\pi^2$,10 m$_\pi^2$.
The results for $\tilde e B=0.1 m_\pi^2$ reproduce the vanishing magnetic field results.
 As the chemical potential increases,
 the masses remain constant upto a critical value of quark chemical potential $\mu_c$ and the superconducting gap 
remains zero. At the critical chemical potential there is a first order phase transition and the constituent 
masses drop sharply from their vacuum values and the superconducting gap becomes non-zero. 
For vanishing magnetic field, the isospin symmetry for the light quarks is unbroken and the constituent 
masses of u and d quarks are degenerate. The critical chemical potential,$\mu_c$, is about 340 MeV for 
(almost) vanishing magnetic field. In this case,  the up and the down quark masses decrease from 
their vacuum values of about 368 MeV to about 80 MeV. The strange mass being coupled to other gaps 
via determinant interaction also decreases from 549 MeV to 472 MeV when this first order transition
happens for the light quarks. However, since this $\mu_c$  is still less than the strange mass its 
density remains zero. The superconducting gap rises from 0 MeV to 88.0 MeV at $\mu_c$. As the chemical potential 
is increased beyond $\mu_c$, the superconducting gap shows a mild increase reaching a maximum value
 of 122 MeV at around $\mu_q\sim$ 475 MeV. Beyond this value of $\mu$, the strange quark mass starts decreasing 
rapidly. This leads to the effective diquark coupling G$_D^\prime$=G$_D$+$\frac{K}{4}\langle \bar{s}s\rangle$ 
decreasing resulting in a decrease in the superconducting gap with increasing chemical potential.

\begin{figure}
\includegraphics[width=8cm,height=8cm]{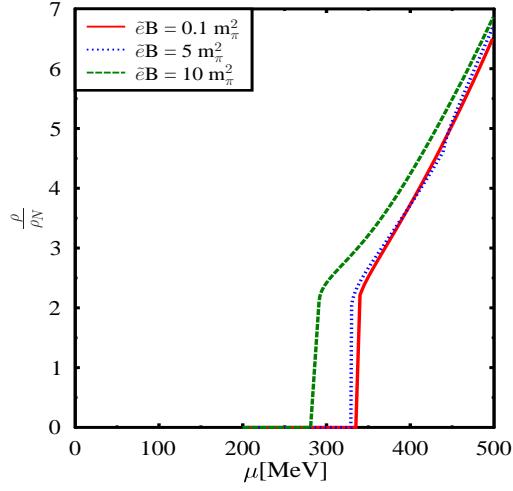}
\caption{Baryon number density in units of nuclear matter density as a function of chemical potential for different strengths of magnetic field at zero temperature.}
\label{fig2}
\end{figure}

\begin{figure}
\includegraphics[width=8cm,height=8cm]{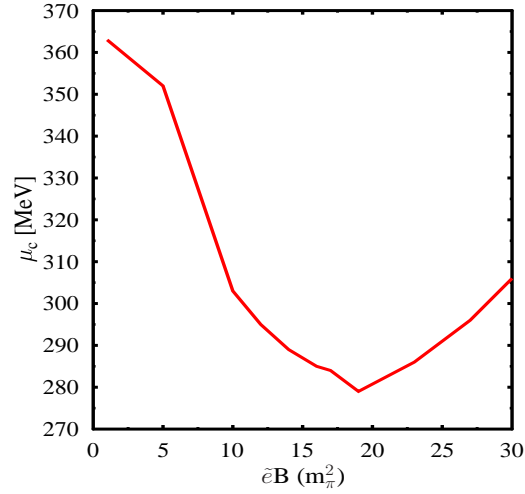}
\caption{Critical chemical potential for chiral transition at zero temperature as a function of magnetic field }
\label{fig3}
\end{figure}

	 In Fig.\ref{fig2}, we have plotted the total baryon number density in units of 
nuclear matter density($\rho_N$=0.17/fm$^{-3}$) as function of quark chemical potential. 
For vanishing magnetic field, at the critical chemical potential $\mu_c \sim$ 340 MeV, the baryon density jumps
 from 0 to 0.38$fm^{-3}$ which is about 2.2 times the nuclear matter density.  

	 Upon increasing the magnetic field, as seen in Fig.\ref{fig1}, the vacuum constituent quark masses 
increase due to magnetic catalysis at zero density. It may also be observed here that 
the $\mu_c$ for chiral transition for the light quarks decreases with the magnetic field. 
Such a phenomenon is known as inverse magnetic catalysis at finite chemical potential.\cite{andreas1}. Let us note 
that in the superconducting phase the $\tilde{e}$ charges of the u and d quarks are identical 
in magnitude while that of unpaired blue quark are different for u and d quarks. This results in the 
color summed scalar condensate $I_s^u$ and $I_s^d$ to be different in presence of magnetic field. This leads to 
difference in constituent masses for the light quarks.  
 For $\tilde{e}$B=10 $m_\pi^2$ the u mass in the chiral symmetry broken phase increases by about 
13.6 percent and strange mass by about 4.7 percent. The critical chemical potential decreases from 
about 340 MeV to about 291 MeV. As seen in the plot, the superconducting gap decreases and the peak value
decreases from 122 MeV to 111 MeV. As may be seen from Eq.(\ref{scgap}) and Eq.(\ref{id1}),
 the superconducting gap depends upon the effective diquark coupling G$_D^\prime$ = G$_D$-$\frac{K}{4} I_{s}^s$. 
With increase in magnetic field the effective coupling G$_D^\prime$ has a slight increase in magnitude as 
the strange quark condensate increases with magnetic field. Therefore, one would have expected an increase in $\Delta$
with magnetic field. However, the variation in $\Delta$ due to the magnetic 
field is essentially decided by Eq.(\ref{id1}). From here also one would have expected an increase in 
$\Delta$  with magnetic field as $\tilde{e}B$ occurs in the numerator in Eq.(\ref{id1}). Infact, this behavior is 
actually seen for high magnetic field, where, only the lowest Landau level contributes to the integral in Eq.(\ref{id1}).
For moderately strong magnetic fields, contributions of the higher Landau levels become relevant 
for the behavior of gap with magnetic field. As long as the contribution of higher Landau levels are non vanishing,
the gap equation can support solution for the gap that decreases with magnetic field. 
We may point out that $\tilde{e}B$=5 m$_\pi^2$ and 10 m$_\pi^2$ the cut off for Landau levels n$_{max}$ equals 3 and 1 
respectively. For $\tilde{e}B \geq $  20 m$_\pi^2$ only the lowest Landau level contributes to the integral in Eq.(\ref{id1}) 
and the gap increases with magnetic field. One may also note that at higher magnetic fields the charge 
asymmetry between the u and d quark becomes apparent in their masses as expected. At 10$m_\pi^2$ the difference is
 about 3.4 percent and at 15$m_\pi^2$ its about 5.7 percent at lower chemical potentials.   

	One may note that below the critical chemical potential $\mu_c$ the u quarks have higher mass compared to 
d quarks as all the three colors are charged for u quarks while for the d quarks, the blue color is chargeless.
 However beyond the critical chemical potential the u quark has a lower mass compared to d quarks. This is because 
with magnetic field the medium contribution to chiral condensate increases. This increase is same for the condensing
pairs of u and d quarks but different for the blue quarks.  The blue up quark has charge $\tilde e=1$  
whereas it is zero for down  blue quark. Therefore the medium contribution from up quark is more than down
 quark and it reduces 
the condensate for up quark and consequently its mass too. As we shall see later, imposing charge neutrality 
requires the d quark chemical potential to be much higher compared to u quarks to balance their larger 
positive charge. This forces the d quark mass to be smaller compared to u quark mass above critical chemical potential .
This results in an opposite behavior for the u and d quark masses with chemical potential, 
beyond $\mu_c$  when charge neutrality condition is imposed vis a vis when such condition is not imposed.

As may be observed from Fig.\ref{fig2}, the baryon number density increases with magnetic field for a given chemical potential. 
This is because for the magnetic fields considered here, the symmetry is restored for lower chemical 
potential at higher magnetic field. Thus for a given chemical potential beyond the critical chemical potential the 
masses become smaller for higher magnetic field leading to larger baryon number density. This is consistent 
with inverse magnetic catalysis. One may note however that  for very large fields, 
there is magnetic catalysis of chiral symmetry breaking in the sense that critical chemical potential 
increases with magnetic field. In Fig.\ref{fig3} we show the behavior of $\mu_c$ as a function of magnetic field. It is
observed that $\mu_c$ is minimum for $\tilde{e}B$=19$m_\pi^2$.	

\begin{figure}[h!]
\includegraphics[width=8cm,height=8cm]{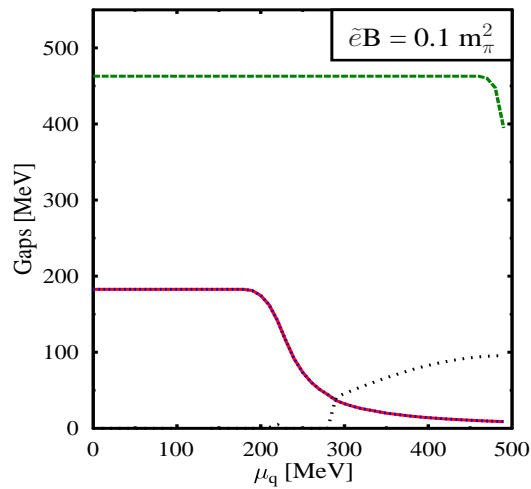}
\caption{Gaps without determinant interaction at zero temperature as a function of quark chemical potential. Solid curve refers to masses of u-d quarks, the dashed curve refers to the mass of strange quark and the dotted curve corresponds to the superconducting gap. }
\label{fig4}
\end{figure}

  	 To examine the effect of flavor mixing determinant interaction, we show in Fig.\ref{fig4}, the variation of the 
masses and the superconducting gap without the determinant interaction. As expected, without the mixing 
of flavors the strange mass remains unaffected when u and d quark masses decrease. This is significantly 
different behavior compared to Fig.\ref{fig1} where the strange mass decreases by about 74 MeV 
beyond $\mu_c$ when there is a first order transition for the light quarks. This also affects the superconducting gap. 
The superconducting gap is smaller as the effective diquark coupling decreases without the determinant interaction term.

\begin{figure}
\vspace{-0.4cm}
\begin{center}
\begin{tabular}{c c }
\includegraphics[width=8cm,height=8cm]{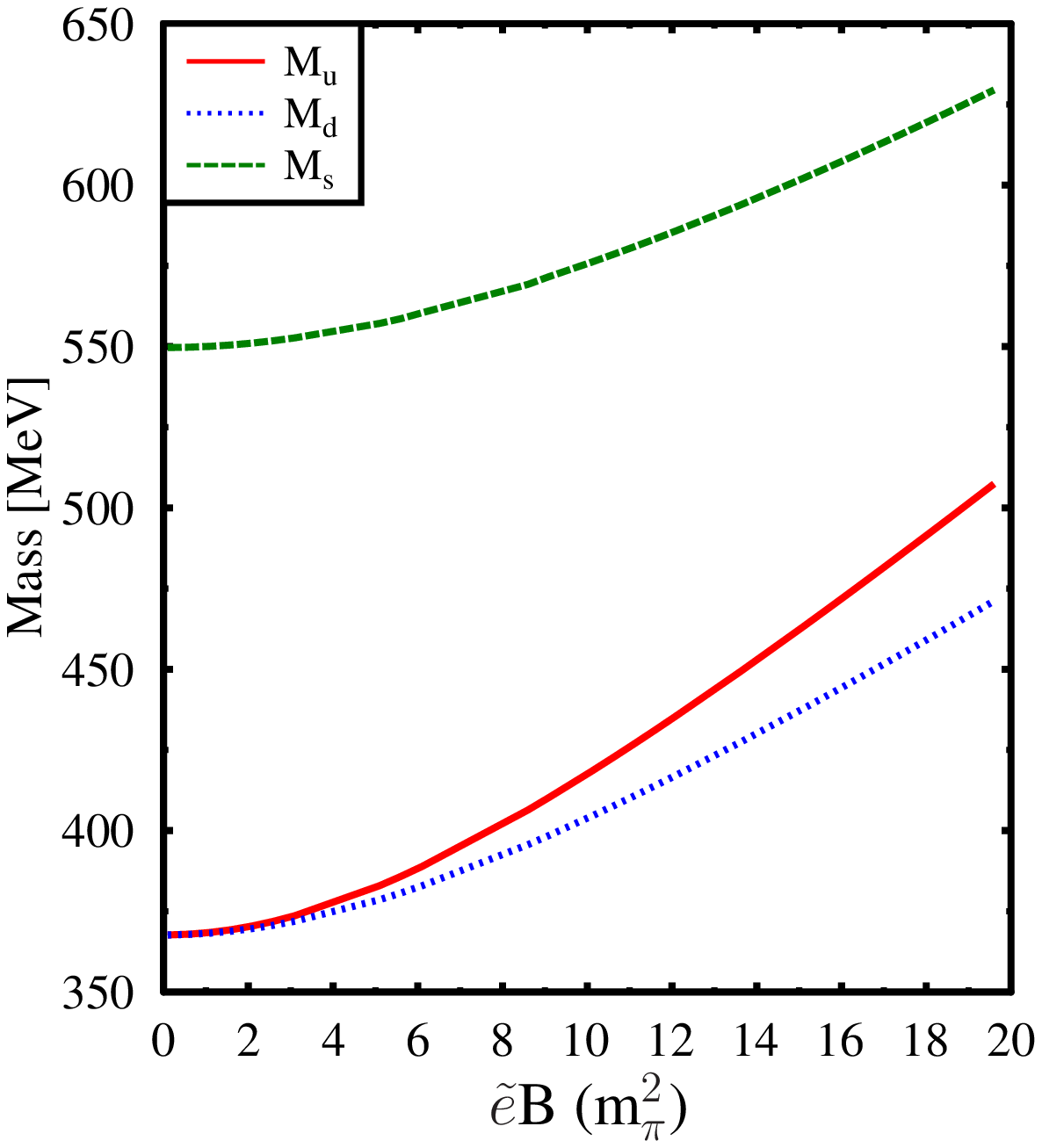}&
\includegraphics[width=8cm,height=8cm]{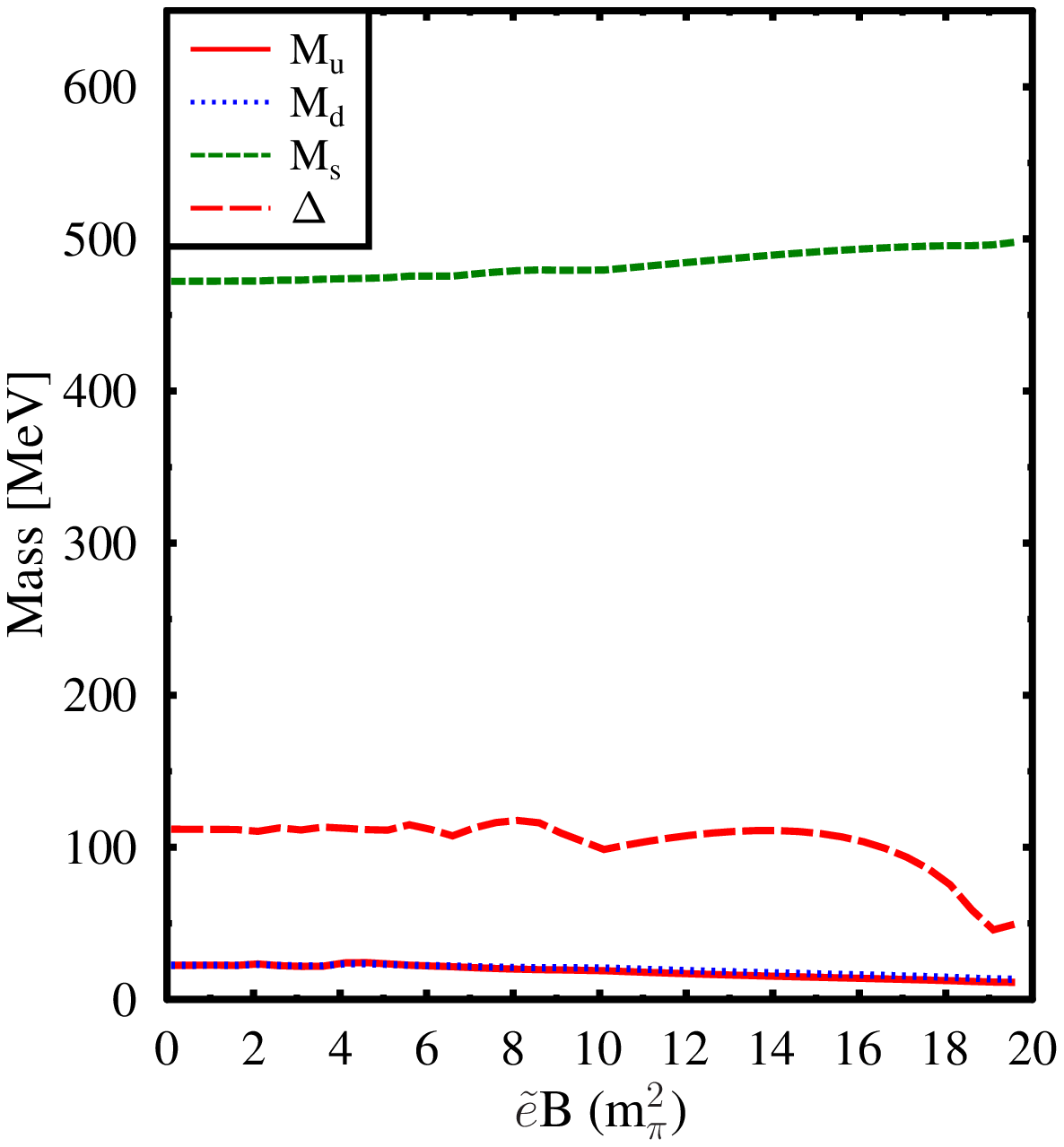}\\
Fig. 5-a & Fig.5-b
\end{tabular}
\end{center}
\caption{Constituent quark masses as a function of magnetic field for T=0.
Fig.5-a shows the masses of the three quarks below the chiral transition for $\mu$=200 MeV.
Fig. 5-b  shows the same for the masses along with the superconducting gap above the chiral transition
for $\mu_q=400$ MeV.
}
\label{fig5}
\end{figure}

	      In Fig.{\ref{fig5}}  we show the variation of the gaps as a function of the magnetic field for $\mu$=200 MeV and 
$\mu$=400 MeV. $\mu$=200 MeV is less than the critical $\mu_c$ for any value of magnetic field considered here.
 Hence the constituent masses are high and the superconducting gap is zero. 
We find that the masses increase 
monotonically with the magnetic field. At $\tilde{e}$B=10 m$_\pi^2$, the u mass increases by 14 percent of its zero field value 
while strange mass increases by 5 percent. Similarly for $\mu$=400 MeV which is larger than the critical chemical
 potential for magnetic fields considered here, one also has finite superconducting gap. However, in this case it is  observed
 that the u and d masses decrease slowly and monotonically with magnetic field while strange quark mass
 remains almost constant. The superconducting gap shows an oscillatory behavior with increase in magnetic field. The 
oscillatory behavior is associated with the  discontinuous changes in the density of states due to Landau quantization 
and is similar to de Hass van Alphen effects for magnetized condensed matter system.
%increases till about $\tilde{e}$B=0.158GeV$^2$ and then shows a slight decrease. 
%-----------------------------------------------

\begin{figure}
\vspace{-0.4cm}
\begin{center}
\begin{tabular}{c c }
\includegraphics[width=8cm,height=8cm]{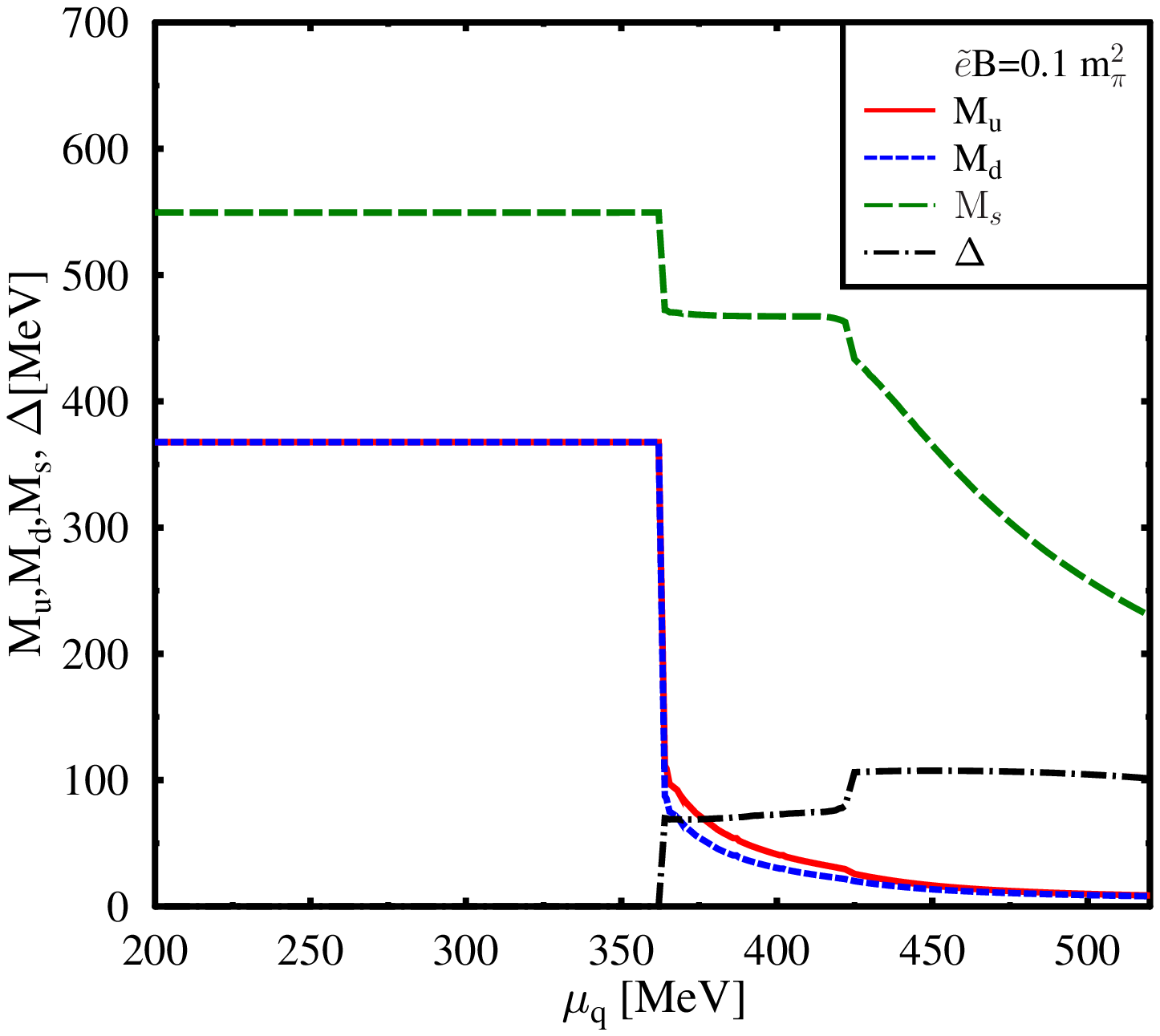}&
\includegraphics[width=8cm,height=8cm]{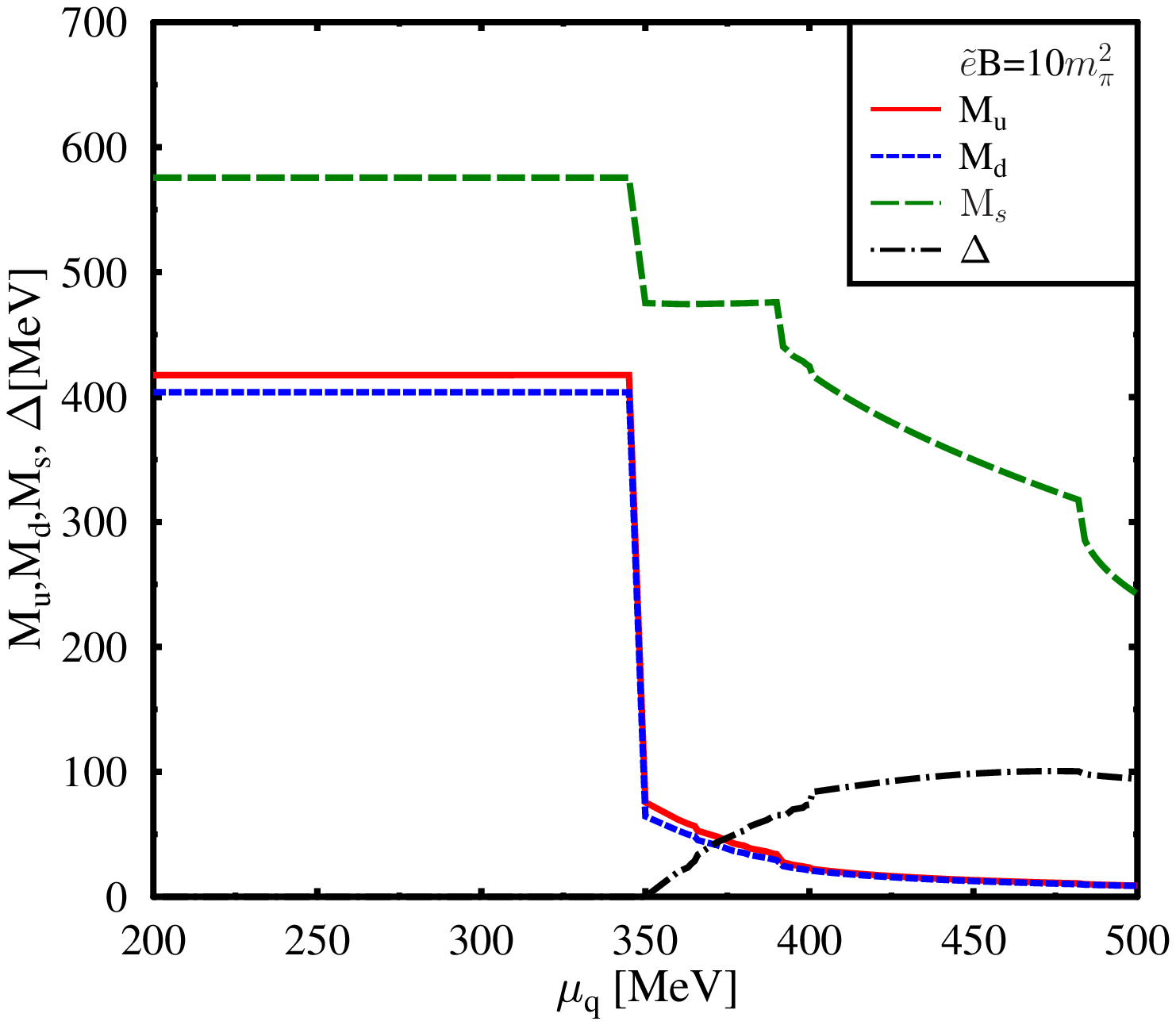}\\
Fig. 6-a & Fig.6-b
\end{tabular}
\end{center}
\caption{Constituent quark masses and superconducting gap when charge neutrality 
conditions are imposed.
Fig.6-a shows the  masses and superconducting gap  at zero temperature 
 as a function of quark chemical potential for magnetic field $\tilde e B=0.1m_\pi^2$
Fig. 6-b  shows the same for $\tilde e B=10m_\pi^2$.
}
\label{fig6}
\end{figure}

\subsection*{Charge neutral magnetized quark matter}

	Next we discuss the consequences of imposing charge neutrality conditions($Q_E=0$,$Q_8=0$). In Fig. \ref{fig6}
we show the results for the masses and the superconducting gaps for strength of the
external magnetic field $\tilde e B=0.1 m_\pi^2$ (Fig 6-a) and $\tilde e B=10m_\pi^2$ (Fig. 6-b).
For small magnetic field($\tilde{e} B=0.1 m_\pi^2$) the masses in symmetry broken phase are the 
same as before but the critical chemical potential is now shifted to around $\mu_c=364 MeV$ as compared to 
$\mu_c= 335 MeV$ when the condition is not imposed. At the transition point with neutrality the u quark mass 
decreases from 367 MeV to 111 MeV and the down quark mass from 367 MeV to 87 MeV. Charge neutrality requires 
d quark number densities to be higher as compared to u quarks. Let us note that near the critical 
chemical potential there are multiple solutions of the gap equations. The solution which is thermodynamically
 preferred when charge neutrality condition is not imposed may no longer be the preferred solution when the constraint 
of charge neutrality is imposed \cite{amhm5}.
	The strange quark mass is higher than the chemical potential at the chiral restoration so its 
density is zero. However due to the determinant interaction the strange mass decreases at the chiral restoration 
from 549 MeV to 472 MeV.  At still higher chemical potential the strange quark density becomes 
non-zero and strange quark also helps in maintaining charge neutrality.  
	The critical baryon density when charge neutrality is imposed is however similar to case when neutrality 
is not imposed. Specifically $\rho_c \sim 2.25 \rho_0$ with charge neutrality while $\rho_c \sim 2.26 \rho_0$ without 
charge neutrality despite the fact that $\mu_c$ is higher ($\mu_c=364$ MeV) for the charge neutral matter
compared when such charge neutrality condition is not imposed ($\mu_c=335$ MeV). This is because the constituent masses at the transition is large ($M_u \sim 111 MeV$ and $M_d \sim 87 MeV$) 
for charge neutral case compared to ($M_u \sim M_d \sim 85 MeV$) without charge neutrality condition. 
	For $\tilde e B=0.1 m_\pi^2$, 
 at the chiral transition $\mu_c=364 MeV$ the superconducting gap increases from zero to 69 MeV. 
As the chemical potential is further increased the superconducting gap increases to 80 MeV till $\mu=\mu_1 \sim $ 420 MeV 
where it shows a sudden jump to 106 MeV. This happens when the gapless modes cease to exist as explained below. 
	As magnetic field is increased to $\tilde{e} B=10 m_\pi^2$, as may be observed in Fig.6-b,
 the critical chemical potential $\mu_c$ for 
the charge neutral matter decreases to 350 MeV similar to the case without charge neutrality condition with 
inverse magnetic catalysis. The superconducting gap on the other hand becomes smaller. One can also
observe that  unlike vanishingly small magnetic field case, 
the superconducting gap increases smoothly with chemical potential from zero initial value to 73 MeV at $\mu=\mu_1 \sim 400$
 MeV where it again jumps to a value of 83 MeV.  

\begin{figure}
\vspace{-0.4cm}
\begin{center}
\begin{tabular}{c c }
\includegraphics[width=8cm,height=8cm]{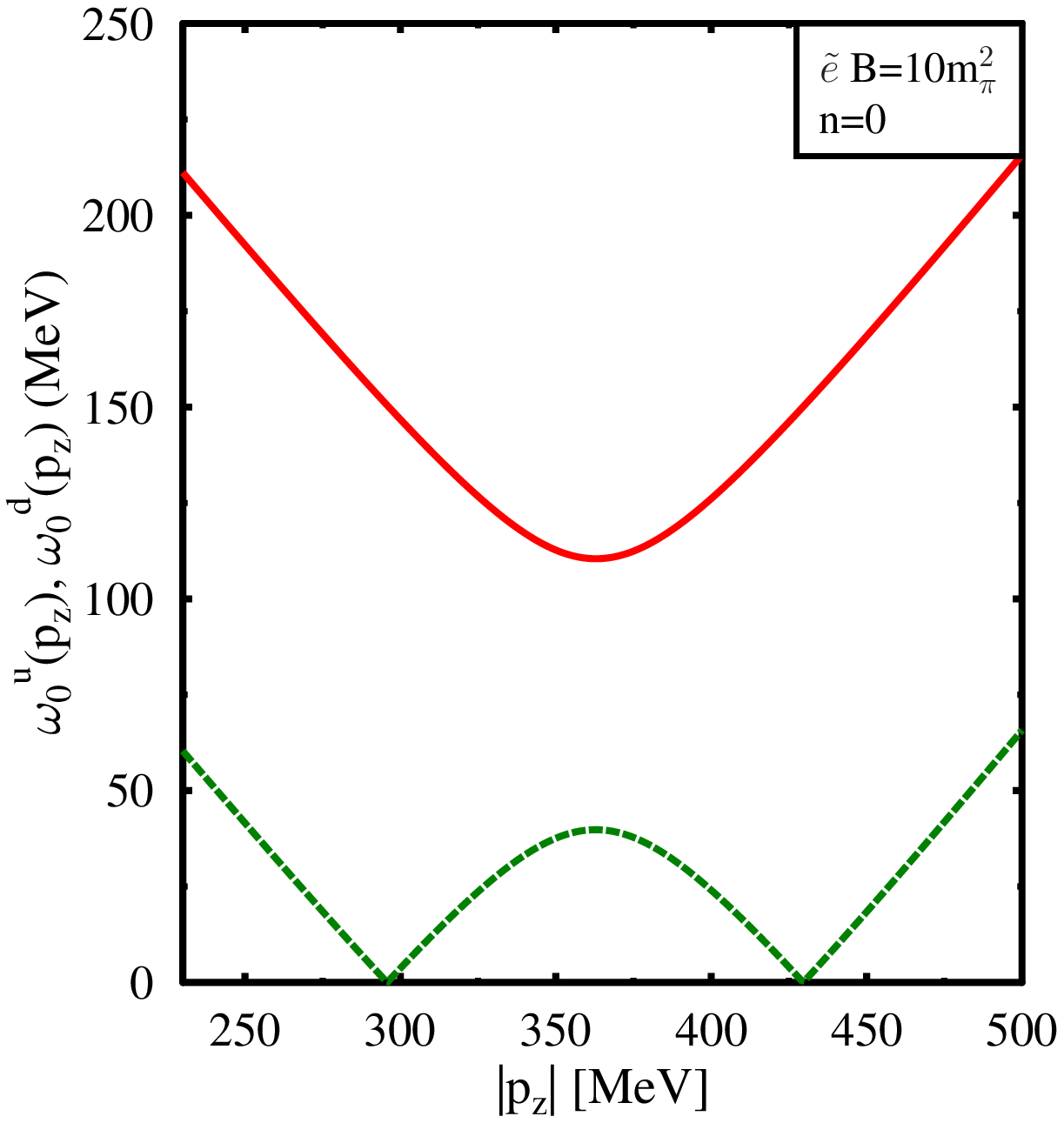}&
\includegraphics[width=8cm,height=8cm]{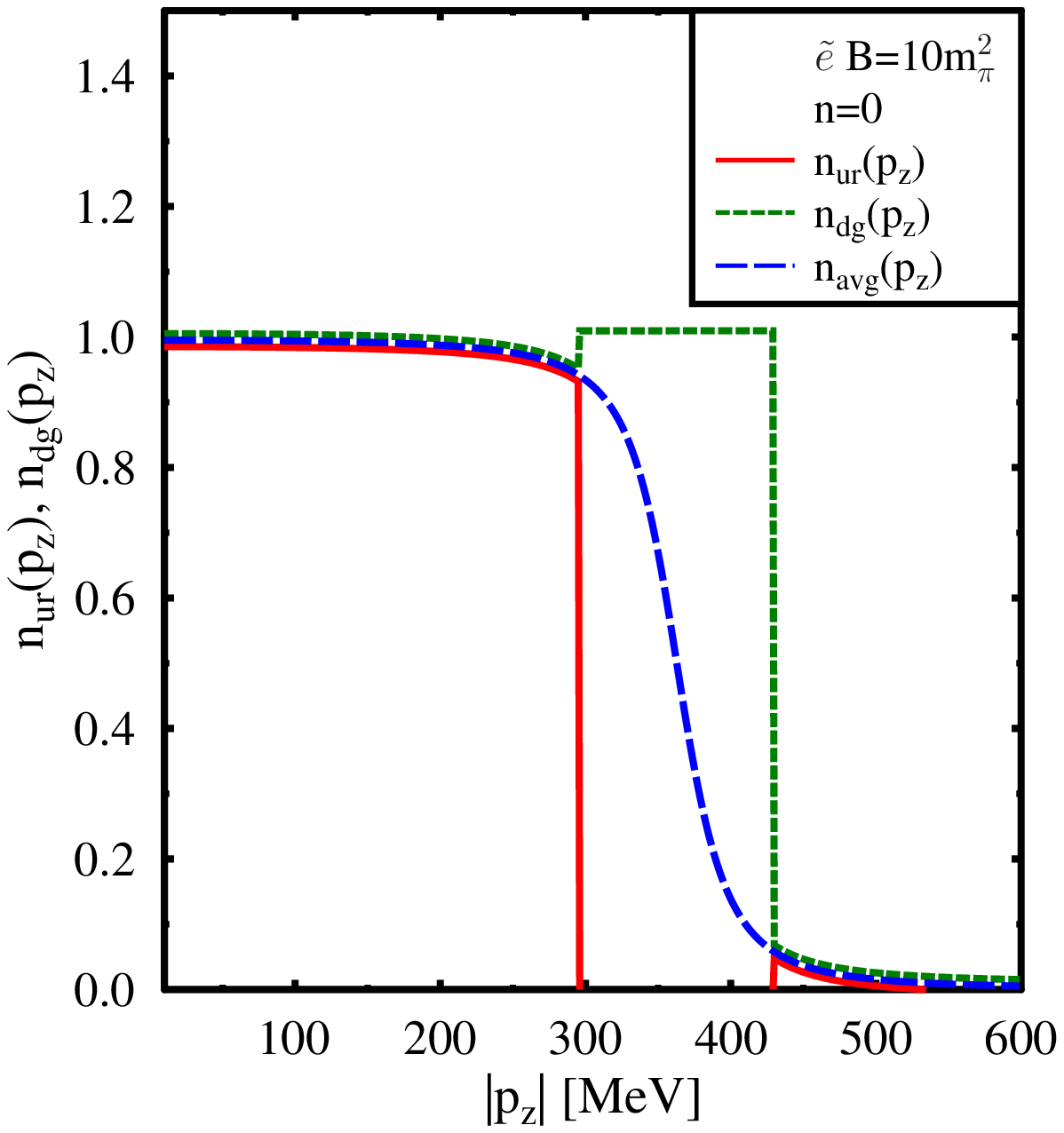}\\
Fig. 7-a & Fig.7-b
\end{tabular}
\end{center}
\caption{Dispersion relation and the occupation number for condensing quarks at T=0,$\mu_q$=340 MeV.
Fig.7-a shows the dispersion relation for the condensing quarks for zeroth Landau level. The upper curve is for u quark and the lower curve corresponds to d quark dispersion relation. 
Fig. 7-b  shows the occupation number as a function of momentum for $\tilde e B=10m_\pi^2$.
}
\label{fig7}
\end{figure}

%\begin{figure}[h!]
%\includegraphics[scale=0.5]{figneutral3.eps}
%\includegraphics[scale=0.5]{figneutral2.eps}
%\includegraphics[scale=0.5]{figneutral3.eps}
%\label{fig7}
%\end{figure}

\subsection*{Gapless modes}

       In the region between $\mu_c$ and $\mu_1$ the system shows gapless mode which we discuss now in some detail. 
Without magnetic field this has earlier been seen for charge neutral matter \cite{igorr,amhm5}.

As discussed earlier, from the dispersion relations  for  Landau levels for the
superconducting matter as given in Eq.(\ref{omgpmu}) and Eq.(\ref{omgpmd}), it is possible to have
zero modes depending upon the values of $\delta\mu$ and $\delta \epsilon_n$.  These quantities are not independent parameters
bu are dependent dynamically on the charge neutrality condition and the gap equations. For charge neutral matter, near $\mu_c$, the d-quark number density is larger so that $\delta\mu=\mu_E/2$ is negative. This renders $\omega_n^u(p_z) > 0$ for any value of momentum $p_z$. On the other hand, for $\delta\mu$ negative, $\omega^d_n $ can vanish for some values of $p_z$.
This defines the fermi surfaces for the superconducting d quarks. 
It is easy to show that the excitation energy of n$^{th}$ Landau level $\omega^d_n$ for the condensing d quarks 
vanishes for momenta $|p_{zn}|$= $\sqrt{\mu_{\pm}^2-2 n \tilde{e}B}$. Here $\mu_{\pm}$=($\bar{\mu} \pm 
\sqrt{\delta \mu^2-\Delta^2}$)$\theta (\delta \mu - \Delta)$. Thus higher Landau levels can also have
 gapless modes so long as $\sqrt{\mu_{\pm}^2-2 n \tilde{e}B}$ is non-negative.
 Gapless modes occur when the chemical potential difference $\delta \mu$ is greater than the superconducting gap. 
In Fig.7-a, we have plotted the dispersion relation i.e. the excitation energy as a function of momentum
for the lowest Landau level for the condensing quarks for $\mu_q=$340 MeV and magnetic field $
\tilde{e}B=10 m_\pi^2$. The superconducting gap turns out to be $\Delta=$35.3 MeV and $\delta \mu=$-74.5 MeV. 
The dispersion for the d quarks is given as $\omega_{0-}^d=\bar{\omega}_{0-}-\delta \epsilon + \delta \mu$ while
the same for u-quark is given as $\omega_{0-}^u=\bar{\omega}_{0-}+\delta \epsilon - \delta \mu$. The average 
chemical potential is $\bar{\mu}$= 366 MeV. Far from the pairing region, $|p_z|\sim\bar\mu=366 MeV$ the spectrum 
looks like usual BCS type dispersion relation. Of the two excitation energies, $\omega_0^u$ shows a minimum at
 $p_z=\bar\mu$ with a value $\omega_{0-}^u(|p_z|=\bar\mu)\sim\Delta-\delta\mu=110$ MeV. On the other hand, $\omega_{0-}^d$
vanishes at momenta $|p_z|=\mu_\pm$. In this breached pairing region one has only unpaired d-quarks and no u-quarks.
This can be seen explicitly as below.

\begin{figure}[h!]
\includegraphics[width=8cm,height=8cm]{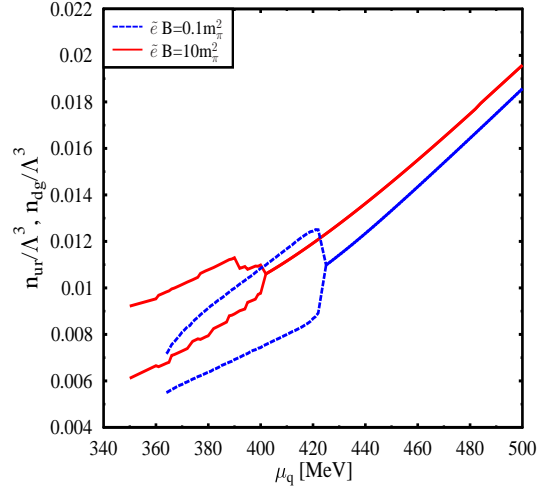}
\caption{Number densities of  up and down quarks participating in the superconductivity  for $\tilde eB=0.1m_\pi^2$
(dashed line) and $\tilde eB=10m_\pi^2$ (solid line)}
\label{fig8}
\end{figure}

       The number densities of u quarks participating in condensation is given by
\be
       \rho_{sc}^u=\rho^{ur}+\rho^{ug}= \sum_n \frac{\alpha_n \tilde{e}B}{(2\pi)^2}\int dp_z \left[\frac{1}{2}
\left(1-\frac{\bar{\zeta}_{n-}}{\bar{\omega}_{n-}}\right)\left(1-\theta (-\omega_n^d)\right)-\frac{1}{2}\left(1-
\frac{\bar{\zeta}_{n+}}{\bar{\omega}_{n+}}\right)\right]
\label{rhoscu}
\ee
       This is because $\omega_{n_-}^u=\bar{\omega}_{n_-}-\delta \mu +\delta \epsilon$ is always positive as 
$\delta \mu$=$\frac{\mu^u-\mu^d}{2}$ is negative and the theta function
 $\theta(-\omega_n^u)$ does not contribute. Similarly the density of d-quarks participating in condensation is given by

\be
       \rho_{sc}^d=\rho^{dr}+\rho^{dg}= \sum_n \frac{\alpha_n \tilde{e}B}{(2\pi)^2}\int dp_z 
\left[\theta (-\omega_n^d) 
+ \frac{1}{2}\left(1-\frac{\bar{\zeta}_{n-}}{\bar{\omega}_{n-}}\right)\left(1-\theta (-\omega_n^d)\right)-
\frac{1}{2}\left(1-\frac{\bar{\zeta}_{n+}}{\bar{\omega}_{n+}}\right)\right]
\label{rhoscd}
\ee
       \noindent For positive $\omega_{n-}^d$, the $\theta$-function contributions vanishes and the distribution 
functions are the BCS distribution function. On the other hand, when $|p_z| \in $ [$P_{n-},P_{n+}$], 
$\omega_n^d$ is negative leading to $\rho^u_{sc}$ to vanish but for the anti-particle contribution. 
In this region of momenta, $\rho^d_{sc}$ is unity. We have plotted in Fig. 7-b the occupation number  
of the up and down quarks that take part in condensation as a function of the magnitude of momentum $p_z$ i.e. 
the integrands of Eq.(\ref{rhoscu}) and Eq.(\ref{rhoscd}) respectively for the lowest Landau level. It is easy to see from 
Eq.(\ref{rhoscu}) and Eq.(\ref{rhoscd}) e.g. for the lowest Landau
level, that except for the interval $(\mu_-,\mu_+)$, the distribution function is like the BCS distribution function.  This
is shown by the blue long-dashed line. The u-quark distribution is shown by the red solid line while the d- quark 
distribution is shown by the
green short dashed line. Indeed, except for the interval $(\mu_-,\mu_+)$, all the three curves overlap with each other. 
In the  'gapless' momentum region, the u-quark occupation vanishes while d-quark occupation is unity. This leads to fact that
the momentum integrated distribution function for the condensing u and d quarks are not the same for the gapless region unlike
the usual BCS phase. We have plotted the number densities for the u- and d- quarks in Fig.\ref{fig8} which shows a
fork structure in the gapless region.

%       The gapless region in momenta $|p_{zn}^+|$-$|p_{zn}^-|$=$\Delta p_z$ is () for $\tilde{e}B=$0.1$m_\pi^2$ and $\Delta p_z^{n=0}$=() for $\tilde{e}B=$10$m_\pi^2$.
%       We have also plotted the dispersion relation $\omega_{n=0}(|p_z|)$ for u and d quarks for $\mu=0$,$\tilde{e}B=$0.1$m_\pi^2$. Because of the gapless modes, the condensing flavors are not the same and show a fork structure in the plot of density as seen in fig. (). Beyond $p_{zn}^+$ the density of the condensing quarks of both flavors are the same as expected in BCS theory.

       Gapless modes have been considered earlier for two flavor quark matter both with\cite{scoccola,digal}
and without magnetic field \cite{igorr,amhm5}. 
However it has been shown\cite{giannakis,huang2004} that in QCD at zero temperature the gapless 2SC phases are unstable. This instability manifests itself in imaginary Meissner mass of some species of the gluons. Finite temperature calculations\cite{fukushima2005} show that at some critical value of temperature the instability vanishes. This value may range from few MeV to tens of MeV. The instability of the gapless phases indicate that there should be other phases of quark matter breaking translational invariance e.g. inhomogenous phase of quark matter like crystalline color superconductivity\cite{rajasharma2006,bowers2002}. One may note that these considerations apply to the case without magnetic field and may change in presence of strong magnetic field. 
%%%%%instability in gapless mode

\begin{figure}
\vspace{-0.4cm}
\begin{center}
\begin{tabular}{c c }
\includegraphics[width=8cm,height=8cm]{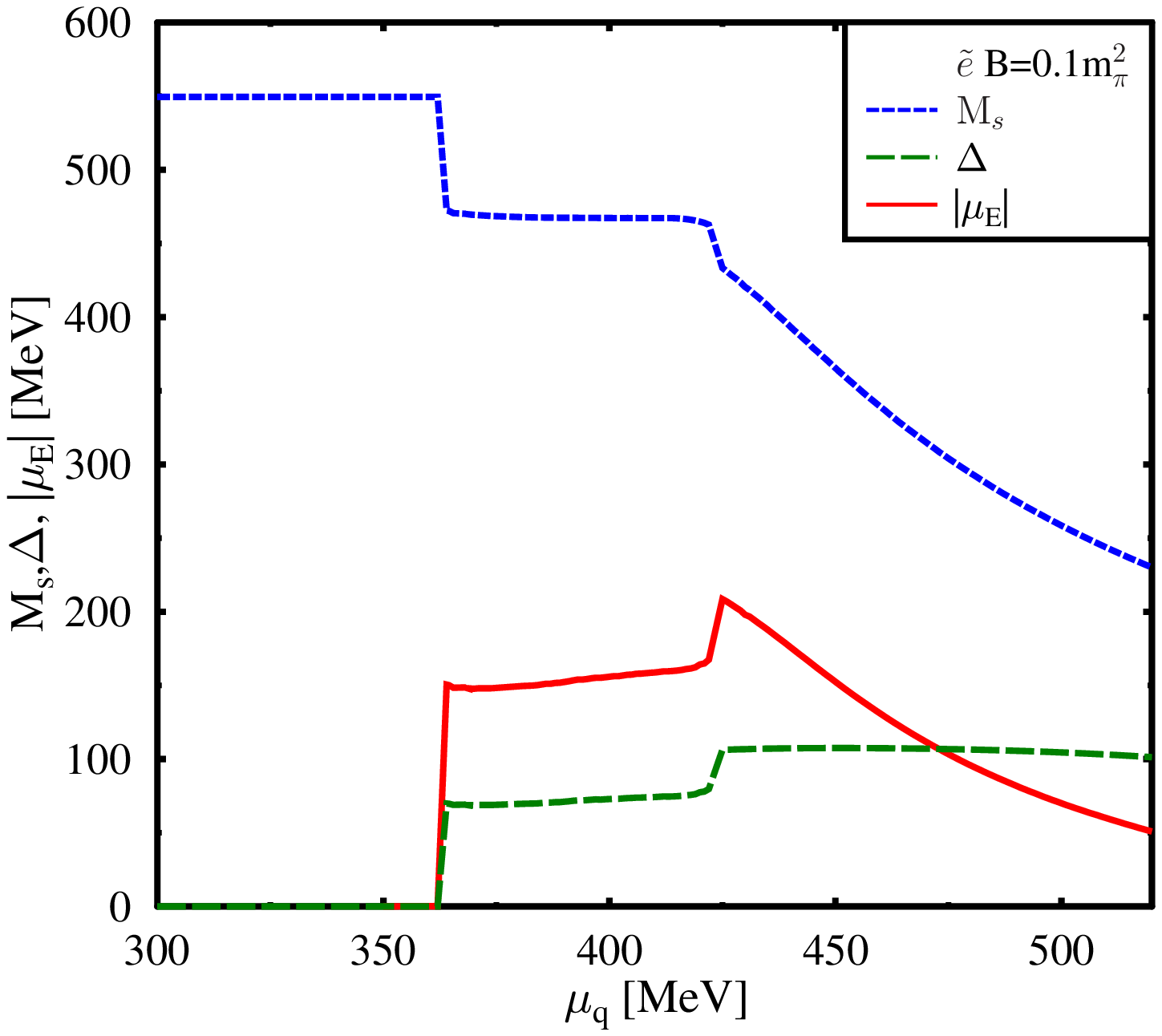}&
\includegraphics[width=8cm,height=8cm]{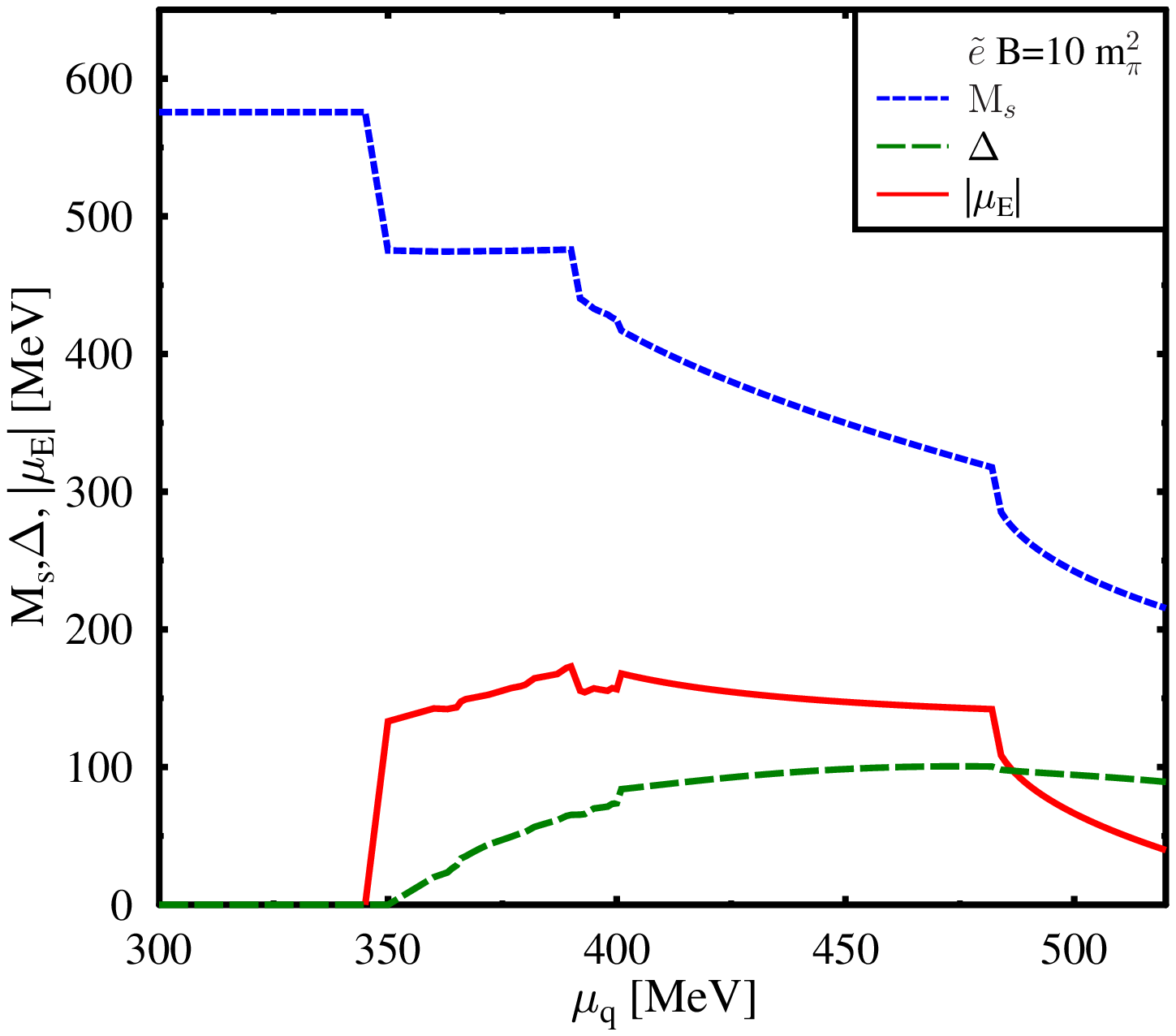}\\
Fig. 9-a & Fig.9-b\\
\includegraphics[width=8cm,height=8cm]{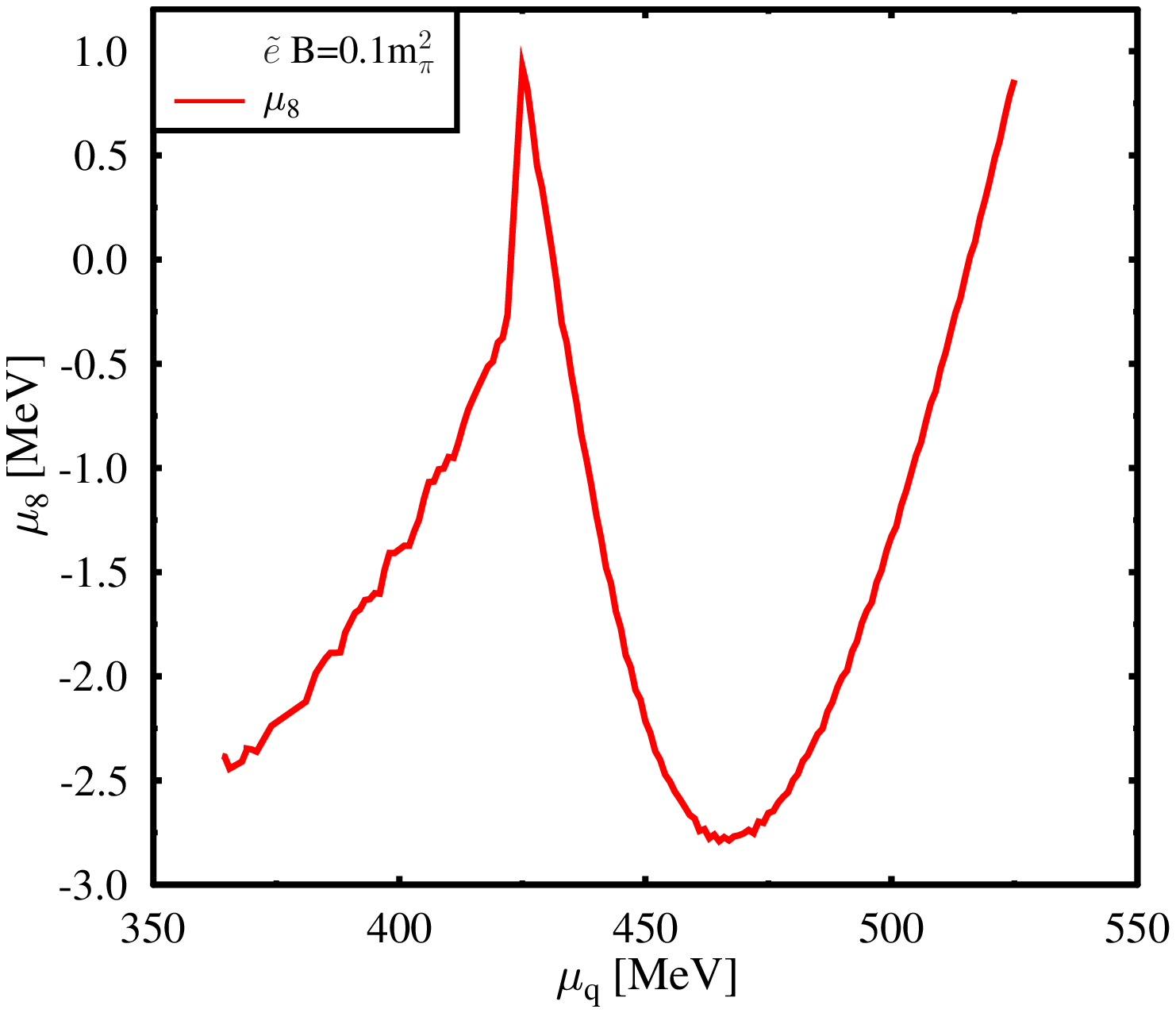}&
\includegraphics[width=8cm,height=8cm]{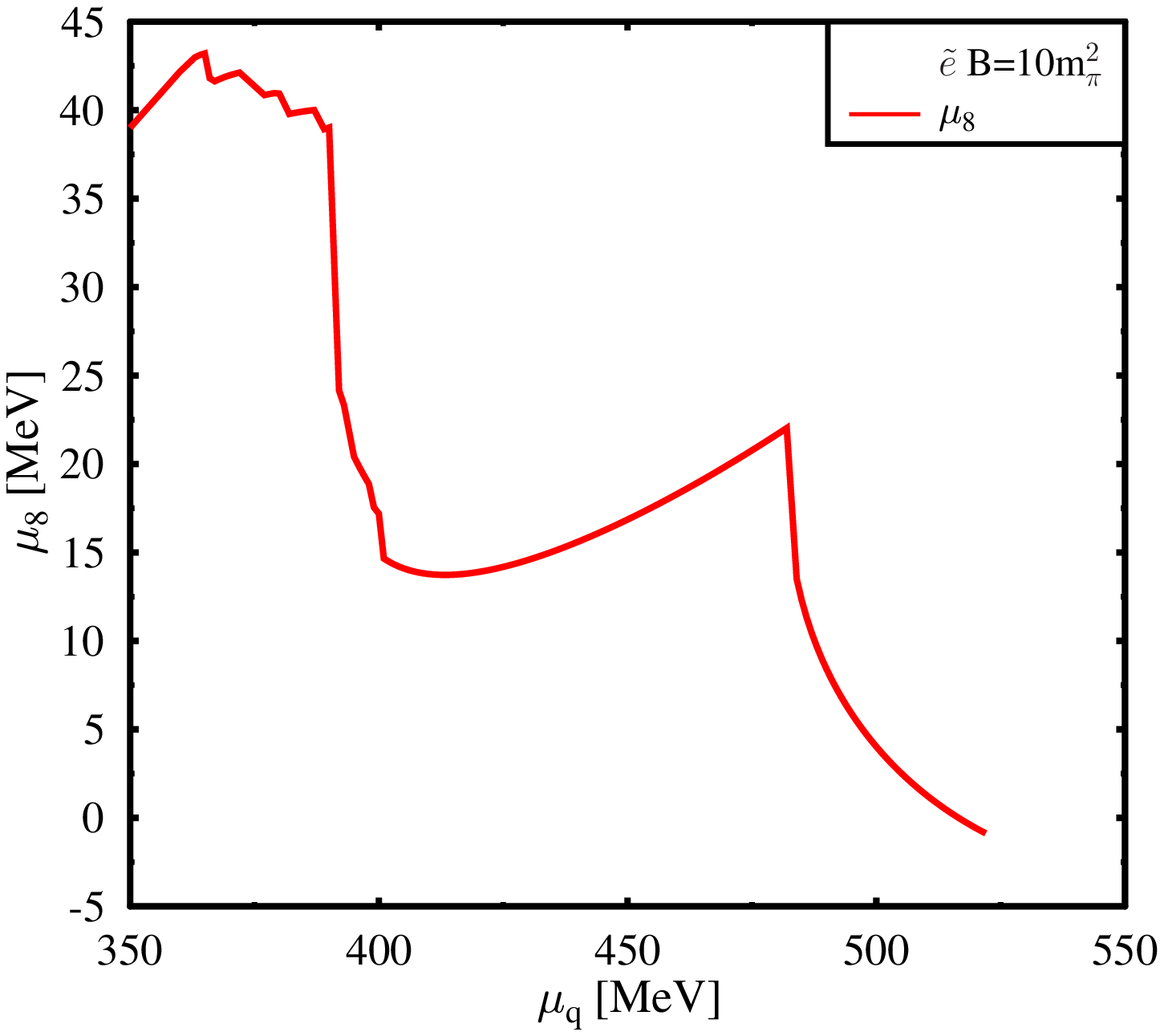}\\
Fig. 9-c & Fig.9-d
\end{tabular}
\end{center}
\caption{ Chemical potential $\mu_E$ and $\mu_8$ for charge neutral quark matter. $|\mu_E|$ is plotted as a function of
quark chemical potential $\mu_q$ for magnetic field $\tilde eB=0.1m_\pi^2$ (Fig.9-a )and for $\tilde eB=10m_\pi^2$ (Fig. 9-b). In Fig.9-a and Fig.9-b we have also plotted the mass of strange quarks and superconducting gap as a function of quark chemical potential to highlight the dependence of charge chemical potential on these two parameters. 
In the lower two plots, 
the color chemical potential $\mu_8$ is plotted as a function $\mu_q$ for $\tilde e B=0.1 m_\pi^2$ (Fig 9-c) and for 
$\tilde e B=10m_\pi^2$.
}
\label{fig9}
\end{figure}

In Fig.\ref{fig9}, we have plotted the electric and color chemical potentials $\mu_E$ and $\mu_8$ to maintain the
electric and color charge neutrality conditions given in Eq.(\ref{neutrality}) and Eq.(\ref{colneutrality}) as
a function of quark chemical potential.
For 2+1 flavor matter, strange quarks play an important role in maintaining charge neutrality. As the the quark 
chemical potential increases, $|\mu_E|$ increases to maintain charge neutrality. When the chemical potential 
becomes large enough for strange quarks to contribute to densities, they also help in maintaining charge neutrality. 
This leads to decrease in electron density or the corresponding chemical potential $|\mu_E|$. This behavior
is reflected in Fig. 9-a and 9-b as the initial slow rise of the $|\mu_E|$. However, as $|\mu_E|$ increases, the
difference $\delta\mu=-\mu_E/2$ also increases and at $\mu_1$, the condition 
 $\delta \mu >$ $\Delta$ for gapless modes to exist ceases to be satisfied. At the gapless to BCS transition point, the
u-quark number density increases while that of d-quarks decreases and both become equal as in the usual BCS pairing phase. 
This leads to an increase in the positive electric charge density. To maintain electrical charge neutrality, 
the electron density increases at this point. Therefore gapless to BCS transition is accompanied with an increase
in $|\mu_E|$. On the other hand, at higher densities when strange quarks start contributing to the density, it is accompanied with a drop in $|\mu_E|$ as strange quarks help in maintaining the charge neutrality along with the electrons. It turns out that
for $\tilde eB=0.1m_\pi^2$, the strange quarks densities become non vanishing after the gapless to BCS transition. This leads to the continuous decrease in the $|\mu_E|$ in the BCS phase as seen in Fig. 9-a. On the other hand, for 
larger fields, e.g. $\tilde e B=10m_\pi^2$, chiral transition occurs at a lower $\mu_c$ due to magnetic catalysis and the strange
quark density starts becoming non vanishing at lower chemical potential. This leads to a decrease in $|\mu_E|$ at $\mu=392$MeV
as may be seen in Fig.9-b. At $\mu=400$ MeV, there is the transition from the gapless to BCS phase and is
accompanied with a rise in $|\mu_E|$ as discussed above. Beyond $\mu=400$ MeV, $|\mu_E|$ starts decreasing monotonically as
strange quark density increase.

\begin{figure}
\vspace{-0.4cm}
\begin{center}
\begin{tabular}{c c }
\includegraphics[width=8cm,height=8cm]{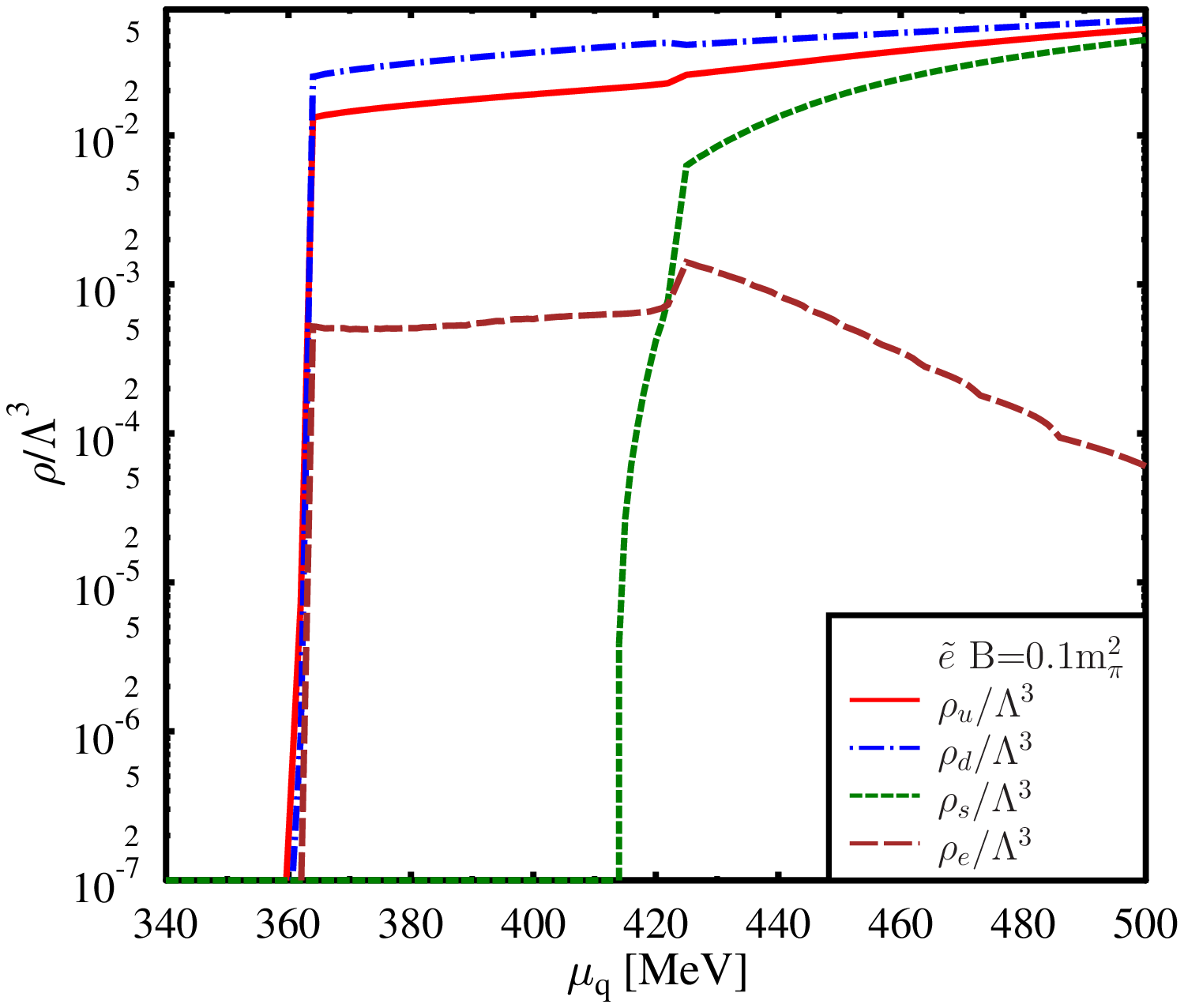}&
\includegraphics[width=8cm,height=8cm]{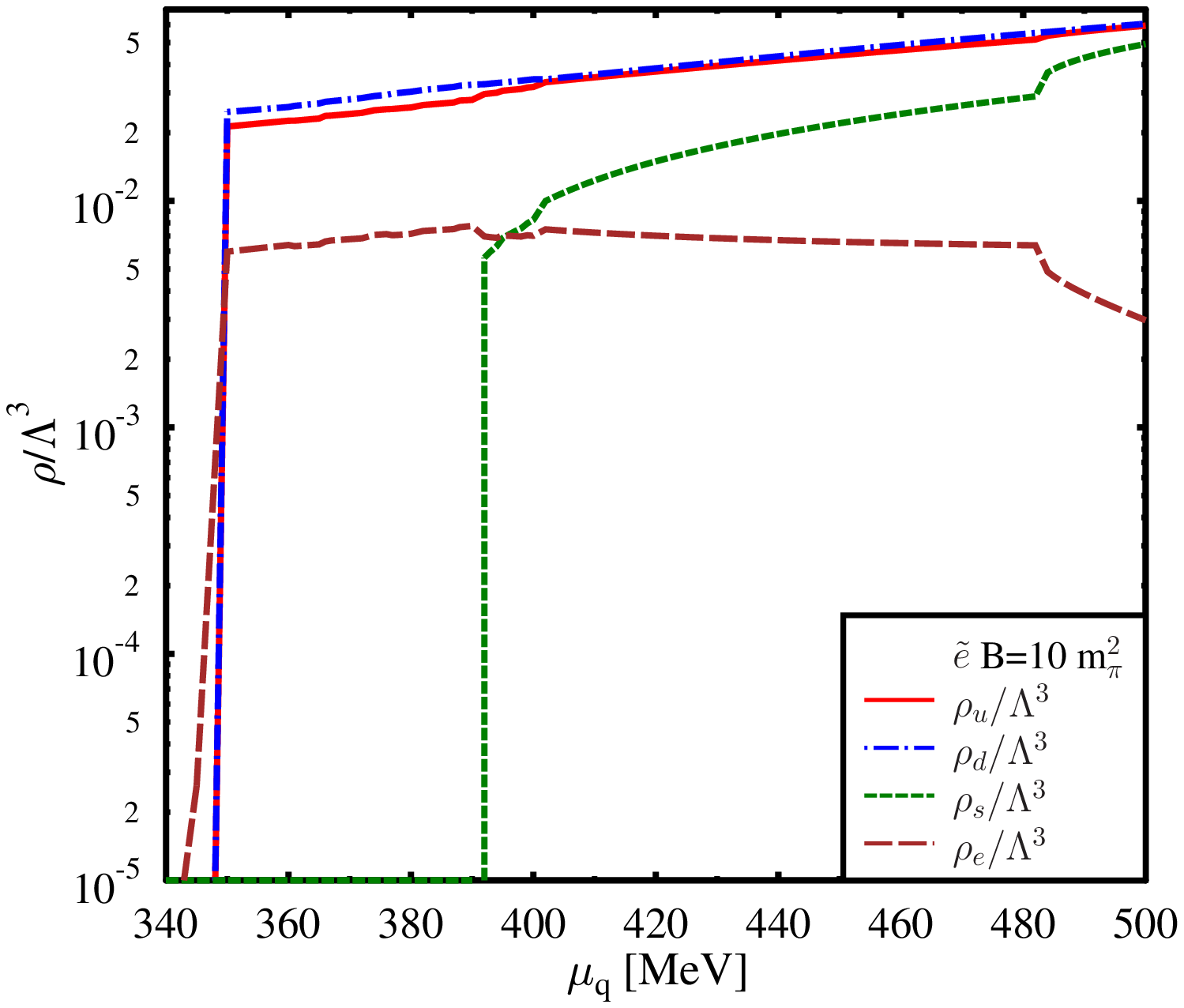}\\
Fig. 10-a & Fig.10-b
\end{tabular}
\end{center}
\caption{ Population of different species for charge neutral quark matter for $\tilde e B=0.1m_\pi^2$ (Fig. 10-a) and
for $\tilde eB=10m_\pi^2$ (Fig. 10-b).
}
\label{fig10}
\end{figure}

In Fig.9-c and Fig.9-d, we have plotted the color chemical potential $\mu_8$. For weak field case, $\mu_8$ is rather small
 (few MeVs) compared to both the electric chemical potential as well as the quark chemical potential which are
 two orders of magnitude larger. For small field, the difference in densities of red and green quarks 
and the blue quarks essentially arises because of the difference in the distribution functions. This results in a
small but finite net color charge. To maintain color neutrality one needs a small $\mu_8$ . On the other hand,
at large magnetic field, the net color charge difference become larger as the $\tilde e$ charges of red and green quarks
and that of blue quarks are different. This requires a somewhat larger $\mu_8$ to maintain color neutrality as seen in Fig.9-d.
In Fig.\ref{fig10} we have plotted the number densities of each species for the charge neutral matter for two different magnetic fields. As may be clear from both the plots the electron number densities gets correlated with the strange quark number densities.

\begin{figure}[h!]
\includegraphics[width=8cm,height=8cm]{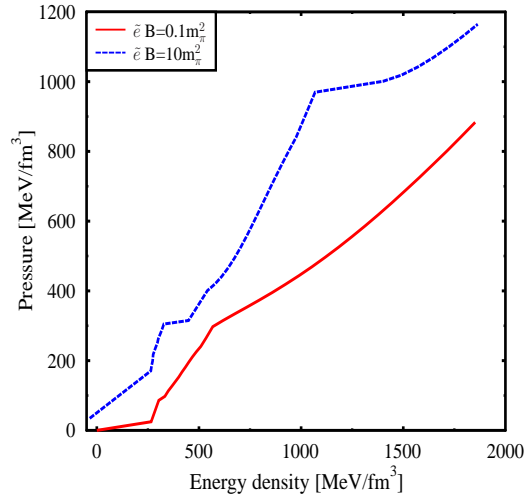}
\caption{Equation of state for $\tilde eB=0.1m_\pi^2$
(dashed line) and $\tilde eB=10m_\pi^2$ (solid line)}
\label{fig11}
\end{figure}
	   Finally, we discuss the equation of state (EOS) for different magnetic fields. In Fig.\ref{fig11} 
 we have plotted pressure as a function of energy for $\tilde{e}$B=0.1$m_\pi^2$ and 10$m_\pi^2$. 
  One can observe that the EOS become stiffer with increase in magnetic field. This can be understood as follows.
For $\mu<\mu_c$, the thermodynamic potential contribution from the field as in Eq. (\ref{omgsc1}), Eq.(\ref{omgub1}),
Eq.(\ref{omgsrg1}) is dominant and decreases with increase in magnetic field. This leads to a higher pressure for
higher magnetic field. As the chemical potential increases, for $\mu>\mu_c$, the medium contribution become dominant.
As the masses decrease with magnetic field, the medium contribution increases with magnetic field. Moreover,
the field contributions also leads to an increase in pressure. Both these effects make the resulting EOS stiffer at higher
magnetic field as may be seen in Fig.\ref{fig11}.

%-------------------------------------------------------------------------------------------------------

%We then study the effect of magnetic field on the equation of state, i.e.
%pressure as a function of energy for the charge neutral matter. This is shown in Fig.
%\ref{fig7} for zero temperature. The effect of Landau quantization shows up
%in the kink structure of the equation of state. For smaller magnetic fields, this effect
%is less visible as the number of filled Landau levels are quite large. Further,
%it may be observed that as the magnetic field is increased, 
%the equation of state
%becomes somewhat stiffer. The reason for this is as follows.
%The contribution to the
%thermodynamic potential due to the
%magnetic field as given in Eq.(\ref{omegafield}) is negative which leads to a decrease
%in the energy density as compared to the zero field case. This means that one has to
%have a larger chemical potential with nonzero magnetic field as compared to zero field 
%to have the same energy density. Now with higher chemical potential, the pressure 
%becomes higher. Therefore the value of pressure
%with field is higher than without the magnetic field for the same 
%energy density,
%leading to a stiffer equation of state in the presence of magnetic field.
%\begin{figure}
%\includegraphics[width=8cm,height=8cm]{eost0del.eps}
%\caption{Equations of state for charge neutral matter at zero temperature
%for different strengths of the magnetic field.}
%\label{fig7}
%\end{figure}

\section{summary}
We have analyzed here the effect of magnetic field and neutrality conditions 
on the chiral as well as diquark condensates within the framework of a three flavor
NJL model. This essentially generalizes the results of Ref.\cite{amhmbhas} to include the
u-d superconductivity in presence of magnetic field.
The methodology
uses an explicit variational construct for the ground state in terms of
quark-antiquark pairing for all the three flavors as well as diquark pairing for the light quarks.
A nice feature of the approach is that the 
four component quark field operator in presence of magnetic field could get expressed
in terms of the ansatz functions that appears for the description of the ground state.
Apart from the methodology being different, we also have new results. Namely,
the present investigations have been done in a three flavor NJL model along 
with a flavor mixing six quark determinant interaction at finite temperature and density
and fields within the same framework. In that sense it generalizes the
two flavor superconductivity in presence of magnetic field considered earlier in Ref.s\cite{digal,iran,ferrerscmag} 
and Ref.\cite{scoccola}.
The gap functions and 
the thermal distribution functions could be determined self consistently for 
given values of the temperature, the quark chemical potential and the 
strength of magnetic field.

For the charge neutral matter the chiral transition is a first order transition and we observe inverse magnetic catalysis at finite density. The chiral condensate for strange quark affects the u-d superconductivity through the flavor mixing determinant interaction. The effective diquark coupling increases in presence of strange quark condensates. On the other hand the diquark condensates contribute to the mass of the strange quark through the determinant interaction. Inverse magnetic catalysis is observed for magnetic fields upto 19 $m_\pi^2$. Beyond it magnetic catalysis is observed for chiral symmetry breaking \cite{andreas1}.

At finite densities, the effects of Landau quantization  get manifested in the 
oscillation of the order parameters
similar to the de Hass van Alphen effect for 
magnetization in metals. However, in the present case of dense quark matter, 
the order parameters, the masses and the superconducting gap themselves are
dependant on the strength of magnetic fields which leads to a non periodic oscillation
of the order parameter. 

Imposition of charge neutrality
condition for the quark matter leads to gapless modes even in presence of magnetic field. The superconducting gaps in gapless modes are smaller compared to the gaps in the BCS phase. The transition from gapless to BCS phase is a sharp transition. Difference in the gap in the two phases at this transition decreases with magnetic field. For charge neutral matter the strange quark plays an important role in maintaining the charge neutrality. This leads to a depletion of electron density at higher chemical potential where strange quarks start to contribute to the densities. The resulting equation of state becomes stiffer with magnetic field.

We have considered here quark anti-quark pairing and diquark pairing in the ansatz for ground state which is homogeneous with zero total momentum. However it is possible that the condensates be spatially inhomogeneous\cite{sreemoyee} with a net total momentum\cite{dunne,frolov,nickel,abuki}. Indeed, 
 the gapless modes for the charge neutral matter leads to instability arising from imaginary Meissener masses for some of the gluons when $\delta \mu > \Delta$\cite{huang2004}. This can be suggestive of having inhomogeneous superconducting phases\cite{bowers2002,rajasharma2006} which are not considered here. The phase structure here would be non-trivial and interesting in presence of two vectors, the magnetic field and non-zero momentum of the condensate. Furthermore, the equation of state derived for charge neutral quark matter combined with same for hadronic matter can be used to study structural properties of neutron star with quark matter core. It will be interesting to see the compatibility of such equation of state which is constrained by astrophysical observations like GW170817\cite{fattoyev}. Some of these investigations are in progress and will be reported elsewhere.

\acknowledgments
The authors would like to thank Amruta Mishra for many discussions.
\appendix
\section{ Evaluation of operator expectation values of some operators}

We give here some details of the evaluation of some operators at finite T,$\mu$ and B in the state 
given in Eq.(\ref{ubt}). As the state is obtained from $|0\rangle$, one can calculate the expectation values  of
different operators. e.g.
\be
\langle q^{ia \dagger}_r(n,k_{\omit x}),q^{jb}_{r^\prime}(n^{\prime},k^{\prime}_{\omit x})\rangle=
\delta^{ij}\delta^{ab}\delta_{rr^\prime}\delta_{nn^\prime}\delta(\zbf k_{\omit x}-\zbf k_{\omit x}^\prime)F^{ia}(\zbf k_{\omit x}).
\label{q1}
\ee
where,
\be 
F^{ia}(\zbf k_{\omit x})=\sin^2\theta_-^{ia}+\sin^2 f \left(1-\sin^2\theta_-^{ia}-|\epsilon^{ij}\epsilon^{ab}\sin^2\theta_-^{jb}\right)(1-\delta^{a3})(1-\delta^{i3}).
\ee
Similarly for the expectation values for the operators involving anti-quarks, we have
\be
\langle \tilde q^{ia \dagger}_r(n,k_{\omit x}),\tilde q^{jb}_{r^\prime}(n^{\prime},k^{\prime}_{\omit x})\rangle=
\delta^{ij}\delta^{ab}\delta_{rr^\prime}\delta_{nn^\prime}\delta(\zbf k_{\omit x}-\zbf k_{\omit x}^\prime)(1-F_1^{ia}(\zbf k_{\omit x}).
\label{tildeq1}
\ee
where,
\be 
F_1^{ia}(\zbf k_{\omit x})=\sin^2\theta_+^{ia}+\sin^2 f_1 \left(1-\sin^2\theta_+^{ia}-|\epsilon^{ij}\epsilon^{ab}\sin^2\theta_+^{jb}\right)(1-\delta^{a3})(1-\delta^{i3}).
\ee
Using the field operator expansion of Eq.(\ref{psip}) and Eq.s (\ref{q1}) and (\ref{tildeq1}), one can evaluate
\be
\langle \psi^{ia\dagger}_\alpha(\zbf x)\psi^{jb}_\beta(\zbf y)\rangle=\sum_{n}\frac{|q_iB|}{(2\pi)^2}\int d
k_{\omit x} e^{i k_{\omit x}\cdot (\zbf x-\zbf y)}{\Lambda_{-}^{ia,jb}}_{\beta\alpha}(n,k_{\omit x})
\label{lambdam}
\ee
with 
\be
\Lambda_-^{ia,jb}=\delta^{ij}\delta^{ab}\left[F^{ia}(n,k_z)U_{\beta r}(n,k_{\omit x})U_{r\alpha}(n,k_{\omit x})^\dagger
+F^{ia}_1(n,k_z)V_{\beta r}(n,-k_{\omit x})V_{r\alpha}(n,-k_{\omit x})^\dagger\right]
\label{lambdam}
\ee

Explicitly,
\bearr
U_r(n,\vec p_{\omit x}) U_r^\dagger (n,\vec p_{\omit x})   &  = &
\frac{1}{2}
\left(\begin{array}{cccc}
(1+\cos\phi)I_n^2 & 0 &\hat p_z\sin\phi I_n^2 & i\hat p_\perp\sin\phi I_nI_{n-1}\\
 0 & (1+\cos\phi)I_{n-1}^2 & -i\hat p_\perp\sin\phi I_nI_{n-1} & -\hat p_z\sin\phi I_{n-1}^2\\
\hat p_z\sin\phi I_n^2 & i\hat p_\perp\sin\phi I_nI_{n-1} & (1-\cos\phi)I_n^2 & 0 \\
 -i\hat p_\perp\sin\phi I_nI_{n-1} & -\hat p_z\sin\phi I_{n-1}^2 & 0 & (1-\cos\phi)I_{n-1}^2  \\
\end{array}\right).\nonumber\\
&=&\frac{1}{2}\bigg[I_n^2(1+\gamma^0\cos\phi)\Pi^++I_{n-1}^2(1+\gamma^0\cos\phi)\Pi^-
+\frac{\hat p_z}{2}\sin\phi\left(\gamma_0\gamma^3(I_n^2+I_{n-1}^2)+\gamma^5(I_n^2-I_{n-1}^2)\right)\nonumber\\
&-&\hat p_\perp\sin\phi\gamma^2\gamma^0\bigg]\nonumber\\
\eearr
where, we have defined $\Pi^{\pm}=(1\pm i\gamma^1\gamma^2)/2$.

Similarly for the anti-quark spinors
\bearr
V_r(n,-\vec p_{\omit x}) V_r^\dagger (n,-\vec p_{\omit x})  &  =&
\frac{1}{2}
\left(\begin{array}{cccc}
(1-\cos\phi)I_n^2 & 0 & -\hat p_z\sin\phi I_n^2 & -i\hat p_\perp\sin\phi I_nI_{n-1}\\
 0 & (1-\cos\phi)I_{n-1}^2 & i\hat p_\perp\sin\phi I_nI_{n-1} & \hat p_z\sin\phi I_{n-1}^2\\
-\hat p_z\sin\phi I_n^2 & -i\hat p_\perp\sin\phi I_nI_{n-1} & (1+\cos\phi)I_n^2 & 0 \\
 i\hat p_\perp\sin\phi I_nI_{n-1} & \hat p_z\sin\phi I_{n-1}^2 & 0 & (1+\cos\phi)I_{n-1}^2  \\
\end{array}\right).\nonumber\\
&=&\frac{1}{2}\bigg[I_n^2(1-\gamma^0\cos\phi)\Pi^++I_{n-1}^2(1-\gamma^0\cos\phi)\Pi^-
-\frac{\hat p_z}{2}\sin\phi\left(\gamma_0\gamma^3(I_n^2+I_{n-1}^2)+\gamma^5(I_n^2-I_{n-1}^2)\right)\nonumber\\
&+&\hat p_\perp\sin\phi\gamma^2\gamma^0\bigg]\nonumber\\
\eearr
This leads to, e.g. for the expectation value of chiral condensate for a given flavor as
\be
I_s^i=\langle\bar\psi^i\psi^i\rangle=-\frac{1}{(2\pi)^2}\sum_n\sum_a\int dp_ydp_z\left(1-F^{ia}-F_1^{ia}\right)\cos\phi_n^i(I_n^2+I_{n-1}^2)
\ee
One can integrate over $dp_y$ to obtain the contribution for the quarks that are charged as
\be
I_s^i=\sum_a\sum_n\frac{\alpha_n}{(2\pi)^2}|q_iB|\int dp_z \left(1-F^{ia}-F_1^{ia}\right)\cos\phi_n^i
\ee
while, the contribution from the quarks that are neutral (down blue strange blue ) is given as
\be
I_s^i=\frac{2}{(2\pi)^3}\int d\zbf p\cos\phi^i(1-\sin^2\theta^{i3}_--\sin^2\theta^{i3}_+)\quad (i=2,3)
\ee

Next, we discuss about the contributions to diquark condensates.Similar to Eq.(\ref{q2}), we have
\bearr
\langle q^{ia }_r(n,k_{\omit x}),q^{jb}_{r^\prime}(n^{\prime},k^{\prime}_{\omit x})\rangle &=&
r\delta_{r,-r^\prime}\epsilon^{ij}\epsilon^{3ab}\delta_{nn^\prime}\delta(\zbf k_{\omit x}+\zbf k_{\omit x}^\prime)
\sin2 f(n,k_z)\left(1-\sin^2\theta_-^{ia}-\sin^2\theta_-^{jb}\right)\nonumber\\ &\equiv&
r\delta_{r,-r^\prime}\epsilon^{ij}\epsilon^{3ab}\delta_{nn^\prime}\delta(\zbf k_{\omit x}+\zbf k_{\omit x}^\prime)
G(k_z,n)
\label{q2}
\eearr
and, for anti-quark operators
\bearr
\langle \tilde q^{ia }_r(n,k_{\omit x}),\tilde q^{jb}_{r^\prime}(n^{\prime},k^{\prime}_{\omit x})\rangle &=&
r\delta_{r,-r^\prime}\epsilon^{ij}\epsilon^{3ab}\delta_{nn^\prime}\delta(\zbf k_{\omit x}+\zbf k_{\omit x}^\prime)
\sin2 f(n,k_z)\left(1-\sin^2\theta_-^{ia}-\sin^2\theta_-^{jb}\right)\nonumber \\ &\equiv&
r\delta_{r,-r^\prime}\epsilon^{ij}\epsilon^{3ab}\delta_{nn^\prime}\delta(\zbf k_{\omit x}+\zbf k_{\omit x}^\prime)
G_1(k_z,n)
\label{tildeq2}
\eearr

For the diquark condensates we have \\

\be
\langle \psi^{ia}_\alpha(\zbf x)\psi^{jb}_\beta(\zbf y)\rangle=\epsilon^{ij} \epsilon^{3ab} \sum_{n}\frac{|q_iB|}{(2\pi)^2}\int d
k_{\omit x} e^{i k_{\omit x}\cdot (\zbf x-\zbf y)} \left [ P_u C \gamma^5 G(k_z,n)+ P_v C \gamma^5 G_1 (k_z,n)\right ]_{\beta \alpha} 
\label{diquark1}
\ee

where $P_u C \gamma^5=\sum_r r U_{\alpha r} U^{'}_{-r \beta}$ and $P_v C \gamma^5=\sum_{r} r V_{\alpha r} V^{'}_{-r \beta}$
and the prime on the spinors denotes a spinor with opposite charge and momentum corresponding to the unprimed spinors. Explicitly,
\bearr
P_u   &  = &
\frac{1}{2}
\left(\begin{array}{cccc}
\cos \frac{\phi}{2}\cos \frac{\phi^\prime}{2} I_n^2 & 0 &\hat p_z \cos \frac{\phi}{2} \sin \frac{\phi^\prime}{2} I_n^2 & i\hat p_\perp \cos \frac{\phi}{2} \sin \frac{\phi^\prime}{2} I_nI_{n-1}\\
 0 & \cos \frac{\phi}{2}\cos \frac{\phi^\prime}{2} I_{n-1}^2 & -i\hat p_\perp \cos \frac{\phi}{2} \sin \frac{\phi^\prime}{2} I_nI_{n-1} & -\hat p_z \cos \frac{\phi}{2} \sin \frac{\phi^\prime}{2} I_{n-1}^2\\
\hat p_z \cos \frac{\phi^\prime}{2} \sin \frac{\phi}{2} I_n^2 & i\hat p_\perp \cos \frac{\phi^\prime}{2} \sin \frac{\phi}{2} I_nI_{n-1} & \sin \frac{\phi}{2} \sin \frac{\phi^\prime}{2} I_n^2 & 0 \\
 -i\hat p_\perp \sin \frac{\phi}{2} \cos \frac{\phi^\prime}{2} I_nI_{n-1} & -\hat p_z \sin \frac{\phi}{2} \cos \frac{\phi^\prime}{2} I_{n-1}^2 & 0 & \sin \frac{\phi}{2} \sin \frac{\phi^\prime}{2} I_{n-1}^2  \\
\end{array}\right)
\eearr

and,

\bearr
P_v   &  = &
\frac{1}{2}
\left(\begin{array}{cccc}
- \sin \frac{\phi}{2}\sin \frac{\phi^\prime}{2} I_n^2 & 0 &\hat p_z \sin \frac{\phi}{2} \cos \frac{\phi^\prime}{2} I_n^2 & i\hat p_\perp \sin \frac{\phi}{2} \cos \frac{\phi^\prime}{2} I_nI_{n-1}\\
 0 & - \sin \frac{\phi}{2}\sin \frac{\phi^\prime}{2} I_{n-1}^2 & -i\hat p_\perp \sin \frac{\phi}{2} \cos \frac{\phi^\prime}{2} I_nI_{n-1} & -\hat p_z \sin \frac{\phi}{2} \cos \frac{\phi^\prime}{2} I_{n-1}^2\\
\hat p_z \cos \frac{\phi}{2} \sin \frac{\phi^\prime}{2} I_n^2 & i\hat p_\perp \cos \frac{\phi}{2} \sin \frac{\phi^\prime}{2} I_nI_{n-1} & -\cos \frac{\phi}{2} \cos \frac{\phi^\prime}{2} I_n^2 & 0 \\
 -i\hat p_\perp \cos \frac{\phi}{2} \sin \frac{\phi^\prime}{2} I_nI_{n-1} & -\hat p_z \cos \frac{\phi}{2} \sin \frac{\phi^\prime}{2} I_{n-1}^2 & 0 & -\cos \frac{\phi}{2} \cos \frac{\phi^\prime}{2} I_{n-1}^2  \\
\end{array}\right)
\eearr

This leads to e.g. for expectation value of the diquark condensate as,

\bearr
I_D&=&\langle\bar\psi_{c}^{ia} \gamma^5 \psi^{jb} \rangle \epsilon^{ij} \epsilon^{3ab}\nonumber \\ &=&\frac{2}{(2\pi)^2}\sum_n \alpha_n |q_i B| \int dp_z \cos \left(\frac{\phi_1 - \phi_2}{2}\right)\bigg[ \sin 2f\left(1-\sin^2 \theta_{-}^{1}-\sin^2 \theta_{-}^{2}\right)\nonumber \\ &+& \sin2f_1 \left(1-\sin^2 \theta_{+}^{1}-\sin^2 \theta_{+}^{2}\right)\bigg]
\eearr

\def \ltg{R.P. Feynman, Nucl. Phys. B 188, 479 (1981); 
K.G. Wilson, Phys. Rev. \zbf  D10, 2445 (1974); J.B. Kogut,
Rev. Mod. Phys. \zbf  51, 659 (1979); ibid  \zbf 55, 775 (1983);
M. Creutz, Phys. Rev. Lett. 45, 313 (1980); ibid Phys. Rev. D21, 2308
(1980); T. Celik, J. Engels and H. Satz, Phys. Lett. B129, 323 (1983)}

\def\berges {J. Berges, K. Rajagopal, {\NPB{538}{215}{1999}}.}% Nucl. Phys. B538, 215, (1999).}
\def \svz {M.A. Shifman, A.I. Vainshtein and V.I. Zakharov,
Nucl. Phys. B147, 385, 448 and 519 (1979);
R.A. Bertlmann, Acta Physica Austriaca 53, 305 (1981)}

\def \spmbst {S.P. Misra, Phys. Rev. D35, 2607 (1987)}

\def \hmgrnv { H. Mishra, S.P. Misra and A. Mishra,
Int. J. Mod. Phys. A3, 2331 (1988)}

\def \snss {A. Mishra, H. Mishra, S.P. Misra
and S.N. Nayak, Phys. Lett 251B, 541 (1990)}

\def \amqcd { A. Mishra, H. Mishra, S.P. Misra and S.N. Nayak,
Pramana (J. of Phys.) 37, 59 (1991). }
\def\qcdtb{A. Mishra, H. Mishra, S.P. Misra 
and S.N. Nayak, Z.  Phys. C 57, 233 (1993); A. Mishra, H. Mishra
and S.P. Misra, Z. Phys. C 58, 405 (1993)}
% pure qcd at zero and finite temperature and finite baryon densities

\def \spmtlk {S.P. Misra, Talk on {\it `Phase transitions in quantum field
theory'} in the Symposium on Statistical Mechanics and Quantum field theory, 
Calcutta, January, 1992, hep-ph/9212287}

\def \hmspmnjl {H. Mishra and S.P. Misra, 
{\PRD{48}{5376}{1993}.}}
%Phys. Rev. D {\bf 48},5376 (1993)}

\def \hmqcd {A. Mishra, H. Mishra, V. Sheel, S.P. Misra and P.K. Panda,
hep-ph/9404255 (1994)}

\def \amcrl {A. Mishra, H. Mishra and S.P. Misra, Z. Phys. C 57, 241 (1993)}

\def \higgs { S.P. Misra, in {\it Phenomenology in Standard Model and Beyond}, 
Proceedings of the Workshop on High Energy Physics Phenomenology, Bombay,
edited by D.P. Roy and P. Roy (World Scientific, Singapore, 1989), p.346;
A. Mishra, H. Mishra, S.P. Misra and S.N. Nayak, Phys. Rev. D44, 110 (1991)}

\def \nmtr {A. Mishra, 
H. Mishra and S.P. Misra, Int. J. Mod. Phys. A5, 3391 (1990); H. Mishra,
 S.P. Misra, P.K. Panda and B.K. Parida, Int. J. Mod. Phys. E 1, 405, (1992);
 {\it ibid}, E 2, 547 (1993); A. Mishra, P.K. Panda, S. Schrum, J. Reinhardt
and W. Greiner, to appear in Phys. Rev. C}
 % nuclear matter

\def \dtrn {P.K. Panda, R. Sahu and S.P. Misra, 
Phys. Rev C45, 2079 (1992)}
\def\hurwitz{E. Elizalde, J. Phys. {\bf A}:Math. Gen. 18,1637 (1985).}

\def \qcd {G. K. Savvidy, Phys. Lett. 71B, 133 (1977);
S. G. Matinyan and G. K. Savvidy, Nucl. Phys. B134, 539 (1978); N. K. Nielsen
and P. Olesen, Nucl.  Phys. B144, 376 (1978); T. H. Hansson, K. Johnson,
C. Peterson Phys. Rev. D26, 2069 (1982)}
% qcd vacuum considered earlier

\def \cornwal {J.M. Cornwall, Phys. Rev. D26, 1453 (1982)}
\def\aichlin {F. Gastineau, R. Nebauer and J. Aichelin,
{\PRC{65}{045204}{2002}}.}
% Phys. Rev. C65, 045204 (2002).}

\def \mndglv {J. E. Mandula and M. Ogilvie, Phys. Lett. 185B, 127 (1987)}

\def \schwinger {J. Schwinger, Phys. Rev. 125, 1043 (1962); ibid,
127, 324 (1962)}

\def \schutte {D. Schutte, Phys. Rev. D31, 810 (1985)}

\def \amspm {A. Mishra and S.P. Misra, 
{\ZPC{58}{325}{1993}}.}
%Z. Phys. C 58, 325 (1993)}

\def \gft{ For gauge fields in general, see e.g. E.S. Abers and 
B.W. Lee, Phys. Rep. 9C, 1 (1973)}

\def \gribov {V.N. Gribov, Nucl. Phys. B139, 1 (1978)}

\def \spm78 {S.P. Misra, Phys. Rev. D18, 1661 (1978); {\it ibid}
D18, 1673 (1978)} 

\def \lopr {A. Le Youanc, L.  Oliver, S. Ono, O. Pene and J.C. Raynal, 
Phys. Rev. Lett. 54, 506 (1985)}

\def \spphi {S.P. Misra and S. Panda, Pramana (J. Phys.) 27, 523 (1986);
S.P. Misra, {\it Proceedings of the Second Asia-Pacific Physics Conference},
edited by S. Chandrasekhar (World Scientific, 1987) p. 369}

\def\spmdif {S.P. Misra and L. Maharana, Phys. Rev. D18, 4103 (1978); 
    S.P. Misra, A.R. Panda and B.K. Parida, Phys. Rev. Lett. 45, 322 (1980);
    S.P. Misra, A.R. Panda and B.K. Parida, Phys. Rev. D22, 1574 (1980)}

\def \spmvdm {S.P. Misra and L. Maharana, Phys. Rev. D18, 4018 (1978);
     S.P. Misra, L. Maharana and A.R. Panda, Phys. Rev. D22, 2744 (1980);
     L. Maharana,  S.P. Misra and A.R. Panda, Phys. Rev. D26, 1175 (1982)}

\def\spmthr {K. Biswal and S.P. Misra, Phys. Rev. D26, 3020 (1982);
               S.P. Misra, Phys. Rev. D28, 1169 (1983)}

\def \spmstr { S.P. Misra, Phys. Rev. D21, 1231 (1980)} 

\def \spmjet {S.P. Misra, A.R. Panda and B.K. Parida, Phys. Rev Lett. 
45, 322 (1980); S.P. Misra and A.R. Panda, Phys. Rev. D21, 3094 (1980);
  S.P. Misra, A.R. Panda and B.K. Parida, Phys. Rev. D23, 742 (1981);
  {\it ibid} D25, 2925 (1982)}

\def \arpftm {L. Maharana, A. Nath and A.R. Panda, Mod. Phys. Lett. 7, 
2275 (1992)}

\def \van {R. Van Royen and V.F. Weisskopf, Nuov. Cim. 51A, 617 (1965)}

\def \rchpi {S.R. Amendolia {\it et al}, Nucl. Phys. B277, 168 (1986)}
% pion charge radius (experimental) .66+-.01 fm; rch2=11.22 Gev-2.

\def \chrl{ Y. Nambu, {\PRL{4}{380}{1960}};
%Phys. Rev. Lett. \zbf 4, 380 (1960);
A. Amer, A. Le Yaouanc, L. Oliver, O. Pene and
J.C. Raynal,{\PRL{50}{87}{1983a}};{\em ibid}
{\PRD{28}{1530}{1983}};
% Phys. Rev. Lett.\zbf  50, 87 (1983);
%ibid, Phys. Rev.\zbf  D28, 1530 (1983); 
M.G. Mitchard, A.C. Davis and A.J.
MAacfarlane, {\NPB{325}{470}{1989}};
%Nucl. Phys. \zbf B325, 470 (1989);
B. Haeri and M.B. Haeri,{\PRD{43}{3732}{1991}};
% Phys. Rev.\zbf  D43, 3732 (1991); 
V. Bernard,{\PRD{34}{1604}{1986}};
% Phys. Rev.\zbf  D34, 1601 (1986);
 S. Schramm and
W. Greiner, Int. J. Mod. Phys. \zbf E1, 73 (1992), 
J.R. Finger and J.E. Mandula, Nucl. Phys. \zbf B199, 168 (1982),
S.L. Adler and A.C. Davis, Nucl. Phys.\zbf  B244, 469 (1984),
S.P. Klevensky, Rev. Mod. Phys.\zbf  64, 649 (1992).}
\def\klevansky{S.P. Klevensky, Rev. Mod. Phys.\zbf  64, 649 (1992).}
\def\sedrakianalford {M. Alford and A. Sedrakian, J Phys. G \zbf 37, 075202, 2010.}

\def \spmijp { S.P. Misra, Ind. J. Phys. 61B, 287 (1987)}

\def \feynman {R.P. Feynman and A.R. Hibbs, {\it Quantum mechanics and
path integrals}, McGraw Hill, New York (1965)}

\def \glstn{ J. Goldstone, Nuov. Cim. \zbf 19, 154 (1961);
J. Goldstone, A. Salam and S. Weinberg, Phys. Rev. \zbf  127,
965 (1962)}

\def \anderson {P.W. Anderson, Phys. Rev. \zbf {110}, 827 (1958)}

\def \nambu{ Y. Nambu, Phys. Rev. Lett. \zbf 4, 380 (1960)}

\def\donogh {J.F. Donoghue, E. Golowich and B.R. Holstein, {\it Dynamics
of the Standard Model}, Cambridge University Press (1992)}

\def\satz {T. Matsui and H. Satz, Phys. Lett. B178, 416 (1986)}

\def\cps {C. P. Singh, Phys. Rep. 236, 149 (1993)}

\def\prliop {A. Mishra, H. Mishra, S.P. Misra, P.K. Panda and Varun
Sheel, Int. J. of Mod. Phys. E 5, 93 (1996)}

\def\hmcor {V. Sheel, H. Mishra and J.C. Parikh, Phys. Lett. B382, 173
(1996); {\it biid}, to appear in Int. J. of Mod. Phys. E}
\def\cort { V. Sheel, H. Mishra and J.C. Parikh, Phys. ReV D59,034501 (1999);
{\it ibid}Prog. Theor. Phys. Suppl.,129,137, (1997).}
% J. Phys. G23,143, (1997).}

\def\surcor {E.V. Shuryak, Rev. Mod. Phys. 65, 1 (1993)} 

\def\stevenson {A.C. Mattingly and P.M. Stevenson, Phys. Rev. Lett. 69,
1320 (1992); Phys. Rev. D 49, 437 (1994)}

\def\mac {M. G. Mitchard, A. C. Davis and A. J. Macfarlane,
 Nucl. Phys. B 325, 470 (1989)} 
\def\tfd
 {H.~Umezawa, H.~Matsumoto and M.~Tachiki {\it Thermofield dynamics
and condensed states} (North Holland, Amsterdam, 1982) ;
P.A.~Henning, Phys.~Rep.253, 235 (1995).}
\def\amph4{Amruta Mishra and Hiranmaya Mishra,
{\JPG{23}{143}{1997}}.}
\def\amhmbhas {Bhaswar Chatterjee, Hiranmaya Mishra and Amruta Mishra,{\PRD{84}{014016}{2011}.}}
\def \neglecor{M.-C. Chu, J. M. Grandy, S. Huang and 
J. W. Negele, Phys. Rev. D48, 3340 (1993);
ibid, Phys. Rev. D49, 6039 (1994)}

\def\revdata {Particle Data Group, Phys. Rev. D 50, 1173 (1994)}

\def\sinp {S.P. Misra, Indian J. Phys., {\bf 70A}, 355 (1996)}
\def\hmparikh{H. Mishra and J.C. Parikh, {\NPA{679}{597}{2001}.}}
% Nucl. Physics A679, 597 (2001).}
\def\krisch {M. Alford and K. Rajagopal, JHEP 0206,031,(2002)}
\def\reddy {A.W. Steiner, S. Reddy and M. Prakash,
{\PRD{66}{094007}{2002}.}}
\def\hmam {Amruta Mishra and Hiranmaya Mishra,
{\PRD{69}{014014}{2004}.}}
% Phys. Rev. D66, 094007, 2002}
\def\hmampp {Amruta Mishra and Hiranmaya Mishra,
in preparation.}
\def\bryman {D.A. Bryman, P. Deppomier and C. Le Roy, Phys. Rep. 88,
151 (1982)}
\def\thooft {G. 't Hooft, Phys. Rev. D 14, 3432 (1976); D 18, 2199 (1978);
S. Klimt, M. Lutz, U. Vogl and W. Weise, Nucl. Phys. A 516, 429 (1990)}
\def\alkz { R. Alkofer, P. A. Amundsen and K. Langfeld, Z. Phys. C 42,
199(1989), A.C. Davis and A.M. Matheson, Nucl. Phys. B246, 203 (1984).}
\def\sarah {T.M. Schwartz, S.P. Klevansky, G. Papp,
{\PRC{60}{055205}{1999}}.}
% Phys. Rev. C60, 055205 (1999)}
\def\wil{M. Alford, K.Rajagopal, F. Wilczek, {\PLB{422}{247}{1998}};
%Phys. Lett. B422,247(1998), 
{\it{ibid}}{\NPB{537}{443}{1999}}.}
%Nucl. Phys. B537,443 (1999).}
\def\sursc{R.Rapp, T.Schaefer, E. Shuryak and M. Velkovsky,
{\PRL{81}{53}{1998}};{\it ibid}{\AP{280}{35}{2000}}.}
% Phys. Rev. Lett.  81, 53(1998),{\it {ibid}},Ann. Phys. 280, 35, 2000.}
\def\pisarski{
D. Bailin and A. Love, {\PR{107}{325}{1984}},
%Phys. Rep. 107 (1984) 325,
D. Son, {\PRD{59}{094019}{1999}}; 
%Phys. Rev. D59 (1999) 094019,
T. Schaefer and F. Wilczek, {\PRD{60}{114033}{1999}};
%Phys. Rev. D60 (1999) 114033,
D. Rischke and R. Pisarski, {\PRD{61}{051501}{2000}}, 
%Phys. Rev. D61 (2000) 051501,
D. K. Hong, V. A. Miransky, 
I. A. Shovkovy, L.C. Wiejewardhana, {\PRD{61}{056001}{2000}}.}
% Phys. Rev. D61 (2000) 056001.}
\def\leblac {M. Le Bellac, {\it Thermal Field Theory}(Cambridge, Cambridge University
Press, 1996).}
\def\bcs{A.L. Fetter and J.D. Walecka, {\it Quantum Theory of Many
particle Systems} (McGraw-Hill, New York, 1971).}
\def\alexander{Aleksander Kocic, Phys. Rev. D33, 1785,(1986).}
\def\bubmix{F. Neumann, M. Buballa and M. Oertel,
{\NPA{714}{481}{2003}.}}
% Nucl. Phys. A714:481-501,2003.}
\def\kunihiro{M. Kitazawa, T. Koide, T. Kunihiro, Y. Nemeto,
{\PTP{108}{929}{2002}.}}
% Prog.  theo. Phys. 108, 929, 2002.}
\def\igor{Igor Shovkovy, Mei Huang, {\PLB{564}{205}{2003}}.}
\def\prasanth{P. Jaikumar and M. Prakash,{\PLB{516}{345}{2001}}.}
\def\igorr{Mei Huang, Igor Shovkovy, {\NPA{729}{835}{2003}}.}
%Phys. Rev. D67, 103004, 2003.}
\def\abrikosov{A.A. Abrikosov, L.P. Gorkov, Zh. Eskp. Teor.39, 1781,
1960}
\def\krischprl{M.G. Alford, J. Berges and K. Rajagopal,
 {\PRL{84}{598}{2000}.}}
%Phys. Rev. Lett. 84, 598, 2000.}
\def\hatmampp{A. Mishra and H.Mishra, in preparation}
\def\blaschke{D. Blaschke, M.K. Volkov and V.L. Yudichev,
{\EPJA{17}{103}{2003}}.}
% Eur. Phys. J. A17,103, 2003}
\def\mei{M. Huang, P. Zhuang, W. Chao,
{\PRD{65}{076012}{2002}}}
% Phys. Rev D65, 076012, 2002}
\def\bubnp{M. Buballa, M. Oertel,
{\NPA{703}{770}{2002}}.}
\def\sarma{G. Sarma, J. Phys. Chem. Solids 24,1029 (1963).}
% Nucl. Phys. A703, 770 (2002);
\def\ebert {D. Ebert, H. Reinhardt and M.K. Volkov,
Prog. Part. Nucl. Phys.{\bf 33},1, 1994.}
\def\ebertmag {D. Ebert, K.G. Klimenko, M.A. Vdovichenko and A.S. Vshivtsev,
{\PRD{61}{025005}{1999}.}}
\def\ferrerscmag{E.J. Ferrer, V. de la Incera and C. Manuel, {\PRL{95}{152002}{2005}};
E.J. Ferrer and V. de la Incera, {\PRL{97}{122301}{2006}}; E.J. Ferrer and
V. de la Incera, {\PRD{76}{114012}{2007}}.}
\def\ferrer{E. J. Ferrer, V. Incera, J. P. Keith and P. Springsteen, {\PRC{82}{065802}{2010}}.}
\def\rehberg{ P. Rehberg, S.P. Klevansky and J. Huefner,
{\PRC{53}{410}{1996}.}}
\def\lutz{M. Lutz, S. Klimt, W. Weise,{\NPA{542}{521}{1992}.}}
% Phys. Rev. C53,410 (1996).}
%\begin{references}
\def\rapid{B. Deb, A.Mishra, H. Mishra and P. Panigrahi,
Phys. Rev. A {\bf 70},011604(R), 2004.}
\def\kriscfl{M. Alford, C. Kouvaris, K. Rajagopal, Phys. Rev. Lett.
{\bf 92} 222001 (2004), arXiv:hep-ph/0406137.}
\def\shovris{S.B. Ruester, I.A. Shovkovy and D.H. Rischke,
arXiv:hep-ph/0405170.}
\def\spmindianj{S. P. Misra, Indian J. Phys. {\bf 70A}, 355 (1996).}
\def\kausik{K. Bhattacharya,arXiv:0705.4275[hep-th]; M. deJ. Aguiano-Galicia,
A. Bashir and A. Raya,{\PRD{76}{127702}{2007}}.}
\def\krisaug{K. Fukushima, C. Kouvaris and K. Rajagopal, arxiv:hep-ph/0408322}.
\def\wilczek{W.V. Liu and F. Wilczek,{\PRL{90}{047002}{2003}},E. Gubankova,
W.V. Liu and F. Wilczek, {\PRL{91}{032001}{2003}.}}
\def\review{For reviews see K. Rajagopal and F. Wilczek,
arXiv:hep-ph/0011333; D.K. Hong, Acta Phys. Polon. B32,1253 (2001);
M.G. Alford, Ann. Rev. Nucl. Part. Sci 51, 131 (2001); G. Nardulli,
Riv. Nuovo Cim. 25N3, 1 (2002); S. Reddy, Acta Phys Polon.B33, 4101(2002);
T. Schaefer arXiv:hep-ph/0304281; D.H. Rischke, Prog. Part. Nucl. Phys. 52,
197 (2004); H.C. Ren, arXiv:hep-ph/0404074; M. Huang, arXiv: hep-ph/0409167;
I. Shovkovy, arXiv:nucl-th/0410191.}
\def\kunihiroo{ M. Kitazawa, T. Koide, T. Kunihiro and Y. Nemoto,
{\PRD{65}{091504}{2002}}, D.N. Voskresensky, arXiv:nucl-th/0306077.}
\def\rupak{S.Reddy and G. Rupak, arXiv:nucl-th/0405054}
\def\ida{K. Iida and G. Baym,{\PRD{63}{074018}{2001}},
Erratum-ibid{\PRD{66}{059903}{2002}}; K. Iida, T. Matsuura, M. Tachhibana 
and T. Hatsuda, {\PRL{93}{132001}{2004}}; ibid,{arXiv:hep-ph/0411356}}
\def\chromo{Mei Huang and Igor Shovkovy,{\PRD{70}{051501}{2004}};
 {\em ibid}, {\PRD{70}{094030}{2004}}}
\def\steiner{A.W. Steiner, {\PRD{72}{054024}{2005}.}}
\def\andreaskris{K. Rajagopal and A. Schimitt, {\PRD{73}{045003}{2006}.}}
\def\amhm5{A. Mishra and H. Mishra, {\PRD{71}{074023}{2005}.}}
\def\leupold{K. Schertler, S. Leupold and J. Schaffner-Bielich,
{\PRC{60}{025801}(1999).}}
\def\bubrep{Michael Buballa, Phys. Rep.{\bf 407},205, 2005.}
\def\hatkun{T. Hatsuda and T. Kunihiro, Phys. Rep.{\bf 247},221, 1994.}
\def\lkw{ M. Lutz, S. Klimt and W. Weise, Nucl Phys. {\bf A542}, 521, 1992.}
\def\ruester{S.B. Ruester, V.Werth, M. Buballa, I. Shovkovy, D.H. Rischke,
arXiv:nucl-th/0602018; S.B. Ruester, I. Shovkovy, D.H. Rischke,
{\NPA{743}{127}{2004}.}}
\def\larrywarringa{D.Kharzeev, L. McLerran and H. Warringa, {\NPA{803}{227}{2008}};
K.Fukushima, D. Kharzeev and H. Warringa,{\PRD{78}{074033}{2008}}.}
\def\skokov{V. Skokov, A. Illarionov and V. Toneev, Int. j. Mod. Phys. A {\bf 24}, 5925,
(2009).}
\def\dima{D. Kharzeev, Ann. of Physics, K. Fukushima, M. Ruggieri and R. Gatto,
{\PRD{81}{114031}{2010}}.}
\def\fraga{A.J. Mizher, M.N. Chenodub and E. Fraga,arXiv:1004.2712[hep-ph].}
\def\maglat{M.D'Elia, S. Mukherjee and F. Sanflippo,{\PRD{82}{051501}{2010}.}}
\def\fukushimaplb{K. Fukushima, M. Ruggieri and R. Gatto, {\PRD{81}{114031}{2010}}.}
\def\igormag{E.V. Gorbar, V.A. Miransky and I. Shovkovy,{\PRC{80}{032801(R)}{2009}};
ibid, arXiv:1009.1656[hep-ph].}
\def\gorbar2000{E.V. Gorbar, {\PRD{62}{014007}{2000}}.}
\def\miranski{V.P. Gusynin, V. Miranski and I. Shovkovy,{\PRL{73}{3499}{1994}};
{\PLB{349}{477}{1995}}; {\NPB{462}{249}{1996}}, E.J. Ferrer and V de la 
Incerra,{\PRL{102}{050402}{2009}}; {\NPB{824}{217}{2010}.}}
\def\providencia{D.P. Menezes, M. Benghi Pinto, S.S. Avancini and C. Providencia
,{\PRC{80}{065805}{2009}}; D.P. Menezes, M. Benghi Pinto, S.S. Avancini , A.P. Martinez
and C. Providencia, {\PRC{79}{035807}{2009}}}
\def\boomsma{J. K. Boomsma and D. Boer, {\PRD{81}{074005}{2010}}}
\def\somenath{D. Bandyopadhyaya, S. Chakrabarty and S. Pal, {\PRL{79}{2176}{1997}};
S. Chakrabarty, S. Mandal, {\PRC{75}{015805}{2007}}}
\def\klimenko{D. Ebert and K.G. Klimenko,{\NPA{728}{203}{2003}}}
\def\fukuwarringa{K. Fukushima and H. J. Warringa, {\PRL{100}{032007}{2008}}.}
\def\noronah{J. Noronah and I. Shovkovy,{\PRD{76}{105030}{2007}}.}
\def\armendirk{X.G. Huang, M. Huang, D.H. Rischke and A. Sedrakian, 
{\PRD{81}{045015}{2010}}.}
\def\metlitsky{M. A. Metlitsky and A. R. Zhitnitsky,{\PRD{72}{045011}{2005}}.}
\def\digal{T. Mandal, P. Jaikumar and S. Digal, arXiv:0912.1413 [nucl-th]; T. Mandal and P. Jaikumar,
{\PRC{87}{045208}{2013}i}; T. Mandal and P. Jaikumar, {\PRD{94}{074016}{2016}}.}
\def\dunc{ R. C. Duncan and C. Thompson, Astrophys. J. 392, L9 (1992).}
\def\duncc {C. Thompson and R. C. Duncan, Astrophys. J. 408, 194 (1993).}
 \def\dunccc{C. Thompson and R. C. Duncan, Mon. Not. R. Astron.  Soc. 275, 255 (1995).}
 \def\duncccc{C. Thompson and R. C. Duncan, Astrophys. J. 473, 322 (1996).}
 \def\kouvel{C. Kouveliotou et al., Astrophys. J. 510, L115 (1999).}
 \def\lat{C. Y. Cardall, M. Prakash, and J. M. Lattimer, Astrophys.  J. 554, 322 (2001).}
 \def\broder{A. E. Broderick, M. Prakash, and J. M. Lattimer, {\PLB{531}{167}{2002}}.}
 \def\lai{D. Lai and S. L. Shapiro, Astrophys. J. 383, 745 (1991).}
\def\dune{G.Baser, G. Dunne and D. Kharzeev, {\PRL{104}{232301}{2010}}.}
\def\frolov{I.E. Frolov, V. Ch. Zhukovsky and K.G. Klimenko, {\PRD{82}{076002}{2010}}.}
\def\nickel{D. Nickel,{\PRD{80}{074025}{2009}}.}
\def\gato{R. Gatto and M. Ruggieri, {\PRD{82}{054027}{2010}}.}
\def\iran{Sh. Fayazbakhsh and N. Sadhooghi, {\PRD{82}{045010}{2010}};{\em ibid} {\PRD{83}{025026}{2011}.}}
\def\hongmag{Deog Ki Hong, arXiv:1010.3923[hep-th].}
\def\bhamhmppn{B. Chatterjee, H. Mishra and A. Mishra; in preparation}
\def\amhmdet{Amruta Mishra and Hiranmaya Mishra, {\PRD{74}{054024}{2006}}.}
\def\fattoyev{F.J. Fattoyev, J. Piekarewicz and C.J. Horowitz, {\PRL{120}{172702}{2018}}.}
\def\buballarev{M. Buballa, {\PR{407}{205}{2005}}.}
\def\berges{M.G. Alford, J. Berges and K. Rajagopal, {\NPA{39}{6349}{2006}}.}
\def\scoccola{M. Coppola, P. Allen, A.G. Grunfeld and K. N.N. Scoccola, {\PRD{96}{056013}{2017}}.}
\def\andreas1{F. Preis, A. Rebhan, and A. Schmitt, J. High Energy Phys.{\bf 03},033 (2011)}
\def\dunne{G. Baser, G. Dunne, and D. Kharzeev,{\PRL{104}{232301}{2010}}}
\def\abuki{H. Abuki, {\PRD{98}{054006}{2018}}}
\def\huang2004{M. Huang and I. Shovkovy,{\PRD{70}{094030}{2004}}}
\def\rajasharma2006{K. Rajagopal and R. Sharma,{\PRD{74}{094019}{2016}}}
\def\bowers2002{J.A. Bowers and K. Rajagopal,{\PRD{66}{065002}{2002}}}
\def\giannakis{I. Giannakis and H.-C. Ren, {\PLB{611}{137}{2005}}}
\def\fukushima2005{K. Fukushima,{\PRD{72}{074002}{2005}}}
\def\sreemoyee{S. Sarkar and R. Sharma,{\PRD{96}{094025}{2017}}}

\end{document}